\documentclass[secnum,seceqn,secthm]{elsart}

\usepackage{natbib}
\usepackage{epsfig}
\usepackage{amssymb}

\def\4R{{}^{(4)}R}
\def\Lie{{\mathcal L}}

\makeatletter
 \@addtoreset{figure}{section}
 
 \@addtoreset{table}{section}
 
\makeatother

\begin{document}

\begin{frontmatter}

\title{Numerical Relativity and Compact Binaries}

\author[ad1,ad2]{Thomas W. Baumgarte} and
\author[ad2,ad3]{Stuart L. Shapiro}

\address[ad1]{Department of Physics and Astronomy, 
	Bowdoin College, Brunswick, ME 04011}

\address[ad2]{Department of Physics, University of Illinois
	at Urbana-Champaign, Urbana, IL 61820}

\address[ad3]{Department of Astronomy and NCSA, University of Illinois
	at Urbana-Champaign, Urbana, IL 61820}

\begin{abstract}
Numerical relativity is the most promising tool for theoretically
modeling the inspiral and coalescence of neutron star and black hole
binaries, which, in turn, are among the most promising sources of
gravitational radiation for future detection by gravitational wave
observatories.  In this article we review numerical relativity
approaches to modeling compact binaries.  Starting with a brief
introduction to the 3+1 decomposition of Einstein's equations, we
discuss important components of numerical relativity, including the
initial data problem, reformulations of Einstein's equations,
coordinate conditions, and strategies for locating and handling black
holes on numerical grids.  We focus on those approaches which
currently seem most relevant for the compact binary problem.  We then
outline how these methods are used to model binary neutron stars and
black holes, and review the current status of inspiral and coalescence
simulations.
\end{abstract}


\end{frontmatter}

\tableofcontents

\section{Introduction}
\label{sec1}

Promising to open a new window to the universe, a new generation of
laser interferometer gravitational wave detectors will soon search for
gravitational radiation.  The Japanese instrument TAMA is already in
operation \citep{tetal01}, and the construction of the two LIGO sites
in the US and the European instruments GEO and VIRGO is well advanced.
Among the most promising sources for these detectors are the inspiral
and merger of compact binaries, i.e.~binaries of black holes and
neutron stars.  Even for these sources, the signal strength is
expected to be much less than the detectors' noise, so that
sophisticated data analysis techniques will be required to extract the
signal from the noise (e.g.~\citet{cetal93}).  One such technique is
matched filtering, in which the detector output is cross-correlated
with a catalog of theoretically predicted waveforms.  The
cross-correlation between a signal present in the data and a member of
the catalog allows observers to detect signals that are otherwise
overwhelmed by noise.  Clearly, the chances of detecting a generic
astrophysical signal depend critically on the size and quality of the
signal catalog \citep{fh98}.  The success of the new gravitational
wave interferometers therefore depends on accurate theoretical models
of compact binary inspiral.

For our purposes, the entire inspiral of compact binaries can be
separated into three different phases.  By far the longest phase is
the initial quasi-equilibrium {\it inspiral} phase, during which the
separation between the stars decreases adiabatically as energy is
carried away by gravitational radiation.  As the separation decreases,
the frequency and amplitude of the emitted gravitational radiation
increases.  Since gravitational radiation tends to circularize binary
orbits, and since we will be mostly interested in binaries at very
small separation, it is reasonable to focus on quasi-circular orbits.
These quasi-circular orbits become unstable at the innermost stable
circular orbit (ISCO), where the inspiral gradually enters a {\it
plunge and merger} phase.  The merger and coalescence of the stars
happens on a dynamical timescale.  The final stage of the evolution is
the {\it ringdown} phase, during which the merged object settles down
to equilibrium.

Different techniques are commonly employed to model the binary in the
different phases.  The early inspiral phase, for large binary
separations, can be modeled very accurately with post-Newtoni\-an
methods \citep[and references therein]{bdiww95,djs00}.  It is
generally accepted that the plunge and merger phase has to be
simulated by means of a fully self-consistent numerical relativity
simulation.  During the late ringdown phase, the merged object can be
approximated as a distorted equilibrium object, so that perturbative
techniques can be applied \citep{pp94,bbclt01,bclt02}.  In addition to
simulating the dynamical plunge and merger phase, numerical relativity
may also be required for the late quasi-adiabatic inspiral phase, just
outside of the ISCO, where finite-size and relativistic effects may
become large enough for post-Newtonian point-mass techniques to break
down.

A number of review articles have recently appeared on various aspects
of numerical relativity, including \citet{l01} on numerical relativity
in general, \citet{c00} on initial data, \citet{mm99} and
\citet{f00} on numerical hydrodynamics in special and general
relativity, \citet{rs99} on the coalescence of binary neutron stars,
\citet{r98} on hyperbolic methods of Einstein's equations, and 
\citet{n02} on the generation of gravitational waves from gravitational
collapse.  While we will refer to these articles in many places, the
perspective and focus of this article is very different.  It is
intended as a review of different numerical relativity approaches to
the compact binary problem, and we will focus on those methods of
numerical relativity that currently seem most promising for these
purposes.  Specifically, we will only discuss the so-called Cauchy
approach in ``3+1'', i.e.~three spatial dimensions (``3D'') plus
time\footnote{We will provide examples of spherically symmetric
configurations to illustrate concepts and techniques.  These examples,
which can be treated in 1+1 dimensions, are clearly marked and set in
{\it italics}.}, and we limit matter sources, if present, to perfect
fluids.  This means that many other promising techniques and important
results of numerical relativity will not be covered, including
characteristic methods\footnote{Which in fact may be a very promising
alternative to the Cauchy approach even for the modeling of compact
binaries; see, e.g.~\citet{getal98}.}, collisionless matter and scalar
wave sources, critical phenomena, and results in spherical or
axisymmetry.  Simultaneously, we will review numerical relativity
approaches to the compact binary problem only, and will not discuss
post-Newtonian or perturbative methods.

Even with this fairly focused scope of our article we will omit some
important aspects.  Perhaps most importantly, we will not cover wave
extraction and boundary conditions for outer boundaries, since in most
current applications in 3+1 these are only implemented fairly crudely.
Some relevant references to more rigorous treatments include
\citet{bghmpw96,aetal98,fn99,sgbw00,clt01,ssw02}.  While we will cast
equations into a form that is suitable for numerical implementation,
we will hardly discuss numerical methods that can be used for their
solution.  We will similarly ignore related computer science issues,
including memory and CPU requirements and computational resources on
current parallel computers.  Some discussion of these aspects and
references can be found in \citet{l01}.

Loosely speaking, this article is organized into two parts.  The first
part, Sections~\ref{sec2} through \ref{sec6}, introduces concepts and
techniques of numerical relativity, both traditional approaches and
more recent developments.  The second part, Sections~\ref{sec7}
through \ref{sec10}, reviews their applications to binary black holes
and neutron stars.

The first part starts by introducing the $3+1$ decomposition and the
so-called ADM equations in Section~\ref{sec2}.  We will see that, like
Maxwell's equations, these equations separate into constraint and
evolution equations.  In Section~\ref{sec3} we discuss strategies for
solving the constraint equations and the construction of initial data.
In Section~\ref{sec4} we will revisit the ADM equations, and will show
that the evolution equations can be brought into a form that is better
suited for numerical integrations.  Before these evolution equations
can be integrated, certain coordinate conditions have to be imposed,
which we will discuss in Section \ref{sec5}.  In Section \ref{sec6} we
review the definition of black hole horizons, and show how these
horizons can be located in numerically generated spacetimes.

In the second part we discuss the construction of binary black holes
and neutron stars.  Specifically, we review the initial data problem
and evolution simulations of binary black holes in Sections \ref{sec7}
and \ref{sec8} and binary neutron stars in Sections \ref{sec9} and
\ref{sec10}.  We briefly summarize in Section \ref{sec11}.

Also included are three short appendixes.  Appendix \ref{appA}
explains our notation, Appendix \ref{appB} shows how the flat vector
Laplace operator, which is encountered in several places in this
article, can be solved, and Appendix \ref{appC} discusses some
arguments which have been made in favor or against the approximation
of spatial conformal flatness.

As a last remark we would like to apologize to all those authors and
colleagues whose publications we may have missed.  While we have put
considerable effort into being as complete as possible, without doubt
we have missed some important references.  We again apologize for
these unintended omissions.

\section{Decomposing Einstein's Equations}
\label{sec2}

In this Section we briefly summarize the Arnowitt-Deser-Misner or
``ADM'' decomposition of Einstein's equations.  We will state only the
most important relations and results and refer to the literature for
more complete and rigorous treatments.  In addition to the original
article by \citet{adm62} such presentations can be found, for example,
in the article by \citet{y79}, the dissertations by \citet{s75},
\citet{e84} and \citet{c90}, and in the lecture notes of
\citet{bsa98}.

\subsection{Foliations of Spacetime}
\label{sec2.1}

The unification of space and time into spacetime is central to general
relativity and is one of its greatest aesthetic appeals.  For a 
numerical treatment, however, and similarly for many mathematical
treatments, it is more desirable to reverse this unification and 
recast general relativity into a so-called ``3+1'' formulation, in 
which a time coordinate is explicitly split from three spatial
coordinates.  Putting it differently, the four-dimensional spacetime
is ``carved up'' into a family of three-dimensional spatial ``slices''.

In more technical terms, we assume that the spacetime $(M,g_{ab})$ can
be foliated into a family of non-intersecting, spacelike,
three-dimensional hypersurfaces $\Sigma$.  At least locally, these
timeslices $\Sigma$ form level surfaces of a scalar function $t$,
which we will later identify with the coordinate time.  The 1-form
\begin{equation}
{\bf \Omega} = {\bf d} t
\end{equation}
is obviously closed (${\bf d \Omega} = 0$) and has the norm
\begin{equation}
|\Omega|^2 = g^{ab} \nabla_a t \nabla_b t \equiv - \alpha^{-2},
\end{equation}
where the {\em lapse function} $\alpha$ is strictly positive (see
Appendix \ref{appA} for a summary of our notation).  This implies that
the surfaces $\Sigma$ are spacelike.  We now define the timelike 
unit normal vector $n^a$ as
\begin{equation}
n^a \equiv - \alpha g^{ab} \Omega_b = - \alpha g^{ab} \nabla_b t.
\end{equation}
Here the negative sign has been chosen so that $n^a$ points into the
direction of increasing $t$.  The four-dimensional metric $g_{ab}$ now
induces the {\em spatial metric}
\begin{equation} \label{2_def_gamma}
\gamma_{ab} \equiv g_{ab} + n_a n_b
\end{equation}
on the hypersurfaces $\Sigma$.  

Any four-dimensional tensor can now be decomposed into spatial parts,
which live in the hypersurfaces $\Sigma$, and timelike parts, which
are normal to $\Sigma$ and hence aligned with $n^a$.  The spatial part
can be found by contracting with the projection operator
\begin{equation}
\gamma^a_{~b} = g^{ac} \gamma_{cb} = 
g^a_{~b} + n^a n_b = \delta^a_{~b} + n^a n_b,,
\end{equation}
and the timelike part by contracting with
\begin{equation}
N^a_{~b} = - n^a n_b.
\end{equation}

We now define the three-dimensional covariant derivative of a spatial
tensor by projecting all indices of a four-dimensional covariant
derivative into $\Sigma$; for example
\begin{equation}
D_a T^b_{~c} \equiv
\gamma_a^{~d} \gamma_e^{~b} \gamma_c^{~f} \nabla_d T^e_{~f}.
\end{equation}
It is easy to show that this derivative is compatible with the spatial
metric, $D_a \gamma_{bc} = 0$, as it is supposed to be.  The
three-dimensional covariant derivative can be expressed in terms of
three-dimensional connection coefficients, which, in a coordinate
basis, are given by
\begin{equation} \label{2_connection}
\Gamma^a_{bc} = \frac{1}{2} \gamma^{ad}(
\gamma_{db,c} + \gamma_{dc,b} - \gamma_{bc,d}).
\end{equation}
The three-dimensional Riemann tensor associated with $\gamma_{ij}$ is
defined by requiring that
\begin{equation}
2 D_{[a} D_{b]} w_c = R^d_{~cba} w_d~~~~~~~~~~~~R^d_{~cba} n_d = 0 
\end{equation}
for every spatial $w_d$.  In terms of coordinate components, Riemann
can be computed from
\begin{equation} \label{2_riemann}
R_{abc}^{~~~d} = \Gamma^{d}_{ac,b} - \Gamma^{d}_{bc,a} 
	+ \Gamma^e_{ac}\Gamma^d_{eb} - \Gamma^e_{bc} \Gamma^d_{ea}.
\end{equation} 
Contracting the Riemann tensor yields the three-di\-men\-sional Ricci
tensor $R_{ab} = R^c_{~acb}$ and the three-dimensional scalar
curvature $R = R^a_{~a}$.

It is intuitive that casting Einstein's equations into a $3+1$ form
will necessitate expressing the four-dimensional Riemann tensor
$\4R^d_{~abc}$ in terms its three-dimensional cousin $R^d_{~abc}$.  It
is also clear, however, that the latter cannot contain all the
relevant information: $R^d_{~abc}$ is a purely spatial object (and can
be computed from spatial derivatives of the spatial metric alone),
while $\4R^d_{~abc}$ is a spacetime creature which also contains
time-derivatives of the four-dimensional metric.  This missing
information is expressed by a tensor called {\em extrinsic curvature},
which describes how the slice $\Sigma$ is embedded in the spacetime
$M$.  The extrinsic curvature can be defined as
\begin{equation}
K_{ab} \equiv - \gamma_a^{~c} \gamma_b^{~d} \nabla_{(c} n_{d)}.
\end{equation}
Note that $K_{ab}$ is spatial and symmetric by construction.  An
alternative expression can be found in terms of the ``acceleration''
of normal observers $a_a = n^b \nabla_b n_a$,
\begin{equation}
K_{ab} = - \nabla_a n_b - n_a a_b.
\end{equation}
Since $a^a n_a = 0$, we immediately find for the trace of the 
extrinsic curvature
\begin{equation} \label{2_trace_K}
K \equiv g^{ab} K_{ab} = - \nabla^a n_a.
\end{equation}
In yet another equivalent expression, the extrinsic curvature can be
written in terms of the Lie derivative of the spatial metric along the
normal vector $n^a$
\begin{equation} \label{2_ext_curv}
K_{ab} = - \frac{1}{2} \Lie_{\bf n} \gamma_{ab}.
\end{equation}
The Lie derivative in the above equation may be thought of as the
geometric generalization of the partial time derivative $\partial_t$.
Introductions to the Lie derivative can be found, for example, in
\citet{s80,w84,d92,bsa98}. Formally, the Lie derivative along a vector
field $X^a$ measures by how much the changes in a tensor field along
$X^a$ differ from a mere infinitesimal coordinate transformation
generated by $X^a$.  For a scalar $f$, the Lie derivative reduces to
the partial derivative
\begin{equation}
\Lie_{\bf X} f = X^b D_b f = X^b \partial_b f;
\end{equation}
for a vector field $v^a$ the Lie derivative is the commutator
\begin{equation} \label{Lie_vec}
\Lie_{\bf X} v^a = X^b D_b v^a - v^b D_b X^a = [X,v]^a,
\end{equation}
and for a 1-form $\omega_a$ the Lie derivative is given by
\begin{equation}
\Lie_{\bf X} \omega_a = X^b D_b \omega_{a} + \omega_b D_a X^b.
\end{equation}
It then follows that for a tensor $T^a_{~b}$ of rank $({}^1_1 )$ the Lie 
derivative is
\begin{equation} \label{2_lie}
\Lie_{\bf X} T^a_{~b} = X^c \partial_c T^a_{~b} - T^c_{~b} \partial_c X^a +
 T^a_{~c} \partial_b X^c.
\end{equation}
Generalization to tensors of arbitrary rank follows naturally.  It is
easy to verify that in all of the above expressions for the Lie
derivative one may replace the partial derivative with a covariant
derivative, since all connection coefficients cancel each other.

Equation (\ref{2_ext_curv}) most clearly illustrates the
interpretation of $K_{ab}$ as a time-derivative of the spatial metric.
It is therefore not surprising that spatial projections of
$\4R^a_{~bcd}$ will involve the extrinsic curvature and its time
derivative.

Given its symmetry properties, $\4R^a_{~bcd}$ can be projected in
three different ways.  Projecting all four indices into $\Sigma$
yields, after some manipulations, {\em Gauss' equation},
\begin{equation} \label{2_gauss}
R_{abcd} + K_{ac} K_{bd} - K_{ad} K_{bc} 
= \gamma^p_{~a} \gamma^q_{~b} \gamma^r_{~c} \gamma^s_{~d} \4R_{pqrs}.
\end{equation}
Three spatial projections and a contraction with $n^a$ yields the {\em
Codazzi equation}
\begin{equation} \label{2_codazzi}
D_a K_{bc} - D_b K_{ac}  = \gamma^r_{~b} \gamma^p_{~a} \gamma^q_{~c} n^s 
\4R_{rpqs}.
\end{equation}
Finally, two spatial projections and two contractions with $n^a$ yield
{\em Ricci's equation}
\begin{equation} \label{2_ricci}
\Lie_{\bf n} K_{ab} = n^d n^c \gamma^q_{~a} \gamma^r_{~b} \4R_{rdqc}
	- \frac{1}{\alpha} D_a D_b \alpha - K^c_{~b} K_{ac}.
\end{equation}
In this equation the derivative of the lapse entered through the 
identity
\begin{equation} \label{2_a}
a_a = D_a \ln \alpha.
\end{equation}

\subsection{The ADM equations}
\label{sec2.2}

In the last Section we simply recorded geometrical identities relating
the geometry of the three-dimensional hypersurfaces $\Sigma$ to the
geometry of the embedding four-dimensional spacetime $M$.  According
to general relativity, the geometry of the latter is governed
dynamically by Einstein's equation
\begin{equation} \label{2_einstein}
{}^{(4)}G_{ab} \equiv \4R_{ab} - \frac{1}{2} \4R\, g_{ab} = 8 \pi T_{ab},
\end{equation}
where ${}^{(4)}G_{ab}$ is the Einstein tensor and $T_{ab}$ the
stress-energy tensor.  We will now take projections of Einstein's
equations into $\Sigma$ and $n^a$ and will use the Gauss, Codazzi and
Ricci equations to eliminate the four-dimensional Ricci tensor
$\4R_{ab}$.  The result will be the ADM equations, which relate
three-dimensional curvature quantities to projections of the
stress-energy tensor.  One relation that is very useful in these
derivations is
\begin{equation} \label{2_proj_deriv}
D_a V^b = \gamma_a^{~c} \nabla_c V^b + K_{ac} V^c n^b,
\end{equation}
which holds for any spatial vector $V^a$.

Contracting Gauss' equation (\ref{2_gauss}) twice with the spatial
metric and inserting~(\ref{2_def_gamma}) yields
\begin{equation}
2 n^a n^b \, {}^{(4)}G_{ab} = R + K^2 - K_{ab} K^{ab}.
\end{equation}
We now define the total energy density as measured by a normal
observer $n^a$ as
\begin{equation} \label{2_rho}
\rho \equiv n^a n^b T_{ab}
\end{equation}
and, using Einstein's equation (\ref{2_einstein}), find the 
{\em Hamiltonian constraint}
\begin{equation} \label{2_ham_1}
R + K^2 - K_{ab} K^{ab} = 16 \pi \rho.
\end{equation}

We can similarly contract the Codazzi equation~(\ref{2_codazzi}) once
to find the {\em momenum constraint}
\begin{equation} \label{2_mom_1}
D_b K^b_{~a} - D_a K = 8 \pi j_a,
\end{equation}
where
\begin{equation} \label{2_j}
j_a \equiv - \gamma^b_{~a} n^c T_{bc}
\end{equation}
is the momentum density (mass current) as measured by a normal
observer $n^a$.  The Hamiltonian and momenum constraints are called
constraint equations because they only involve spatial quantities and
their spatial derivatives.  They therefore have to hold on each
individual spatial slice $\Sigma$ -- in fact they are the necessary
and sufficient integrability conditions for the embedding of the
spatial slices $(\Sigma,\gamma_{ab},K_{ab})$ in the spacetime
$(M,g_{ab})$.

Evolution equations that describe how data $\gamma_{ab}$ and $K_{ab}$
evolve in time, from one spatial slice to the next, can be found from
equation (\ref{2_ext_curv}) and the Ricci equation (\ref{2_ricci}).
However, the Lie derivative along $n^a$, $\Lie_{\bf n}$, is not a
natural time derivative orthogonal to the spatial slices, since $n^a$
is not dual to the surface 1-form ${\bf \Omega}$, i.e.~their dot
product is not unity but rather
\begin{equation}
n^a \Omega_a = - \alpha g^{ab} \nabla_b t \nabla_a t = \alpha^{-1}.
\end{equation}
Instead, the vector
\begin{equation} \label{2_t}
t^a = \alpha n^a + \beta^a
\end{equation}
is dual to ${\bf \Omega}$ for any spatial {\em shift vector}
$\beta^a$.  The lapse $\alpha$ and the shift $\beta^a$ together
determine how the coordinates evolve from one slice $\Sigma$ to the
next.  The lapse determines how much proper time elapses between
timeslices along the normal vector $n^a$, while the shift determines
by how much spatial coordinates are shifted with respect to the normal
vector.

The Lie derivative along $t^a$ is a natural time derivative, because
the duality
\begin{equation}
t^a \Omega_a = t^a \nabla_a t = 1 
\end{equation}
implies that the integral curves of $t^a$ are naturally parametrized
by $t$.  As a consequence, all (infinitesimal) vectors $t^a$
originating on one spatial slice $\Sigma_1$ will end on the same
spatial slice $\Sigma_2$ (unlike the corresponding vectors $n^a$,
which generally would end on different slices).  This also implies
that the Lie derivative of any spatial tensor along $t^a$ is again
spatial (see, e.g., Problem 8.14 in \citet{lppt75} for an
illustration).

Rewriting equation~(\ref{2_ext_curv}) in terms of $t^a$ yields
the evolution equation for the spatial metric
\begin{equation} \label{2_gdot_1}
\Lie_{\bf t} \gamma_{ab} = 
-2 \alpha K_{ab} + \Lie_{\beta} \gamma_{ab}.
\end{equation}
The evolution equation for the extrinsic curvature can be found by
combining Ricci's equation~(\ref{2_ricci}) with Einstein's
equations~(\ref{2_einstein})
\begin{equation} \label{2_Kdot_1}
\begin{array}{rcl} 
\Lie_{\bf t} K_{ab} & = &  - D_a D_b \alpha + 
\alpha (R_{ab} - 2 K_{ac} K^c_{~b} + K K_{ab}) \\ 
& & - \alpha 8 \pi (S_{ab} - \frac{1}{2} \gamma_{ab}(S - \rho)) +
\Lie_{\bf \beta} K_{ab},
\end{array}
\end{equation}
where $S_{ab}$ is the spatial projection of the stress-energy tensor
\begin{equation} \label{2_S}
S_{ab} \equiv \gamma_{ac} \gamma_{bd} T^{cd},
\end{equation}
and $S$ its trace $S \equiv \gamma^{ab} S_{ab}$. 

While the two constraint equations~(\ref{2_ham_1}) and~(\ref{2_mom_1})
constrain $\gamma_{ab}$ and $K_{ab}$ on every spatial slice $\Sigma$,
the evolution equations~(\ref{2_gdot_1}) and~(\ref{2_Kdot_1}) describe
how these quantities evolve from one slice to the next.  It can be
shown that the evolution equations preserve the constraint equations,
meaning that if the constraints hold on one slice, they will continue
to hold on later slices.  This structure is very similar to that
of Maxwell's equations, as we will discuss in Section \ref{sec2.4}.

So far, we have made no assumptions about the choice of coordinates,
and have expressed all quantities in a coordinate-independent way.  It
is quite intuitive, though, that things will simplify if we adopt a
coordinate system that reflects our 3+1 split of spacetime.  To do so,
we introduce a basis of spatial vectors $e_i$ (where $i = 1,2,3$, see
Appendix \ref{appA} for a summary of our notation) that span each
slice $\Sigma$, so that $\Omega_a (e_i)^a = 0$.  It can be shown that
this condition is preserved if the spatial vectors are Lie dragged
along $t^a$.  As the fourth basis vector we pick $(e_0)^a = t^a$, or,
in components, $t^a = (1,0,0,0)$.  As an immediate consequence, the
Lie derivative along $t^a$ reduces to a partial derivative with
respect to $t$.  Since $n_i = n_a (e_i)^a =
\alpha \Omega_a (e_i)^a = 0$, the covariant, spatial components of the
normal vector vanish.  Since contractions of any spatial tensor with
the normal vector are zero, this implies that the zeroth components of
contravariant spatial tensors have to vanish.  For the shift vector,
for example, $\beta^a n_a = \beta^0 n_0 = 0$, so that we can write
\begin{equation}
\beta^a = (0,\beta^i).
\end{equation}
Solving (\ref{2_t}) for $n^a$ then yields the contravariant components
of the shift vector
\begin{equation}
n^a = \frac{1}{\alpha} (1, -\beta^i)
\end{equation}
and, since $n^a n_a = -1$,
\begin{equation}
n_a = (- \alpha, 0,0,0).
\end{equation}
From the definition of the spatial metric (\ref{2_def_gamma}) we find
$\gamma_{ij} = g_{ij}$.  Since zeroth components of contravariant,
spatial tensors are zero, we also have $\gamma^{a0} = 0$.  The inverse
metric can therefore be expressed as
\begin{equation} \label{2_cont_metric}
g^{ab} = 
\left( \begin{array}{ccc}
- \alpha^{-2}  		& ~~ & \alpha^{-2}\beta^i \\
\alpha^{-2}\beta^j 	& ~~ &\gamma^{ij} - \alpha^{-2}\beta^i \beta^j
\end{array}
\right)
\end{equation}
Using $n_i = 0$, $\gamma_{ij} = g_{ij}$ and $\gamma^{a0} = 0$ it is
also possible to show that $\gamma_{ij}$ and $\gamma^{ij}$ are 
inverses $\gamma^{ik}\gamma_{kj} = \delta^i_{~j}$, so that they can
be used to raise and lower indices of spatial tensors.  
Inverting~(\ref{2_cont_metric}) we find the components of the 
four-dimensional metric
\begin{equation} \label{2_metric}
g_{ab} = 
\left( \begin{array}{ccc}
- \alpha^{2} + \beta_k\beta^k  		& ~~ & \beta_i \\
\beta_j 	& ~~ &\gamma_{ij}
\end{array}
\right)
\end{equation}
and equivalently the line element
\begin{equation} \label{2_final_metric}
ds^2 = - \alpha^2 dt^2 + \gamma_{ij} (dx^i + \beta^i dt) (dx^j + \beta^j dt).
\end{equation}

The entire content of a spatial tensor is encoded in its spatial
components alone.  We can therefore restrict the constraint and
evolution equations to spatial components.  Moreover, since in our
coordinate system the zeroth components of contravariant, spatial
tensors are zero, we can also restrict all contractions to spatial
components.  The connection coefficients~(\ref{2_connection}) reduce
to
\begin{equation} \label{2_connection_2}
\Gamma^i_{jk} = \frac{1}{2} \gamma^{il}(
\gamma_{lj,k} + \gamma_{lk,j} - \gamma_{jk,l}),
\end{equation}
and expressing the Ricci tensor~(\ref{2_riemann}) in terms of second
derivatives of the metric yields
\begin{equation} \label{2_ricci_2}
\begin{array}{rcl}
R_{ij} & = & \displaystyle \frac{1}{2} \gamma^{kl} 
        \Big( \gamma_{kj,il} + \gamma_{il,kj} 
                - \gamma_{kl,ij} - \gamma_{ij,kl} \Big) \\
        & & + \gamma^{kl} \Big( \Gamma^m_{il} \Gamma_{mkj}
        - \Gamma^m_{ij} \Gamma_{mkl} \Big)
\end{array}
\end{equation}

With these simplifications, the Hamiltonian constraint~(\ref{2_ham_1})
now becomes
\begin{equation} \label{2_ham_2}
R + K^2 - K_{ij} K^{ij} = 16 \pi \rho,
\end{equation}
the momenum constraint 
\begin{equation} \label{2_mom_2}
D_j K^j_{~i} - D_i K = 8 \pi j_i,
\end{equation}
the evolution equation for the spatial metric~(\ref{2_gdot_1})
\begin{equation} \label{2_gdot_2}
\partial_t \gamma_{ij} = - 2 \alpha K_{ij} + D_i \beta_j + D_j \beta_i,
\end{equation}
and the evolution equation for the extrinsic curvature~(\ref{2_Kdot_1})
\begin{equation} \label{2_Kdot_2}
\begin{array}{rcl} 
\partial_t K_{ij} & = &  - D_i D_j \alpha + 
\alpha (R_{ij} - 2 K_{ik} K^k_{~j} + K K_{ij}) \\ 
& & \displaystyle
- \alpha 8 \pi (S_{ij} - \frac{1}{2} \gamma_{ij}(S - \rho)) \\
& &  +
\beta^k D_k K_{ij} + K_{ik} D_j \beta^k + K_{kj} D_i \beta^k.
\end{array}
\end{equation}
The shift terms in
the last two equations arise from the Lie derivatives $\Lie_{\beta}
\gamma_{ij}$ and $\Lie_{\beta} K_{ij}$.  In (\ref{2_gdot_2}) it is
convenient to express the Lie derivative in terms of the covariant
derivative $D_i$ which eliminates the term $\beta^k D_k \gamma_{ij}$,
but in (\ref{2_Kdot_2}) the covariant derivative may be replaced with
the partial derivative $\partial_i$ in these terms.

Equations (\ref{2_ham_2}) to (\ref{2_Kdot_2}) are commonly referred to
as the ADM equations\footnote{\citet{adm62} originally derived these
equations in terms of the conjugate momenta $\pi_{ij}$ instead of the
extrinsic curvature.}.  As it turns out, these equations are
not yet in a form that is generally well suited for numerical
implementation.  In Section \ref{sec4} we will see that the stability
properties of numerical implementations can be improved by introducing
certain new auxiliary functions and rewriting the ADM equations in
terms of these functions.  Most current numerical relativity codes are based on
such reformulations of the ADM equation.

Note that the ADM equations only determine the spatial metric
$\gamma_{ij}$ and the extrinsic curvature $K_{ij}$, but not the lapse
$\alpha$ or the shift $\beta^i$.  The latter determine how the
coordinates evolve from one timeslice to the next and reflect the
coordinate freedom of general relativity.  Choosing coordinates that
are suitable for the situation that one wishes to simulate is central
to the success of the simulation.  In Section~\ref{sec5} we will
discuss strategies for both choosing and numerically implementing
various coordinate conditions.

For later purposes it is useful to take the traces of the evolution
equations  (\ref{2_gdot_2}) and (\ref{2_Kdot_2}).  Since
$\partial_t \ln \gamma = \gamma^{ij} \partial_t \gamma_{ij}$, we 
find
\begin{equation} \label{2_trgamma}
\partial_t \ln \gamma^{1/2} = - \alpha K + D_i \beta^i,
\end{equation}
for the determinant of the metric $\gamma$, and combining the
Hamiltonian constraint (\ref{2_ham_2}) with the trace of
(\ref{2_Kdot_2}) yields
\begin{equation} \label{2_trK}
\partial_t K = - D^2 \alpha + \alpha \left(
K_{ij} K^{ij} + 4 \pi (\rho + S) \right) + \beta^i D_i K.
\end{equation}
Here $D^2 \equiv \gamma^{ij} D_i D_j$ is the covariant Laplace
operator associated with $\gamma_{ij}$.

\subsection{Electrodynamics}
\label{sec2.4}

To discuss the structure of the ADM equations it is very instructive
to compare them to the equations of electrodynamics.  Maxwell's
equations in flat space naturally split into a set of {\em constraint}
equations
\begin{eqnarray}
D_i E^i & = & 4 \pi \rho_e \label{2_Econst}\\
D_i B^i & = & 0         \label{2_Bconst}
\end{eqnarray}
which hold at each instant of time, and a set of {\em evolution}
equations
\begin{eqnarray}
\partial_t E_i & = & \epsilon_{ijk} D^j B^k - 4 \pi J_i \label{2_Edot1} \\
\partial_t B_i & = & - \epsilon_{ijk} D^j E^k          \label{2_Bdot1}
\end{eqnarray}
which describe the evolution of the electric field $E_i$ and the
magnetic field $B_i$ from one instant of time to the next.  Here
$\rho_e$ and $J_i$ are the charge density and current.  Note that the
evolution equations preserve the constraints so that, if they are
satisfied at any time, they are automatically satisfied at all times.

It is often useful to introduce the vector potential
$A_a = (-\Phi,A_i)$ and write $B^i$ as a curl of $A_i$
\begin{equation}
B^i = \epsilon^{ijk} D_j A_k
\end{equation}
so that $B^i$ satisfies the constraint~(\ref{2_Bconst}) automatically.
Maxwell's equations can now be rewritten as two evolution equations
for $A_i$ and $E_i$
\begin{eqnarray}
\partial_t A_i & = & - E_i - D_i \Phi \label{2_Adot2} \\[1mm]
\partial_t E_i & = & - D^j D_j A_i + D_i D^j A_j
 - 4 \pi J_i \label{2_Edot2}
\end{eqnarray}
together with the constraint equation (\ref{2_Econst}) (see, e.g.,
\citet{j75}).  The gauge quantity $\Phi$ is, like the lapse and shift
in the ADM equations, undetermined by the equations, and has to be
chosen independently.

Interestingly, the evolution equations (\ref{2_Adot2}) and
(\ref{2_Edot2}) are quite similar to the ADM evolution equations
(\ref{2_gdot_2}) and (\ref{2_Kdot_2}), which can be seen by
identifying $A_i$ with $\gamma_{ij}$ and $E_i$ with $K_{ij}$.  Both
right hand sides of (\ref{2_gdot_2}) and (\ref{2_Adot2}) involve a
field variable and a derivative of a gauge variable, while both right
hand sides of (\ref{2_Kdot_2}) and (\ref{2_Edot2}) involve matter
sources as well as second derivatives of the second field variable
(which in (\ref{2_Kdot_2}) are hidden in the Ricci tensor
(\ref{2_ricci_2})).  In Section \ref{sec4.1} we will further explore
these similarities.

\subsection{An Illustration in Spherical Symmetry}
\label{sec2.5}

To illustrate the concepts introduced in this Section, it is useful to
work out some of the expressions in spherical symmetry.  Throughout
this Article we will return to this example.

\begin{exmp} \label{2_exmp_1}
The general form of a spherically symmetric spacetime metric is
\begin{equation}
ds^2 = - A dt^2 + 2 B dtdr + C dr^2 + D (d\theta^2 +
\sin^2\theta d\phi^2),
\end{equation}
where the coefficients $A$, $B$, $C$ and $D$ are functions of time $t$ and
radius $r$ only (see, e.g., equation (14.29) in~\citet{d92}).  We now
introduce a new radial coordinate by the transformation 
\begin{equation} \label{ss_coor_trans}
r \rightarrow \tilde r = (D/C)^{1/2}, 
\end{equation}
which, after dropping all tildes, brings the metric into the {\em
isotropic} form
\begin{equation}
ds^2 = - A dt^2 + 2 B dtdr + C (dr^2 + r^2 d^2 \Omega ),
\end{equation}
where $d^2 \Omega = d\theta^2 + \sin^2\theta d\phi^2$.
Comparing with the metric~(\ref{2_final_metric}), we can now identify
$B$ with $\beta_r$ and, since $\beta^r$ has to be the only
non-vanishing component of the shift vector, $A$ with $\alpha^2 -
\beta_r \beta^r$.  Following convention, we will also rewrite $C$ as
$\psi^4$, where $\psi$ is called the ``conformal factor'' (compare
Section~\ref{sec3}).  Defining $\beta$ as the contravariant component
$\beta^r$, we have
\begin{equation} \label{ss_metric}
ds^2 = -(\alpha^2 - \psi^4 \beta^2) dt^2 + 2 \psi^4 \beta dr dt
	+ \psi^4 (dr^2 + r^2 d^2 \Omega).
\end{equation}

The only non-vanishing, spatial connection coefficients can now be
found to be
\begin{equation} \label{ss_connection}
\begin{array}{ccc}
\Gamma^r_{rr} = \displaystyle 2 \frac{\psi'}{\psi} 
~~&~~ 
\Gamma^r_{\theta\theta} = \displaystyle 
- 2 \frac{r^2 \psi'}{\psi} - r  \\
\Gamma^r_{\phi\phi} = \sin^2 \theta \, \Gamma^r_{\theta\theta}
~~ & ~~
\Gamma^{\theta}_{\phi\phi} = - \sin\theta \cos\theta \\	
\Gamma^{\theta}_{r\theta} = \Gamma^{\theta}_{\theta r} = 
\Gamma^{\phi}_{r\phi} = \Gamma^{\phi}_{\phi r} = 
\displaystyle 2 \frac{\psi'}{\psi} + \frac{1}{r} 
~~ & ~~
\Gamma^{\phi}_{\theta\phi} = \Gamma^{\phi}_{\phi\theta} = \cot \theta.
\end{array}
\end{equation}
Here and in all following examples a prime denotes partial derivative
with respect to radius $r$, and a dot will denote partial derivative
with respect to time $t$.  As expected, the connection reduces to that
of a flat metric in spherical symmetry for $\psi = 1$.  The
non-vanishing components of the extrinsic curvature can be computed
from equation~(\ref{2_gdot_2}) according to
\begin{equation} \label{ss_extrinsic_curvature}
\begin{array}{rcl}
K^r_{~r} & = & \displaystyle
	- \frac{2}{\alpha} \left(\frac{\dot \psi}{\psi} -
	\beta \frac{\psi'}{\psi} - \frac{\beta'}{2} \right) \\
K^{\theta}_{~\theta} & = & K^{\phi}_{~\phi} = \displaystyle
	- \frac{2}{\alpha} \left(\frac{\dot \psi}{\psi} -
	\beta \frac{\psi'}{\psi} - \frac{\beta}{2r} \right),
\end{array}
\end{equation}
and the trace of the extrinsic curvature is
\begin{equation} \label{ss_K}
\alpha K = - \frac{6}{\psi} (\dot \psi - \beta \psi') 
+ \frac{1}{r^2} (\beta r^2)'.
\end{equation}

Note that subtracting the two equations in~(\ref{ss_extrinsic_curvature})
yields an equation for the shift
\begin{equation}
r\beta' - \beta = \alpha (K^r_{~r} - K^{\theta}_{~\theta})\,r
\end{equation}
which can be integrated to yield
\begin{equation} \label{ss_shift}
\beta = r \int \frac{\alpha(K^r_{~r} - K^{\theta}_{~\theta})}{r} \, dr.
\end{equation}
It may seem surprising that we find an equation for the shift, even
though we have emphasized on several occasions that Einstein's
equations do not determine the lapse and the shift, and that,
representing the coordinate freedom in general relativity, they have
to be chosen independently.  However, we have already used up the
spatial coordinate freedom in equation~(\ref{ss_coor_trans}), by
bringing the metric into the isotropic form~(\ref{ss_metric}).  The
shift condition~(\ref{ss_shift}) ensures that the metric remains
isotropic for all times, and it can therefore no longer be chosen
freely.  The lapse, however, is still undetermined.

One could carry the exercise further, and derive, for example, the
Ricci tensor $R_{ij}$ from the connection~(\ref{ss_connection}), which
would be straight-forward but quite lengthy.  Instead we will postpone
this until we develop a formalism in Section~\ref{sec3} that will
simplify this exercise significantly.
\end{exmp}

\section{Constructing Initial Data}
\label{sec3}

In this Section we will present strategies for solving the constraint
equations~(\ref{2_ham_2}) and~(\ref{2_mom_2}).  Since a review of the
initial value problem for numerical relativity has appeared very
recently \citep{c00}, we will focus only on the the most important
results that are often employed for the construction of binary black
holes and neutron stars.  In this Section we will review the most
common formalisms, and will defer the discussion of particular
solutions for binary black holes and neutron stars to Sections
\ref{sec7} and \ref{sec9}.  Parts of this Section are based on the
lecture notes of \citet{bsa98}.

\subsection{Conformal Decompositions}
\label{sec3.1}

Most approaches to solving the relativistic initial value problem
involve a conformal decomposition, in which the physical metric
$\gamma_{ij}$ is written as a product of a {\em conformal factor}
$\psi$ and an auxiliary metric, usually referred to as the {\em
conformally related} or {\em background metric} $\bar \gamma_{ij}$,
\begin{equation} \label{3_conf_1}
\gamma_{ij} = \psi^4 \bar \gamma_{ij}
\end{equation}
\citep{l44,y71,y72}.  Taking $\psi$ to the fourth power turns out
to be convenient, but is otherwise arbitrary.  For other purposes it
is advantageous to consider conformal transformations of the
four-dimensional spacetime metric $g_{ab}$, but here we will only
consider conformal transformations of the spatial metric
$\gamma_{ij}$.

Superficially, we can think of the decomposition (\ref{3_conf_1}) as a
mathematical trick, namely writing one unknown as a product of two
unknowns, which makes the solving of some equations a little easier.
At a deeper level, the conformal transformation (\ref{3_conf_1})
defines an equivalence class of manifolds and metrics, which are all
related by the {\em conformal metric} $\bar \gamma_{ij} =
\gamma^{-1/3} \gamma_{ij}$, where $\gamma$ is the determinant of the
metric $\gamma_{ij}$.  In this natural definition $\bar \gamma = 1$,
but other normalizations can be chosen.  A metric that is conformally
related to the flat spatial metric, $\gamma_{ij} = \psi^4
\eta_{ij}$, is called {\em conformally flat}.

From (\ref{3_conf_1}), the connection coefficients transform according
to
\begin{equation} \label{3_connection_1}
\Gamma^i_{jk} = \bar \Gamma^i_{jk}
+ 2 (     \delta^i_{~j} \partial_k \ln \psi 
	+ \delta^i_{~k} \partial_j \ln \psi
	- \bar \gamma_{jk} \bar \gamma^{il} \partial_l \ln \psi),
\end{equation}
from which it is easy to verify that $\bar D_k \bar \gamma_{ij} = 0$.
For the Ricci tensor (\ref{2_ricci_2}) we find
\begin{equation} \label{3_ricci_1}
\begin{array}{rcl}
R_{ij}  & = & \bar R_{ij} - 2\, \Big( \bar D_i \bar D_j \ln \psi +
 \bar \gamma_{ij} \bar \gamma^{lm} \bar D_l \bar D_m \ln \psi \Big)
 \\
 & & + 4\, \Big( (\bar D_i \ln \psi)(\bar D_j \ln \psi)
 - \bar \gamma_{ij} \bar \gamma^{lm} (\bar D_l \ln \psi)
 (\bar D_m \ln \psi) \Big),
\end{array}
\end{equation}
and for the curvature scalar
\begin{equation} \label{3_ricci_scalar_1}
R = \psi^{-4} \bar R - 8 \psi^{-5} \bar D^2 \psi.
\end{equation}
Here $\bar D^2 = \bar \gamma^{ij} \bar D_i \bar D_i$, $\bar R_{ij}$
and $\bar R$ are the covariant Laplace operator, Ricci tensor and
scalar curvature associated with $\bar \gamma_{ij}$.  $\bar R_{ij}$
can be computed by inserting $\bar
\gamma_{ij}$ into equation (\ref{2_ricci_2}).

\begin{exmp} \label{3_exmp_1}
Returning to Example \ref{2_exmp_1}, we now see that the spatial part
of the metric (\ref{ss_metric}) is a conformal factor $\psi^4$ times
the flat metric $\eta_{ij}$ in spherical coordinates.  Since any
spherically symmetric metric can be brought into the form
(\ref{ss_metric}) without loss of generality, we have shown that any
spherically symmetric metric is (spatially) conformally flat.

Instead of computing curvature quantities from the spatial metric
$\gamma_{ij}$, it is now much easier to employ the above formalism.
The connection coefficients (\ref{ss_connection}), for example, can be
found by adding the connection coefficients $\bar
\Gamma^i_{jk}$ of a flat metric in spherical coordinates to spatial
derivatives of $\psi$ according to (\ref{3_connection_1}).  Since
$\bar \gamma_{ij} = \eta_{ij}$ is flat, the conformally related Ricci
tensor vanishes $\bar R_{ij} = 0$ and the physical Ricci tensor is
given by the derivatives of $\psi$ in (\ref{3_ricci_1}) alone.
The Ricci scalar, finally, reduces to
\begin{equation} \label{ss_ricci_scalar}
R = - 8 \psi^{-5} \bar D^2 \psi,
\end{equation}
a remarkably simple expression.
\end{exmp}

It is also convenient to split the extrinsic curvature $K_{ij}$ 
into its trace $K$ and a traceless part $A_{ij}$
\begin{equation} \label{3_A_0}
K_{ij} = A_{ij} + \frac{1}{3} \, \gamma_{ij} K.
\end{equation}
Solving the initial value problem usually proceeds by further
decomposing the traceless part $A_{ij}$, which is done slightly
differently in different approaches.  In the following we will briefly
discuss the transverse-traceless and the thin-sandwich decomposition.

\subsection{The Conformal Transverse-Traceless Decomposition}
\label{sec3.2} 

In the conformal transverse-traceless decomposition, we first 
introduce the conformal traceless extrinsic curvature by defining
\begin{equation} \label{3_A_1}
A^{ij} = \psi^{-10} \bar A^{ij}
\end{equation}
and accordingly $A_{ij} = \psi^{-2} \bar A_{ij}$ (see, e.g.,
\citet{y79}).  Other choices for the exponent of the conformal factor
are possible as well -- and in fact we will use a different scaling in
Section \ref{sec4.3} -- but for our purposes here the exponent $-10$
is particularly convenient since any symmetric traceless tensor
$S^{ij}$ satisfies $D_j S^{ij} = \psi^{-10}
\bar D_j (\psi^{10} S^{ij})$.

Any symmetric, traceless tensor can be split into a
transverse-traceless tensor, which is divergenceless, and a
longitudinal part, which can be written as a symmetric, traceless
gradient of a vector \citep{y73}.  We can therefore further decompose
$\bar A^{ij}$ as
\begin{equation}
\bar A^{ij} = \bar A^{ij}_{TT} + \bar A^{ij}_{L},
\end{equation}
where the transverse part is divergenceless
\begin{equation}
\bar D_j \bar A^{ij}_{TT} = 0
\end{equation}
and where the longitudinal part satisfies
\begin{equation} \label{3_LO}
\bar A^{ij}_{L} = \bar D^i W^j + \bar D^j W^i 
	- \frac{2}{3} \bar \gamma^{ij} \bar D_k W^k
	\equiv (\bar L W)^{ij}.
\end{equation}
Here $W^i$ is a vector potential, and it is easy to see that the {\em
longitudinal operator} or {\em vector gradient} $\bar L$ produces a
symmetric, traceless tensor.  The divergence of $\bar A^{ij}$ can be
written as
\begin{equation}
\begin{array}{rcl}
\bar D_j \bar A^{ij} & = & \bar D_j \bar A^{ij}_{L}
= \bar D_j (\bar L W)^{ij} = \bar D^2 W^i + \frac{1}{3} \bar D^i (\bar D_j W^j)
+ \bar R^i_{~j} W^j \\
& \equiv & (\bar \Delta_L W)^i,
\end{array}
\end{equation}
where $\bar \Delta_L$ is the vector Laplacian.

Inserting the conformally related quantities into (\ref{2_ham_2})
now yields the Hamiltonian constraint
\begin{equation} \label{3_ham_1}
8 \, \bar D^2 \psi - \psi \bar R - \frac{2}{3} \psi^5 K^2 
+ \psi^{-7} \bar A_{ij} \bar A^{ij} = - 16 \pi \psi^5 \rho,
\end{equation}
while the momentum constraint (\ref{2_mom_2}) becomes
\begin{equation} \label{3_mom_1}
(\bar \Delta_L W)^i - \frac{2}{3} \psi^6 \bar \gamma^{ij} \bar D_j K
= 8 \pi \psi^{10} j^i.
\end{equation}

\begin{exmp} \label{3_exmp_1a}
As a consequence of the Bianchi identities, not all of Einstein's
equations are independent.  This redundancy can be exploited in
numerical evolution calculations, where some quantities can be solved
for using either constraint or evolution equations.  The conformal
factor $\psi$, for example, satisfies an evolution equation (which in
spherical symmetry can be found by solving equation (\ref{ss_K}) for
$\dot \psi$) and the Hamiltonian constraint (\ref{3_ham_1}).  In fact,
the evolution equations can be completely eliminated in spherical
symmetry, which reflects the fact that in spherical symmetry the
gravitational fields do not carry any dynamical degrees of freedom.
In {\em constrained evolution} codes some of the evolution equations
are replaced by constraint equations, while in {\em unconstrained
evolution} codes the gravitational fields are computed from the
evolution equations alone, while the constraint equations are
typically used to monitor the quality of the numerical solution.
Constrained evolution has been very popular in spherical and
axial symmetry.  In full 3D, however, solving the elliptic constraint
equation is very expensive computationally, so that most current
codes in $3+1$ use unconstrained evolution.  
\end{exmp}

Given a solution to the constraint equations, one can find 
the mass, linear and angular momenta from the asymptotic behavior
of the solution.  For asymptotically flat solutions, the 
total (ADM) energy is
\begin{equation} \label{3_mass}
M = - \frac{1}{2 \pi} \oint_{\infty} \bar D^i \psi d^2 S_i,
\end{equation}
the linear momentum is 
\begin{equation} \label{3_p}
P^i = \frac{1}{8 \pi} \oint_{\infty} \bar K^{ij} d^2 S_j,
\end{equation}
and, in Cartesian coordinates, the angular momentum is
\begin{equation} \label{3_j}
J_i = \frac{\epsilon_{ijk}}{8 \pi} \oint_{\infty} x^j K^{kl} d^2 S_l
\end{equation}
(see, e.g., \citet{oy74,by80,y80}).

At this point it is instructive to count degrees of freedom.  We
started out with six independent variables in both the spatial metric
$\gamma_{ij}$ and the extrinsic curvature $K_{ij}$.  Splitting off the
conformal factor $\psi$ left five degrees of freedom in the
conformally related metric $\bar \gamma_{ij}$ (specifying $\bar
\gamma$).  Of the six independent variables in $K^{ij}$ we moved one
into its trace $K$, two into $A^{ij}_{TT}$ (which is symmetric,
traceless, and divergenceless), and three into $A^{ij}_{L}$ (which is
reflected in its representation by a vector).  Of the twelve original
degrees of freedom, the constraint equations determine only four,
namely the conformal factor $\psi$ (Hamiltonian constraint) and the
longitudinal part of the traceless extrinsic curvature (momentum
constraint).  Four of the remaining degrees of freedom are associated
with the coordinate freedom -- three spatial coordinates hidden in the
spatial metric and one that determines the evolution in time that is
often identified with $K$.  This leaves four physical degrees of
freedom undetermined -- two in the conformally related metric $\bar
\gamma_{ij}$, and two in the transverse part of traceless extrinsic
curvature $A^{ij}_{TT}$.  These two freely specifiable degrees of
freedom carry the dynamical degrees of freedom of the gravitational
fields.  All others are either fixed by the constraint equations or
represent coordinate freedom.

It is obvious that the background data $\bar \gamma_{ij}$, $\bar
A^{ij}_{TT}$ and $K$ have to be chosen before the constraints can be
solved for $\psi$ and $\bar A^{ij}_{L}$.  The choice of background
data has to be made in accordance with the physical or astrophysical
situation that one wants to represent.  Physically, the choice
reflects the amount of gravitational wave content in the background.
In many situations, as for example for close compact binaries, it is
not a priori clear what choices correspond to the desired
astrophysical scenario and reflect the past inspiral history.
Arguments have then be made that the background data can be
approximated reasonably well by conformal flatness $\bar
\gamma_{ij} = \eta_{ij}$, maximal slicing $K = 0$ (see 
Section~\ref{sec5.2}) and $\bar A^{ij}_{TT} = 0$.  While this choice
is somewhat controversial (see the discussion in Appendix \ref{appC}),
it has the great benefit of dramatically simplifying the constraint
equations (\ref{3_ham_1}) and (\ref{3_mom_1}).

It is quite remarkable that the momentum constraint (\ref{3_mom_1}) is
a linear equation for the vector potential $W^i$.  For maximal slicing
$K = 0$ and in vacuum $j = 0$ it also decouples from the Hamiltonian
constraint, and can be solved independently of a solution for $\psi$.
For conformal flatness, the operator $\bar \Delta_L$ simplifies
significantly (see also Appendix \ref{appB}), and in fact we will
discuss analytic solutions in Section \ref{sec7.1}.

Assuming conformal flatness, the Hamiltonian constraint
(\ref{3_ham_1}) also simplifies significantly, since $\bar D^2$
reduces to the flat Laplace operator and $\bar R = 0$.  Maximal 
slicing further simplifies the equation.

\begin{exmp} \label{3_exmp_2}
As we have seen in Example \ref{3_exmp_1}, the metric
(\ref{ss_metric}) is conformally flat.  If we are further interested
in time-symmetric vacuum solutions (whereby $K_{ij} = - K_{ij} = 0$
and $\rho = 0$), the Hamiltonian constraint (\ref{3_ham_1}) reduces to
the simple Laplace equation
\begin{equation} \label{ss_ham_ts}
\Delta^{\rm flat} \psi = 0
\end{equation}
in spherical symmetry (where $\Delta^{\rm flat}$ is the flat Laplace
operator).  Assuming asymptotic flatness, so that $\psi
\rightarrow 1$ as $r \rightarrow \infty$, we can write the solution as
\begin{equation} \label{ss_psi} 
\psi = 1 + \frac{m}{2r}
\end{equation}
where $m$ is a constant.  Evaluating the ADM mass (\ref{3_mass}) of
this solution shows that $M = m$.  We have therefore rediscovered the
spatial part of the Schwarzschild solution in isotropic coordinates.
We will see in Example \ref{6_exmp_1} that the surface $r = m/2$ is
the event (and apparent) horizon of a black hole.

It can be shown that the coordinate transformation $x^i \rightarrow
(m/2)^2 x^i/r^2$ maps every point inside $r = m/2$ into a point
outside $r = m/2$.  Moreover, the transformation maps the metric into
itself, making it an {\em isometry}.  Applying the coordinate
transformation twice yields the identity transformation, and points on
the horizon $r = m/2$ are mapped into themselves.  We can therefore
think of the transformation as a reflection across $r = m/2$.  The
geometry close to the origin is therefore identical to the geometry
near infinity, which suggests that we can think of the geometry as two
asymptotically flat sheets, or else as two separate but identical
universes, which are connected by a {\em throat} or {\em
Einstein-Rosen bridge} \citep{er35}.  See Figure 31.5 in \citet{mtw73}
for an embedding diagram.

It is remarkable that equation~(\ref{ss_ham_ts}) is linear.  As an
immediate consequence, time-symmetric initial data containing multiple
black holes can be constructed by simply adding several terms of the
form (\ref{ss_psi}).  We will return to the construction of binary
black hole initial data in Section \ref{sec7}.
\end{exmp}

\subsection{The Thin-Sandwich Decomposition} 
\label{sec3.3}

Solving the conformal transverse-traceless decomposition yields data
$\gamma_{ij}$ and $K_{ij}$ intrinsic to one spatial slice $\Sigma$,
but this solution does not tell us anything about how it will evolve
in time away from $\Sigma$, nor does the formalism allow us to
determine any such time-evolution.  In some circumstances, for example
when we are interested in constructing equilibrium or
quasi-equilibrium solutions, we would like to construct data such that
they do have a certain time-evolution.  The {\em thin-sandwich}
approach offers an alternative that does allow us to determine the
evolution of the spatial metric.  Instead of providing data for
$\gamma_{ij}$ and $K_{ij}$ on one timeslice, it provides data for
$\gamma_{ij}$ on two timeslices, or, in the limit of infinitesimal
separation of the two slices, data for $\gamma_{ij}$ and its time
derivative.  The original thin-sandwich conjecture goes back to
\citet{bsw62} (see also the discussion in \citet{mtw73}), but here
we will focus on a more recent, conformal formulation \citep{y99,c00}.
A similar formulation has been developed independently (and earlier) by
\citet{wm89,wm95} (see Section \ref{sec10.2}).

Following \citet{c00}, to whom the reader is referred for a more
detailed treatment, we start by defining $u_{ij}$ as the traceless
part of the time derivative of the spatial metric,
\begin{equation} \label{3_u}
u_{ij} \equiv \gamma^{1/3} \partial_t (\gamma^{-1/3} \gamma_{ij}),
\end{equation}
in terms of which the evolution equation (\ref{2_gdot_2}) becomes
\begin{equation} \label{3_u_1}
u^{ij} = - 2 \alpha A^{ij} + (L \beta)^{ij}.
\end{equation}
Here $L$ is the vector gradient defined in (\ref{3_LO}), except in
terms of the physical metric $\gamma_{ij}$.  Using\footnote{This
follows from the identifications $\bar u_{ij} = \partial_t \bar
\gamma_{ij}$ and $\bar \gamma^{ij} \bar u_{ij} = 0$.}
\begin{equation}
u_{ij} = \psi^4 \bar u_{ij}
\end{equation}
together with the conformal scaling (\ref{3_A_1}) and the identity $(L
\beta)^{ij} = \psi^{-4} (\bar L \beta)^{ij}$, we can rewrite 
equation (\ref{3_u_1}) as
\begin{equation} \label{3_A_2}
\bar A^{ij} = \frac{\psi^6}{2 \alpha} \left( (\bar L \beta)^{ij} 
- \bar u^{ij} \right).
\end{equation}
This equation relates $\bar A^{ij}$ to the shift vector
$\beta^i$.  Inserting this equation into the momentum constraint
(\ref{2_mom_2}), we can therefore derive an equation for the shift
\begin{equation} \label{3_mom_2}
\begin{array}{rcl}
& & (\bar \Delta_L \beta)^i 
- (\bar L \beta)^{ij} \bar D_j \ln (\alpha \psi^{-6}) = \\
& & \displaystyle
~~~~~~~~~~\alpha \psi^{-6} \bar D_j (\alpha^{-1} \psi^6 \bar u^{ij})
+ \frac{4}{3} \alpha \bar D^i K  
+ 16 \pi \alpha \psi^4 j^i.
\end{array}
\end{equation}

A solution of the thin-sandwich formulation can now be constructed as
follows.  We choose the background metric $\bar \gamma_{ij}$ as well
as its time derivative $\bar u_{ij}$.  Given choices for the lapse
$\alpha$ and the trace of the extrinsic curvature $K$, we can then
solve the Hamiltonian constraint (\ref{3_ham_1}) and the momentum
constraint (\ref{3_mom_2}) for the conformal factor $\psi$ and the
shift $\beta^i$\footnote{Strictly speaking, the ``densitized'' lapse
$\hat \alpha = \gamma^{-1/2} \alpha$ (see also equation
(\ref{4_dens_lapse}) below) should be fixed instead of the lapse
$\alpha$ \citep{y99}.  This distinction is unnecessary if the lapse is
determined through maximal slicing.}.  With these solutions, we can
then construct $\bar A^{ij}$ from (\ref{3_A_2}) and finally the
physical quantities $\gamma_{ij}$ and $K_{ij}$.

It is again instructive to count the degrees of freedom, and to
compare with the transverse-traceless decomposition of Section
\ref{sec3.2}.  There, we found that of the twelve independent
variables in $\gamma_{ij}$ and $K_{ij}$, four were determined by the
constraint equations, four were related to the coordinate freedom, and
four represented the dynamical degrees of freedom of general
relativity.  The latter eight can be chosen freely.  In the
thin-sandwich formalism, we count a total of sixteen independent
variables, of which we can freely choose twelve: five each in $\bar
\gamma_{ij}$ and $\bar u_{ij}$, and one each for $\alpha$ and $K$.
The four remaining variables, $\psi$ and $\beta^i$, are then
determined by the constraint equations.  The four new independent
variables are accounted for by the lapse $\alpha$ and the shift
$\beta^i$, which are absent from the transverse-traceless
decomposition.  That approach only deals with quantities intrinsic to
one spatial slice $\Sigma$, and hence only requires coordinates on
$\Sigma$.  The thin-sandwich approach, on the other hand, also takes
into account the evolution of the metric away from that slice, and
therefore requires coordinates in a neighborhood of $\Sigma$.  As a
consequence, the lapse $\alpha$ and the shift $\beta^i$, which
describe the evolution of the coordinates away from $\Sigma$, appear
in the thin-sandwich approach, but not in the the transverse-traceless
decomposition.  The four new free degrees of freedom hence reflect the
time derivatives of the coordinates.

The thin-sandwich approach is particularly useful for the construction
of equilibrium or quasi-equilibrium data, for which it is natural
to choose
\begin{equation}
\bar u_{ij} = 0.
\end{equation}
For equilibrium data it is also natural to choose $K = 0$ and
$\partial_t K=0$, which from equation (\ref{2_trK}) yields the maximal
slicing condition
\begin{equation}
D^2 \alpha = \alpha \left( A_{ij} A^{ij} + 4 \pi (\rho + S)
\right).
\end{equation}
We will discuss maximal slicing in more detail in Section \ref{sec5.2}.
Expressing $D^2$ in terms of $\bar D^2$ and using the conformal
transformations of Section~\ref{sec3.1}, we find that this equation
can be combined with the Hamiltonian constraint to yield
\begin{equation} \label{3_ms_2}
\bar D^2 (\alpha \psi) = 
\alpha \psi \left(\frac{7}{8} \psi^{-8} \bar A_{ij} \bar A^{ij} 
+ \frac{1}{8} \bar R + 2 \pi \psi^4 (\rho + 2 S) \right).
\end{equation}
Equations (\ref{3_ham_1}), (\ref{3_mom_2}) and (\ref{3_ms_2}) now
determine the solutions for $\psi$, $\beta^i$ and $\alpha$.  The 
equations further simply if we assume conformal flatness, in which
case they reduce to
\begin{eqnarray}
\Delta^{\rm flat} \psi & = & - \frac{1}{8} 
\psi^{-7} \bar A_{ij} \bar A^{ij} -  2 \pi \psi^5 \rho 
	\label{3_ts_ham}\\
(\Delta_L^{\rm flat} \beta)^i & = &  
2 \bar A^{ij} \bar D_j (\alpha \psi^{-6}) +
16 \pi \alpha \psi^4 j^i \label{3_ts_md}\\
\Delta^{\rm flat} (\alpha \psi) & = &
\alpha \psi \left(\frac{7}{8} \psi^{-8} \bar A_{ij} \bar A^{ij} 
+ 2 \pi \psi^4 (\rho + 2 S) \right). \label{3_ts_ms}
\end{eqnarray}
Here $\Delta^{\rm flat}$ and $\Delta_L^{\rm flat}$ are the flat Laplacian and
vector Laplacian.  Strategies for solving the flat vector Laplacian
will be discussed in Appendix \ref{appB}.  Interestingly, we will
re-discover the shift condition (\ref{3_ts_md}) in Section \ref{sec5.4}
and will find that it is identical to minimal distortion.  
The thin-sandwich formalism therefore reduces to the Hamiltonian
constraint for the conformal factor, the minimal distortion condition
for the shift, and the maximal slicing condition for the lapse.
 
If initial data for a time evolution calculation are constructed from
the trans\-verse-traceless decomposition, then the lapse and shift
have to be chosen independently of the construction of initial data.
The thin-sandwich formalism, on the other hand, provides a lapse and a
shift together with the initial data $\gamma_{ij}$ and $K_{ij}$.
Obviously, once the initial data are determined, 
the lapse and shift can always be chosen freely in performing
subsequent evolution calculations.   However, the original relation
between the time derivative of $\gamma_{ij}$ and $\bar u_{ij}$ 
only applies when the lapse and shift of the thin-sandwhich solution
are employed.

\section{Rewriting the ADM evolution equations}
\label{sec4}

The ADM equations as presented in Section~\ref{sec2.2} form the basis
for most 3+1 decompositions of the Einstein equations and their
numerical implementations.  As it turns out, the evolution equations
are not yet in their most desirable form, and straight-forward
implementations in three spatial dimensions typically develop
instabilities very quickly.

In this Section we will discuss the draw-backs of the ADM equations
and will present possible alternatives.  We will first illustrate some
of the relevant issues with an electrodynamics analogy, and will then
present two different re-formulations of the ADM equations.

\subsection{Rewriting Maxwell's equations}
\label{sec4.1}

In Section \ref{sec2.4} we showed that Maxwell's equations, when
written as the two evolution equations (\ref{2_Adot2}) and
(\ref{2_Edot2})
\begin{eqnarray}
\partial_t A_i & = & - E_i - D_i \Phi \label{4_Adot2} \\[1mm]
\partial_t E_i & = & - D^j D_j A_i + D_i D^j A_j
 - 4 \pi J_i \label{4_Edot2}
\end{eqnarray}
and the constraint equation (\ref{2_Econst})
\begin{equation} \label{4_Econst}
D^i E_i = 4 \pi \rho_e,
\end{equation}
share some of the structure of the ADM equations.  In fact they also
share some of their draw-backs.  To illustrate these, take a time
derivative of (\ref{4_Adot2}) and insert (\ref{4_Edot2}) to form a
single equation for the vector potential $A_i$
\begin{equation} \label{4_Awave}
- \partial^2_t A_i + D^j D_j A_i - D_i D^j A_j =  
	D_i \partial_t \Phi - 4 \pi J_i.
\end{equation}
It is obvious that this equation would be a simple wave equation for
the components $A_i$ if it weren't for the mixed derivative term $D_i
D^j A_j$.  In general relativity the situation is very similar, since
the Ricci tensor $R_{ij}$ (\ref{2_ricci_2}) on the right hand side of
equation (\ref{2_Kdot_2}) also contains mixed derivative terms in
addition to a Laplace operator acting on $\gamma_{ij}$.  Without these
mixed derivatives, the ADM equations could be written as wave
equations for the components of the spatial metric and would be
manifestly hyperbolic (see, e.g., \citet{f96} for a discussion).  This
is unfortunate, since it would be desirable in many ways to deal with
a hyperbolic system of equations.  Mathematical theorems would
guarantee the existence and uniqueness of solutions, and for numerical
purposes one could bring the equations into a form that allows for the
application of flux-conservative schemes that have been developed and
tested in other fields of computational physics, and quite in general
one might feel more comfortable that numerical implementations will
produce stable evolutions.

These considerations suggest that it might be desirable to eliminate
the mixed derivative terms.  In electrodynamics, three different
approaches can be taken to eliminate the $D_i D^j A_j$ term: one can
make a special {\em gauge choice}, one can bring Maxwell's equations
into an {\em explicitly hyperbolic} form, or one can introduce an {\em
auxiliary variable}.  For the remainder of this section, we will
briefly discuss each one of these three strategies.

The most straightforward approach is to choose a {\em gauge} such that
the term $D_i D^j A_j$ disappears.  This can be achieved, for example,
in the Lorentz gauge
\begin{equation}
\partial_t \Phi = - D^i A_i
\end{equation}
for which these two terms cancel in equation (\ref{4_Awave}), reducing
it to a wave equation for the vector potential $A_i$ (see
\citet{j75}).  Another possibility is the Coulomb gauge $D^i A_i = 0$,
which results in an elliptic equation for $\Phi$.

In general relativity, an analogous approach can be taken by choosing
harmonic coordinates (see Section \ref{sec5.3}), which bring the
equations into a manifestly hyperbolic form.  This was first realized
by \citet{d21} and \citet{l22}, and many recent hyperbolic
formulations of Einstein's equations are based on this gauge choice
(e.g. \citet{c52,c62,fm72}).  For the purpose of numerical simulations
this strategy does not seem very promising, since harmonic coordinates
may not be the optimal coordinate choice for the astrophysical
situation at hand (but see \citet{lt99}, where the traditional ADM
approach together with harmonic coordinates and finely tuned finite
differencing has been used to simulate the merger of binary neutron
stars; see also \citet{g02}).  Moreover, harmonic coordinates may
develop pathologies which could prematurely end a numerical simulation
\citep{a97,am98}.  It is therefore more desirable to preserve the
coordinate freedom, and to adopt a different approach to re-writing
the equations.

An alternative, gauge-covariant approach to bringing Maxwell's
equations into an {\em explicitly hyperbolic} form is to take a time
derivative of (\ref{4_Edot2}) instead of (\ref{4_Adot2}), which yields
\citep{ay97}
\begin{equation}
\partial^2_t E_i = D_i D^j (- E_j - D_j \Phi) - D_j D^j (- E_i - D_i \Phi)
	- \partial_t J_i.
\end{equation}
Using the constraint (\ref{4_Econst}) we can eliminate the first term
and find a wave equation for $E_i$
\begin{equation} \label{4_Ewave}
- \partial^2_t E_i + D_j D^j E_i = \partial_t J_i + 4 \pi D_i \rho_e.
\end{equation}
Interestingly, the gauge dependent quantities $A_i$ and $\Phi$ have
disappeared from this equation, and, quoting \citet{ay97}, ``the
dynamics of electromagnetism have been cleanly separated from the
gauge-dependent evolution of the vector and scalar potentials.''  
We will discuss similar approaches in general relativity in Section
\ref{sec4.2}.

While equation (\ref{4_Ewave}) is aesthetically very appealing, it
also reveals some potential disadvantages for simulations that involve
matter.  In evolution calculations of neutron stars, for example, the
non-smoothness of the matter and fields on the stellar surface or
across shocks always pose numerical difficulties.  It is to be
expected that these would only get worse if further derivatives of the
matter variables have to be taken, as in equation (\ref{4_Ewave}).

This suggests a third approach to re-writing Maxwell's equations,
namely by introducing an {\em auxiliary variable}
\begin{equation} \label{4_Gamma}
\Gamma = D^i A_i.
\end{equation}
Inserting this into (\ref{4_Edot2}) yields
\begin{equation} \label{4_Edot}
\partial_t E_i = - D_j D^j A_i + D_i \Gamma - 4 \pi J_i.
\end{equation}
We now elevate $\Gamma$'s status to that of a new independent
variable, and derive its evolution equation from (\ref{4_Adot2})
\begin{equation} \label{4_Gammadot} 
\begin{array}{rcl}
\partial_t \Gamma & = & \partial_t D^i A_i = D^i \partial_t A_i =
- D^i E_i -  D_i D^i \Phi \\
& = & - D_i D^i \Phi - 4 \pi \rho_e.
\end{array}
\end{equation}
Note that we have used the constraint equation (\ref{4_Econst}) in the
last equality, similar to how we used the same constraint to arrive at
the wave equation (\ref{4_Ewave}).  We will see in equation
(\ref{4_constwave}) below that this step is crucial for stabilizing the
system.

Equations (\ref{4_Adot2}), (\ref{4_Edot}) and (\ref{4_Gammadot}) are
now the evolution equations in this new formulation, and equations
(\ref{4_Econst}) and (\ref{4_Gamma}) are the constraint equations.  In
this formulation the mixed derivative term $D_i D^j A_j$ has been
eliminated without using up the gauge freedom, and without introducing
derivatives of the matter terms.  In Section \ref{sec4.3} we will
introduce an analogous re-formulation of the ADM equations.

\subsection{Hyperbolic formulations} 
\label{sec4.2}

In light of the disadvantages of the ADM system, a large number of
$3+1$ hyperbolic formulations of general relativity have been
developed recently (see the recent review article by \citet{r98} for
an extensive survey and a more complete list of references).  The
first formulations that departed from the assumption of harmonic
coordinates \citep{c52,c62,fm72} were based on a spin-frame formalism
\citep{f81a,f81b,f85,f86a,f86b}.  Other formulations introduced
partial derivatives of the metric and other quantities as new
independent variables (e.g. \citet{bm92,fr94,fr96,bmss95,bmss97}).  In
analogy to the electromagnetic example in Section \ref{sec4.1},
\citet{cr83} and \citet{aacy95} took a time derivative of 
equation (\ref{2_Kdot_2}) to derive a hyperbolic system that is
sometimes referred to as the ``Einstein-Ricci'' system (see also
\citet{ay97}).  Alternatively,
\citet{f96} and \citet{acy97} used the Bianchi identities to derive
another hyperbolic system for general relativity, sometimes called the
``Einstein-Bianchi'' system.  \citet{ay99} developed the
``Einstein-Christoffel'' system by introducing additional
``connection'' variables.  Other recent hyperbolic formulations are
based on frame or tetrad formalisms \citep{ve96,erw97}, the Ashtekar
formulation \citep{ys99,ys00,ys01a,sy00}, a conformal decomposition
\citep{abms99}, and a so-called $\lambda$-system that embeds Einstein's
equations in a larger symmetric hyperbolic system with the constraint
surface of Einstein's equations as an attractor of the evolution
\citep{bfhr99}.

It is clearly beyond the scope of this article to review all of these
formulations.  Some of these systems have features that are not very
desirable numerically, in that they restrict the gauge freedom,
introduce extra derivatives of the matter variables, or introduce a
large number of auxiliary variables.  Only few of these
formulations have been implemented numerically, and most of these
implementations assumed certain simplifying symmetry conditions
(e.g.~spherical symmetry).  Some of these implementations showed
advantages over the ADM formalism (e.g.~\citet{bm92}), but others also
revealed additional problems.  \citet{sbcst97,sbcst98}, for example,
found that a particular equation in the ``Einstein-Ricci'' system
produced an instability, which could be removed in spherical symmetry,
but not in more general 3D simulations.  Very few hyperbolic systems
have been implemented in 3D, including that of Bona and Mass\'{o} 
\citep{bm92,bmss95,bmss97}, and generalized versions of the 
``Einstein-Christoffel'' system \citep{ay99}.  The latter have been
implemented numerically by \citet[see also \citet{kstcc00}]{kst01}
using spectral methods.  Since this formulation is particularly
elegant and currently seems like the most promising hyperbolic
formulation, we provide a brief summary.  Some properties of this
formulation have been analyzed by \citet{clt01,cpst02} and 
\citet{ls02a}.

Starting with the ADM formalism as presented in Section \ref{sec2},
and adopting the notation of \citet{kstcc00}, we define the new
variables
\begin{equation} \label{4_fconst}
f_{kij} = \Gamma_{(ij)k} 
	+ \gamma_{ki} \gamma^{lm} \Gamma_{[lj]m} 
	+ \gamma_{kj} \gamma^{lm} \Gamma_{[li]m}.
\end{equation}
These functions are now promoted to independent functions.  It can
then be shown that the evolution equations (\ref{2_gdot_2}) and
(\ref{2_Kdot_2}) can be rewritten as
\begin{equation} \label{4_cs}
\begin{array}{rcl}  
d_t \gamma_{ij} & = & - 2 \alpha K_{ij} \\
d_t K_{ij} + \alpha \gamma^{kl} \partial_l f_{kij} & = & \alpha M_{ij}\\
d_t f_{kij} + \alpha \partial_k K_{ij} & = & \alpha N_{kij}.
\end{array}
\end{equation}
Here we have used the abbreviation
\begin{equation}
d_t \equiv \partial_t - \Lie_{\beta}
\end{equation}
and the source terms $M_{ij}$ and $N_{ijk}$ are given by
\begin{equation}
\begin{array}{rcl}
M_{ij} & = & \gamma^{kl} (K_{kl} K_{ij} - 2 K_{ki} K_{lj})
	+ \gamma^{kl} \gamma^{mn} (4 f_{kmi} f_{[ln]j} \\
& & 	+ 4 f_{km[n} f_{l]ij} - f_{ikm} f_{jln} + 
	8 f_{(ij)k} f_{[ln]m} + 4 f_{km(i} f_{j)ln} \\ 
& & 	- 8 f_{kli} f_{mnj} + 20 f_{kl(i} f_{j)mn} -
	13 f_{ikl} f_{jmn}) \\
& &	- \partial_i \partial_j \ln \hat \alpha - 
	(\partial_i \ln \hat \alpha) (\partial_j \ln \hat \alpha) 
	+ 2 \gamma_{ij} \gamma^{kl} \gamma^{mn} 
	(f_{kmn} \partial_l \ln \hat \alpha \\
& &	- f_{kml} \partial_n \ln \hat \alpha) + 
	\gamma^{kl} \Big( (2 f_{(ij)k} - f_{kij}) \partial_l \ln \hat \alpha \\
& &	+ 4 f_{kl(i} \partial_{j)} \ln \hat \alpha - 
	3 (f_{ikl} \partial_j \ln \hat \alpha +
	f_{jkl} \partial_i \ln \hat \alpha ) \Big) \\
& & 	- 8 \pi S_{ij} + 4 \pi \gamma_{ij} T
\end{array}
\end{equation}
and
\begin{equation}
\begin{array}{rcl}
N_{kij} & = & \gamma^{mn} \Big( 4 K_{k(i} f_{j)mn} 
	- 4 f_{mn(i} K_{j)k} + K_{ij} ( 2 f_{mnk} - 3 f_{kmn} ) \Big) \\
& & 	+ 2 \gamma^{mn} \gamma^{pq} \Big( K_{mp} ( 
	\gamma_{k(i} f_{j)qn} - 2 f_{qn(i} \gamma_{j)k} ) \\ 
& &	+ \gamma_{k(i} K_{j)m} ( 8 f_{npq} - 6 f_{pqn} ) 
	+ K_{mn} ( 4 f_{pq(i} \gamma_{j)k} - 5 \gamma_{k(i} f_{j)pq}) \Big) \\
& &	- K_{ij} \partial_k \ln \hat \alpha 
	+ 2 \gamma^{mn} (K_{m(i} \gamma_{j)k} \partial_n \ln \hat \alpha
	- K_{mn} \gamma_{k(i} \partial_{j)} \ln \hat \alpha ) \\
& &	+ 16 \pi \gamma_{k(i} j_{j)}.
\end{array}
\end{equation}
We have also used the ``densitized'' lapse function
\begin{equation} \label{4_dens_lapse}
\hat \alpha = \gamma^{-1/2} \alpha
\end{equation}
and, in addition to the matter terms defined in Section \ref{sec2},
the four-dimensional trace of the stress energy tensor
\begin{equation}
T = g^{ab} T_{ab}.
\end{equation}

The above equations form the ``Einstein-Christoffel'' system of
\citet{ay99}.  Evolutions of a single black hole using a spectral
implementation of this system are still unstable (compare Section
\ref{sec8.1}), but \citet{kst01} were able to show that 
the lifetime of these simulation can be extended to late times by a
certain generalization of the equations.  This generalization involves
the redefinition of variables and an addition of constraints, and
embeds the above equations into a 12-parameter family of hyperbolic
formulations. The stability properties of the system depend strongly on
the choice of the free parameters, which can be understood
analytically in terms of energy norm arguments \citep{ls02a}.

The first order symmetric hyperbolic system (\ref{4_cs}) is equivalent
to the original set of evolution equations (\ref{2_gdot_2}) and
(\ref{2_Kdot_2}).  Since the $f_{kij}$ are evolved as independent
functions, their original definition (\ref{4_fconst}) can be
considered as a new constraint equation in addition to (\ref{2_ham_2})
and (\ref{2_mom_2}).  This system is particularly elegant because the
source terms $M_{ij}$ and $N_{ijk}$ on the right hand sides do not
contain any derivatives of the fundamental variables (other than the
arbitrary lapse function $\alpha$).  Equations (\ref{4_cs}) can be
combined to yield a wave equation for the components of the spatial
metric $\gamma_{ij}$ in which the right hand sides appear as sources.

Hyperbolic systems have the great advantage that their characteristic
structure can be analyzed.  In the system (\ref{4_cs}), all
characteristic fields propagate either along the light cone or normal
to the spatial foliation.  This knowledge can be used for the
construction of boundary conditions, both at the outer boundaries and
on inner boundaries if the interior of black holes is excised (see,
e.g., \citet{kstcc00} and Section \ref{sec8.1}).

\subsection{The BSSN formulation} 
\label{sec4.3}

Following the electrodynamic example of Section \ref{sec4.1} we can
also eliminate the mixed second derivatives in the Ricci tensor with
the help of auxiliary variables \citep{nok87}.  In addition, the
conformal factor and the trace of the extrinsic curvature are evolved
separately, which follows the philosophy of separating transverse from
longitudinal, or radiative from non-radiative degrees of freedom (see
Section \ref{sec3}).

We follow here the formulation of \citet{bs99}, which is based on that
of \citet{sn95}.  We start by writing the conformal factor as $\psi =
e^\phi$ so that
\begin{equation}
\bar \gamma_{ij} = e^{- 4 \phi} \gamma_{ij},
\end{equation}
and by choosing it such that the determinant of the conformally
related metric $\bar \gamma_{ij}$ is unity, $\phi = (\ln
\gamma)/12$.  As in equation (\ref{3_A_0}) we split the trace from 
the extrinsic curvature and conformally rescale the traceless part
$A_{ij}$.  Following \citet{sn95} and \citet{bs99} again, we choose a
conformal rescaling that is different from (\ref{3_A_1}) and instead
rescale $A_{ij}$ like the metric itself
\begin{equation} \label{4_A_rescaling}
\tilde A_{ij} = e^{- 4 \phi} A_{ij}.
\end{equation}
We will use tildes as opposed to the bars used in Section~\ref{sec3}
as a reminder of this different rescaling.  Indices of $\tilde A_{ij}$
will be raised and lowered with the conformal metric $\bar
\gamma_{ij}$, so that $\tilde A^{ij} = e^{4 \phi} A^{ij}$

Evolution equations for $\phi$ and $K$ can now be found from 
equation (\ref{2_trgamma}), yielding
\begin{equation} \label{4_phidot3}
\partial_t \phi = - \frac{1}{6} \alpha K + \beta^i \partial_i \phi
	+ \frac{1}{6} \partial_i \beta^i
\end{equation}
and (\ref{2_trK})
\begin{equation} \label{4_Kdot3}
\partial_t K = - \gamma^{ij} D_j D_i \alpha + 
	\alpha(\tilde A_{ij} \tilde A^{ij}
	+ \frac{1}{3} K^2) + 4 \pi \alpha (\rho + S) 
	+ \beta^i \partial_i K.
\end{equation}
Subtracting these from the evolution equations (\ref{2_gdot_2}) and
(\ref{2_Kdot_2}) yields the traceless evolution equations for
$\bar \gamma_{ij}$
\begin{equation} \label{4_gdot3}
\partial_t \bar \gamma_{ij} =  
	- 2 \alpha \tilde A_{ij} 
	+ \beta^k \partial_k \bar \gamma_{ij} 
	+ \bar \gamma_{ik} \partial_j \beta^k
	+ \bar \gamma_{kj} \partial_i \beta^k
	- \frac{2}{3} \bar \gamma_{ij} \partial_k \beta^k.
\end{equation}
and $\tilde A_{ij}$
\begin{equation} \label{4_Adot3}
\begin{array}{rcl}
\partial_t \tilde A_{ij} & = & e^{- 4 \phi} \left( 
	- ( D_i D_j \alpha )^{TF}  +
	\alpha ( R_{ij}^{TF} - 8 \pi S_{ij}^{TF} ) \right)  \\
	& & + \alpha (K \tilde A_{ij} - 2 \tilde A_{il} \tilde A^l_{~j}) \\
	& & 
	+ \beta^k \partial_k \tilde A_{ij} 
	+ \tilde A_{ik} \partial_j \beta^k
	+ \tilde A_{kj} \partial_i \beta^k
	- \frac{2}{3} \tilde A_{ij} \partial_k \beta^k.
\end{array}
\end{equation}
In the last equation, the superscript $TF$ denotes the trace-free part
of a tensor, e.g. $R_{ij}^{TF} = R_{ij} - \gamma_{ij} R/3$.  Note also
that in equations (\ref{4_phidot3}) through (\ref{4_Adot3}) the shift
terms arise from Lie derivatives of the respective variable.  The
divergence of the shift, $\partial_i \beta^i$, appears in the Lie
derivative because the choice $\bar \gamma = 1$ makes $\phi$ a tensor
density of weight $1/6$, and $\bar \gamma_{ij}$ and $\tilde A_{ij}$
tensor densities of weight $-2/3$.

According to (\ref{3_ricci_1}) we can split the Ricci tensor into two
terms
\begin{equation}
R_{ij} = \bar R_{ij} + R_{ij}^{\phi},
\end{equation}
where $R_{ij}^{\phi}$ can be found by inserting $\phi = \ln \psi$ into
(\ref{3_ricci_1}).  The conformally related Ricci tensor $\bar R_{ij}$
could be computed by inserting $\bar \gamma_{ij}$ into
(\ref{2_ricci_2}), which would introduce the mixed second derivatives
similar to those in the electrodynamics illustration of Section
\ref{sec4.1}.  In analogy to the new variable $\Gamma$ that we used
to eliminate those mixed derivatives there, we can now define the
``conformal connection functions''
\begin{equation} \label{4_ccf}
\bar \Gamma^i \equiv \bar \gamma^{jk} \bar \Gamma^{i}_{jk}
	= - \bar \gamma^{ij}_{~~,j},
\end{equation}
where the $\bar \Gamma^{i}_{jk}$ are the connection coefficients
associated with $\bar \gamma_{ij}$, and where the last equality holds
because $\bar \gamma = 1$ (see Problem 7.7f in \citet{lppt75}).  In
terms of these, the Ricci tensor can be written
\begin{equation} \label{4_ricci}
\begin{array}{rcl}
\bar R_{ij} & = & -  \frac{1}{2} \bar \gamma^{lm}
	\bar \gamma_{ij,lm} 
	+ \bar \gamma_{k(i} \partial_{j)} \bar \Gamma^k
	+ \bar \Gamma^k \bar \Gamma_{(ij)k}  + \\
	& & \bar \gamma^{lm} \left( 2 \bar \Gamma^k_{l(i} 
	\bar \Gamma_{j)km} + \bar \Gamma^k_{im} \bar \Gamma_{klj} 
	\right).
\end{array}	
\end{equation}
The only second derivatives of $\bar \gamma_{ij}$ left over in this
operator is the Laplace operator $\bar \gamma^{lm} \bar
\gamma_{ij,lm}$ -- all others have been absorbed in first derivatives
of $\bar \Gamma^i$.  This property of the contraction of the
Christoffel symbols has been known for a long time \citep{d21,l22},
and has been used widely to write Einstein's equations in a hyperbolic
form (e.g.~\citet{c52, fm72}). 

We now promote the $\bar \Gamma^i$ to independent functions, and hence
need to derive their evolution equation.  This can be done in complete
analogy to (\ref{4_Gammadot}) by permuting a time and space derivative
in the definition (\ref{4_ccf})
\begin{equation} 
\partial_t \bar \Gamma^i
=  - \partial_j \Big( 2 \alpha \tilde A^{ij} 
	- 2 \bar \gamma^{m(j} \beta^{i)}_{~,m}
	+ \frac{2}{3} \bar \gamma^{ij} \beta^l_{~,l} 
	+ \beta^l \bar \gamma^{ij}_{~~,l} \Big).
\end{equation}
The divergence of the extrinsic curvature can now be eliminated with 
the help of the momentum constraint (\ref{2_mom_2}), which yields the
evolution equation\footnote{Note that the shift terms enter with
the wrong sign in equation (24) of \citet{bs99}.}
\begin{equation} \label{4_Gammadot3}
\begin{array}{rcl}
\partial_t \bar \Gamma^i & = &
	- 2 \tilde A^{ij} \partial_j \alpha + 2 \alpha \Big(
	\bar \Gamma^i_{jk} \tilde A^{kj} 
	- \frac{2}{3} \bar \gamma^{ij} \partial_j K
	- 8 \pi \bar \gamma^{ij} S_j 
	+ 6 \tilde A^{ij} \partial_j \phi \Big) \\
	& &  + \beta^j \partial_j \bar \Gamma^i 
	- \bar \Gamma^j \partial_j \beta^i
	+ \frac{2}{3} \bar \Gamma^i \partial_j \beta^j 
	+ \frac{1}{3} \bar \gamma^{li} \beta^j_{,jl} 
	+ \bar \gamma^{lj} \beta^i_{,lj}.
\end{array}
\end{equation}

Equations (\ref{4_phidot3}) through (\ref{4_Adot3}) together with
(\ref{4_Gammadot3}) form a new system of evolution equations that is
equivalent to (\ref{2_gdot_2}) and (\ref{2_Kdot_2}).  Since the $\bar
\Gamma^i$ are evolved as independent functions, their original
definition (\ref{4_ccf}) serves as a new constraint equation, in
addition to (\ref{2_ham_2}) and (\ref{2_mom_2}).

\begin{figure}
\begin{center}
\epsfxsize=3in
\epsffile{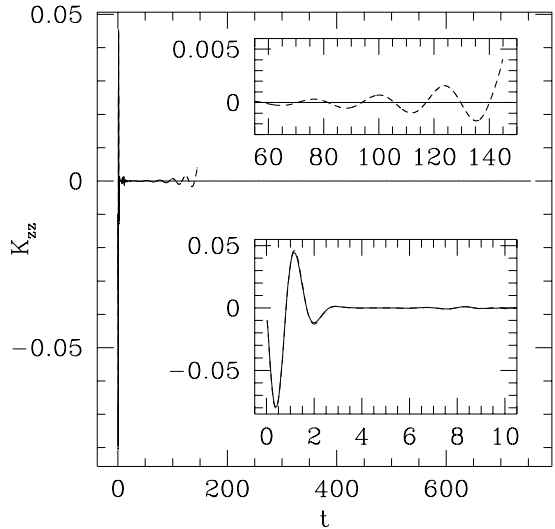}
\end{center}
\caption{Comparison of the evolution of a small amplitude gravitational
wave using the ADM equations (dashed line) and the BSSN equations (solid
line).  The bottom panel shows the evolution of the $K_{zz}$ component
as a function of time for early times, for which both systems agree
very well, while the top panel shows the evolution at a later time,
just before the ADM system crashes.  See text for more details.  (Figure
from \citet{bs99}.)}
\label{fig4.1}
\end{figure}

While the different formulations are equivalent analytically, the
difference in performance of numerical implementations is striking.
In Figure \ref{fig4.1} we show a particular example from \citet{bs99}.
In this example, a small amplitude wave \citep{t82} is evolved with
harmonic slicing (see Section \ref{sec5.3}), zero shift, and a very
simple outgoing wave boundary condition (see also \citet{sn95}).  Both
systems give very similar results early on, but the ADM system crashes
very soon, while the BSSN system remains stable.  Similar improvements
have bee found for many other applications, including strong field
gravitational waves as well as black hole and neutron star spacetimes
\citep{bhs99,aabss00,abdfpsst00,lhg00,ab01}.  It is generally found that
the ADM system is more accurate initially (which is to be expected
given that it uses fewer equations), but that the BSSN system is much
more stable in long term evolution calculations.  Given this success,
the BSSN system, either in the form of \citet{sn95} or \citet{bs99},
is currently used very commonly in numerical relativity and has been
adopted for many recent applications (an incomplete list includes
\citet{s99a,s99b,su00a,sbs00a,sbs00b,aablsst00,abblnst01,
abpst01,dbs01,fgimrssst01}).

Significant effort has also gone into understanding why implementations
of the BSSN system are more stable than those of the ADM system 
\citep[see also \citet{fg00}]{fr99,aabss00,m00,ys02,scpt02}.  
While we are still lacking a complete understanding, several arguments
point to the propagation of the constraints (compare
\citet{f97}). \citet{aabss00} linearized the ADM and BSSN equations on
a flat Minkowski background and showed that modes that violate the
momentum constraint propagate with speed zero in the ADM equations.
They also demonstrate in a model problem that such zero speed modes
lead to instabilities when nonlinear source terms are included.
Furthermore, they show that by adding the momentum constraint in the
derivation of the $\bar \Gamma^i$ evolution equation
(\ref{4_Gammadot3}) of the BSSN system, the momentum constraint
violating modes now propagate with non-zero speed.  Instead of
building up locally, as in the ADM system, constraint violations can
now propagate off the numerical grid\footnote{Assuming that
appropriate boundary conditions are imposed.}, presumably stabilizing
the simulation\footnote{For simulations of black holes it may be
constraint violating modes that propagate along the outward
characteristic as opposed to along the normal that cause
instabilities; see \citet{ls02a}.}.

This effect can be illustrated very easily with the electrodynamic
example of Section \ref{sec4.1} (see \citet{kwb02}).  In the Maxwell
system, the time derivative of a constraint violation
\begin{equation}
{\mathcal C} \equiv D^i E_i - 4 \pi \rho_e
\end{equation}
vanishes identically
\begin{eqnarray}
\partial_t {\mathcal C} & = & \partial_t (D^i E_i - 4 \pi \rho_e) 
= D^i \partial_t E_i - 4 \pi \partial_t \rho_e  \nonumber \\
& = & - D^i D^j D_j A_i + D^i D_i D^j A_j -
	 4 \pi (D^i J_i + \partial_t \rho_e) =  0,
\end{eqnarray}
where we have used the continuity equation $D_i J^i + \partial_t
\rho_e = 0$.  Using the modified system (\ref{4_Adot2}),
(\ref{4_Edot}) and (\ref{4_Gammadot}), on the other hand, it can be
shown that ${\mathcal C}$ now satisfies a wave equation
\begin{eqnarray} \label{4_constwave}
\partial^2_t {\mathcal C} & = & 
	\partial_t D^i \partial_t E_i - 4 \pi \partial^2_t \rho_e 
	= \partial_t D^i (- D_j D^j A_i + D_i \Gamma  - 4 \pi J_i)
	 - 4 \pi \partial^2_t \rho_e \nonumber \\
	& = &  - D^i \Big( D_j D^j \partial_t A_i  
	- D_i \partial_t \Gamma \Big) 
	- 4 \pi \partial_t (D^i J_i + \partial_t \rho_e) \nonumber  \\
	& = &  D^i \Big( D_j D^j (E_i + D_i \Phi) 
	- D_i (D^j D_j \Phi + 4 \pi \rho_e) \Big)  \nonumber  \\
	& = &  D_j D^j (D^i E_i - 4 \pi \rho_e) =  D_j D^j {\mathcal C}.
\end{eqnarray}
The crucial step in achieving this property has been the use of the
constraint (\ref{4_Econst}) to replace $D^i E_i$ with $4 \pi \rho_e$ in
(\ref{4_Gammadot}).  Had we not done that, then the $D^i E_i$ would
cancel in (\ref{4_constwave}), leading again to a zero propagation
speed.  Instead, the two terms now combine to form ${\mathcal C}$,
yielding a wave equation with the speed of light as the characteristic
speed.

If ${\mathcal C} = \partial_t {\mathcal C} = 0$ initially, then the
two systems are equivalent analytically, since both will guarantee
that ${\mathcal C} = 0$ in the domain of dependence of the initial
surface.  Numerically, the two systems behave very differently, since
finite difference error will lead to a constraint violation
$|{\mathcal C}| > 0$.  Solving Maxwell's equations in primitive form,
such a constraint violation will not propagate and will remain
constant.  As the model problem of \citet{aabss00} suggests, this
behavior will lead to instabilities when nonlinear sources are
included.  Using the modified evolution equations (\ref{4_Adot2}),
(\ref{4_Edot}) and (\ref{4_Gammadot}), the constraint violation
${\mathcal C}$ propagates with the speed of light, will leave the
numerical grid very quickly, and will ultimately leave behind
${\mathcal C} = 0$.  This behavior has been verified in numerical
implementations of the two systems
\citep{kwb02}.

The addition of the constraints to the evolution equations is by no
means unique to the BSSN system.  \citet{d87} pointed out that
constraint violations can be controlled by adding the constraints to
the ADM evolution equations.  \citet{f97} demonstrated the importance
of the propagation of constraints in unconstrained evolution
calculations, and showed how this is linked to adding the Hamiltonian
constraint to the evolution equations.  Other groups have experimented
with adding the momentum constraints to the ADM equations and have
found stabilizing effects \citep{d87,ys01b,sy01,kllpsst01}.
Similarly, the derivation of many of the hyperbolic systems in Section
\ref{sec4.2} involve an addition of the constraints to the original
equations.

\section{Choosing Coordinates}
\label{sec5}

Before the evolution equations as derived in Section \ref{sec4} can be
integrated for a set of initial data as constructed in Section
\ref{sec3}, suitable coordinate conditions have to be chosen.  In the
framework of the ADM equations, the coordinate conditions are imposed
with the help of the lapse $\alpha$ and the shift vector $\beta^i$,
which determine how the coordinates evolve from one spatial slice to
the next.  Often, choices for the lapse are referred to as ``slicing
conditions'', while choices for the shift are called ``spatial gauge
conditions''.  In the following we will discuss several different
slicings and spatial gauges that have recently been popular in $3+1$
numerical relativity.  Parts of this Section are based on the lecture
notes of \citet{bsa98}.

\subsection{Geodesic Slicing}
\label{sec5.1}

Since we are free to choose any lapse and shift, we might be tempted
to make the particularly simple choice
\begin{equation} \label{5_geodesic}
\alpha = 1,~~~~~~~~~~~~\beta^i = 0.
\end{equation}
For constant lapse the acceleration of normal observers vanishes
according to equation (\ref{2_a}), and for zero shift normal observers
and coordinate observers coincide.  For this choice of the lapse and
shift, coordinate observers are therefore freely falling and follow
geodesics, which explains the name {\em geodesic
slicing}\footnote{Sometimes these coordinates are also called {\em
Gaussian-normal} coordinates}.  

\begin{exmp} \label{5_exmp_1}
In geodesic slicing, the metric (\ref{ss_metric})
of Example \ref{2_exmp_1} reduces to
\begin{equation}
ds^2 = - dt^2 + \psi^4 (dr^2 + r^2 (d\theta^2 + \sin^2\theta d\phi^2)),
\end{equation}
which can be identified with the flat Robertson-Walker metric in
spherical coordinates.  If one further assumes that a homogeneous and
isotropic perfect fluid is comoving with the coordinate observers, it
is easy to show that the constraint and evolution equations
(\ref{2_ham_2}) to (\ref{2_Kdot_2}) are equivalent to the well-known
Friedmann equations.
\end{exmp}

Unfortunately, geodesic slicing is a particularly poor choice for most
numerical simulations.  In a Schwarzschild spacetime
every coordinate observer, if starting from rest, will fall into the
singularity in a finite time, leaving only very little time until the
numerical simulation would break down (see, e.g., the discussion in
\citet{sy78b}).  To make matters worse, coordinate observers are not
only attracted to physical singularities, but also tend to form
coordinate singularities.  This behavior is well known for the
evolution of even small amplitude gravitational waves.  To illustrate
this property, imagine a small gravitational wave packet located
originally at the origin of the coordinate system.  After a short time
the wave packet will have dispersed, leaving behind a flat spacetimes.
The coordinate observers will initially be attracted by the
mass-energy of the gravitational wave packet, and will continue to
coast toward the origin of the coordinate system even after the
gravitational waves have disappeared.  They will intersect after a
finite time and form a coordinate singularity.

This behavior can also be understood from equation (\ref{2_trK}),
which reduces to
\begin{equation}
\partial_t K = K_{ij} K^{ij} = A_{ij} A^{ij} + \frac{1}{3} K^2
\end{equation}
for geodesic slicing in vacuum.  The right hand side is non-negative,
and once a positive value of $K$ has been induced by any small
perturbation it will continue to grow without bound.  Assuming $K =
K_0$ at $t = 0$ and $A_{ij} = 0$, one finds $K = 3K_0/(3 - K_0 t)$,
indicating that a coordinate singularity forms at $t = 3/K_0$.
Clearly, the usefulness of geodesic slicing is very limited (see also
\citet{sn95,acmsst95}).

\subsection{Maximal Slicing}
\label{sec5.2}

The above considerations suggest that the pathologies of geodesic
slicing can be avoided by imposing a condition on the trace of
the extrinsic curvature $K$.  The most popular such choice is
to set $K = 0$ at all times, which implies $\partial_t K = 0$.
Equation (\ref{2_trK}) then yields an equation for the lapse
$\alpha$
\begin{equation} \label{5_ms}
D^2 \alpha = \alpha \left( K_{ij} K^{ij} + 
4 \pi (\rho + S) \right).
\end{equation}

\begin{exmp} \label{5_exmp_2}
As we have seen in Section \ref{sec3.3}, the maximal slicing condition
(\ref{5_ms}) can be combined with the Hamiltonian constraint
(\ref{3_ham_1}) to yield equation (\ref{3_ms_2}) for the product
$\alpha \psi$.  Returning to Example \ref{3_exmp_2}, we find that all
terms on the right hand side of (\ref{3_ms_2}) vanish: $\bar A_{ij} =
0$ because of time symmetry and equation (\ref{3_A_0}), $\bar R = 0$
because the background metric is flat, and $\rho = S = 0$ because of
vacuum.  The resulting equation
\begin{equation}
\bar D^2 (\alpha \psi) = 0
\end{equation}
can therefore be solved very easily.  Using (\ref{ss_psi}) and
choosing boundary conditions such that $\alpha = 1$ for $r \rightarrow
\infty$ and $\alpha = 0$ on the black hole horizon $r = M/2$ we find
\begin{equation} \label{ss_lapse}
\alpha = \frac{1 - M/(2r)}{1 + M/(2r)}.
\end{equation}
With $\beta^i = 0$, the conformal factor given by (\ref{ss_psi})
and the lapse by (\ref{ss_lapse}), the metric (\ref{ss_metric})
now reduces to
\begin{equation} \label{5_ss}
ds^2 = - \frac{1 - M/(2r)}{1 + M/(2r)} \, dt^2
+ \left(1 + \frac{M}{2r}\right)^4 (dr^2 + r^2 d\Omega^2),
\end{equation}
which, as expected, is the well-known Schwarzschild metric in
isotropic coordinates.
\end{exmp}

It can be shown that maximal slicing extremizes the volume of spatial
slices spanned by a set of normal observers.  A familiar example in
Euclidian, three-dimensional space is a soap film suspended by a wire
loop.  Neglecting gravity, surface tension will minimize the area of
the soap film, and as a consequence the trace of its extrinsic
curvature vanishes (as Problem 9.31 in
\citet{lppt75} demonstrates).

From equation (\ref{2_trace_K}) we find that in maximal slicing the
convergence of normal observers vanishes, $\nabla_a n^a = 0$, implying
that the normal congruence is convergence free.  This property
prevents the focusing of normal observers that we found in
geodesic slicing.  

In strong field regions, this condition will tend to hold back the
evolution of the slice, and will make proper time ``advance more
slowly''.  This property is called {\em singularity avoidance}, and is
very desirable for many numerical applications (see also Section
\ref{sec8.1} and Figure \ref{fig8.1}).  Maximal slices starting at 
$v=0$ in a Kruskal diagram, for example, never reach the singularity
and asymptotically approach a limit surface at $R = 3M/2$ (in
Schwarzschild coordinates, see
\citet{ewcdst73,sy78b,es79}).  Unlike in geodesic slicing, the entire
exterior of the black hole can be covered.

The properties of maximal slicing have also been studied in a simple
model problem \citep{sy78b,y79}.  Among other things, this short
calculation reveals that at late times of an approach to a
singularity, maximal slicing makes the lapse fall off exponentially,
which is commonly referred to as ``the collapse of the lapse''.
Numerically, this collapse of the lapse has been observed in many
calculations, including \citet{e84,st85b,pst85}.

While all these properties are very desirable, maximal slicing also
has disadvantages.  Equation (\ref{5_ms}) is an elliptic equation for
the lapse function $\alpha$, and in particular in three spatial
dimensions solving elliptic equations is very expensive
computationally.  Since maximal slicing is only a coordinate
condition, physical results of a simulation should not be affected if
the maximal slicing condition is solved only approximately.  
This suggests that, instead of solving an elliptic equation, one
could convert that equation into a parabolic equation, which is
much faster to solve numerically \citep{bdsstw96,s99a}.  As a first 
step, it may be advantageous to write the maximal slicing equation
as
\begin{equation} \label{5_driver}
\partial_t K = - cK,
\end{equation}
instead of $\partial_t K$, since this prescription will ``drive'' $K$
back to zero in case numerical errors cause it to deviate from zero in
the course of some evolution.  Here $c$ is a positive number of order
unity.  Inserting (\ref{2_trK}) then yields
\begin{equation}
D^2 \alpha - 
\alpha \left(K_{ij} K^{ij} + 4 \pi (\rho + S) \right)
- \beta^i D_i K  = c K.
\end{equation}
This equation may now be solved by rewriting it as a parabolic
equation by adding a ``time'' derivative of $\alpha$
\begin{equation} \label{5_ams}
\partial_{\lambda} \alpha = D^2 \alpha - 
\alpha \left(K_{ij} K^{ij} + 4 \pi (\rho + S) \right)
- \beta^i D_i K  - c K,
\end{equation}
where $\lambda$ is an appropriately chosen ``time'' parameter.  If
this equation is evolved to ``late'' enough $\lambda$ for every
timestep so that convergence $\partial_{\lambda} \alpha = 0$ is
achieved, then equation (\ref{5_driver}) is solved and $K$ will be
driven toward zero.  This approximate implementation of maximal
slicing is sometimes referred to as ``K-driver'' \citep{bdsstw96} and
sometimes as ``approximate maximal slicing'' or AMS
\citep{s99a,su00a}.  While this approach seems very appealing, it also
requires some fine-tuning of the parameters $c$ and $\lambda$.
Moreover, achieving convergence of the parabolic equation
(\ref{5_ams}) may require a fair number of iteration steps.  Most
elliptic solvers that one would use to solve the elliptic equation
(\ref{5_ms}) directly also involve iterative algorithms, and in order
to save computer time one could limit those to just a few iteration
steps also.  One can therefore think of both approaches as taking a
few iteration steps toward an approximate maximal slicing solution.

\subsection{Harmonic Coordinates and Variations}
\label{sec5.3}

Another particularly simple choice of coordinates is {\em harmonic
coordinates}, for which the coordinates $x^{a}$ are harmonic
functions
\begin{equation}
\nabla^2 x^a = 0.
\end{equation}
It can be shown that this condition is equivalent to 
\begin{equation} \label{5_hc}
{}^{(4)}\Gamma^a \equiv g^{bc} {}^{(4)}\Gamma^a_{bc} = 0,
\end{equation}
where ${}^{(4)}\Gamma^a_{bc}$ is the connection associated with the
spacetime metric $g_{ab}$.  Inserting the metric (\ref{2_cont_metric})
into (\ref{5_hc}) shows that in harmonic coordinates the lapse and
shift satisfy the coupled set of hyperbolic equations
\begin{eqnarray}
(\partial_t - \beta^j \partial_j)\, \alpha & = & - \alpha^2 K \\
(\partial_t - \beta^j \partial_j)\, \beta^i & = & - \alpha^2 
\Big(\gamma^{ij} \partial_j \ln \alpha + \gamma^{jk} \Gamma^i_{jk} \Big)
\end{eqnarray}
(see \citet{y79}).

Harmonic coordinates have played an important role in the mathematical
development of general relativity, since they bring the
four-dimensional Ricci tensor $\4R_{ab}$ into a particularly simple
form.  In the three-dimensional case, we found in Section \ref{sec4.3}
that $R_{ij}$ can be written in the form (\ref{4_ricci}) in terms of the
three-dimensional connection functions $\Gamma^i = \gamma^{jk}
\Gamma^i_{jk}$.  In complete analogy, $\4R_{ab}$ can be expressed as
\begin{equation}
\begin{array}{rcl}
\4R_{ab} & = & - \frac{1}{2} g^{cd} g_{ab,cd}
 + g_{c(a} \partial_{b)} {}^{(4)} \Gamma^c +
 {}^{(4)} \Gamma^c {}^{(4)} \Gamma_{(ab)c} \\[1mm]
& & + 2 g^{ed} {}^{(4)} \Gamma^c_{e(a} {}^{(4)} \Gamma_{b)cd}
 + g^{cd} {}^{(4)} \Gamma^e_{ad} {}^{(4)} \Gamma_{ecb}
\end{array}
\end{equation}
Evidently, in harmonic coordinates where ${}^{(4)} \Gamma^a = 0$
Einstein's equations reduce to a set of nonlinear wave equations
\citep{d21,l22}, which is why all of the early hyperbolic formulations
are based on these coordinates \citep{c52,fm72}.

Completely harmonic coordinates have been adopted in only few
three-di\-men\-sional simulations \citep{lt99,g02}.  More popular is 
so-called {\em harmonic slicing}, in which only the 
time-component ${}^{(4)} \Gamma^0$ is set to zero.  Combining 
harmonic slicing with zero shift yields a particularly simple
equation for the lapse,
\begin{equation} 
\partial_t \alpha = - \alpha^2 K,
\end{equation}
which, after inserting (\ref{2_trgamma}) for $\alpha K$, can be
integrated to
\begin{equation} \label{5_alpha_hs}
\alpha = C(x^i) \gamma^{1/2}.
\end{equation}
Here $C(x^i)$ is a constant of integration that may depend on the
spatial coordinates $x^i$, but not on time.  This condition is
identical to keeping the densitized lapse $\hat \alpha = \gamma^{-1/2}
\alpha$ (see equation (\ref{4_dens_lapse})) constant.

\begin{exmp} \label{5_exmp_3}
For the Schwarzschild metric (\ref{5_ss}) of Example \ref{5_exmp_2},
$K$, $\beta^i$ and $\partial_t \alpha$ are all zero.  It is therefore
obvious that the maximally sliced Schwarzschild spacetime in isotropic
coordinates is simultaneously harmonically sliced.  For numerical
purposes, it may be undesirable that the lapse (\ref{ss_lapse})
vanishes on the event horizon at $r = M/2$.  Well-behaved harmonic
slices, for which the lapse does not vanish on the horizon, have been
constructed by \citet{bm88} and \citet{cs97}.
\end{exmp}

The harmonic slicing condition (\ref{5_alpha_hs}) is just about as
simple as the geodesic slicing condition (\ref{5_geodesic}), but it
provides for a much more stable numerical evolution (see, e.g.,
\citet{sn95,bs99}).  It does not focus coordinate observers and,
as Example \ref{5_exmp_3} suggests, metric components of static
solutions are independent of time in harmonic slicing, hence allowing
for long time evolutions (e.g.~\citet{cs97,bhs99}).  However, there is
no guarantee that harmonic slicing will lead to well-behaved
coordinates in more general situations (see,
e.g.~\citet{a97,am98,kn01}) and it has been pointed out that the
singularity avoidance properties of harmonic slicing are weaker than
those of, for example, maximal slicing (e.g.~\citet{sn95};
cf.~\citet{g02}).

Equation (\ref{5_alpha_hs}) is an example of an algebraic coordinate
condition, in which the lapse can be found algebraically, without
having to solve complicated and computer intensive differential
equations as for maximal slicing.  A generalization of that condition
has been suggested by \citet{bmss95}
\begin{equation} \label{5_bm}
\partial_t \alpha = - \alpha^2 f(\alpha) K,
\end{equation}
where $f(\alpha)$ is a positive but otherwise arbitrary function of
$\alpha$.  For $f = 1$ this condition obviously reduces to harmonic
slicing.  For $f = 0$ (and $\alpha = 1$ initially), it reduces to
geodesic slicing (Section \ref{sec5.1}).  Formally, maximal slicing
(Section \ref{sec5.2}) corresponds to $f \rightarrow \infty$.  
For $f = 2/\alpha$
the condition (\ref{5_bm}) can be integrated to yield 
\begin{equation} \label{5_1pluslog}
\alpha = 1 + \ln \gamma,
\end{equation}
where we have used (\ref{2_trgamma}) and chosen a constant of
integration to be unity.  This quite popular slicing condition is
often called ``$1 + \log$'' slicing \citep{b93,bmss95}, and has been used
in various applications (including
\citet{acmsst95,bmsw98,abblnst01,fgimrssst01}).  As an algebraic
slicing condition it has the virtue of being extremely simple to
implement and fast to solve.  It has also been found to have stronger
singularity avoidance properties than harmonic slicing, which can be
motivated by the fact that $f$ becomes large when $\alpha$ becomes
small, so that it probably behaves more like maximal slicing than
harmonic slicing.  A similar slicing condition has been suggested by
\citet{sn95} and \citet{ons97} with the goal of enhancing the 
singularity avoidance properties of harmonic slicing.

\subsection{Minimal Distortion and Variations}
\label{sec5.4}

In Section \ref{sec3} we found that the conformally related metric
$\bar \gamma_{ij}$ has five independent functions, two of which
correspond to true gravitational degrees of freedom and three to
coordinate freedom.  For a stable and accurate numerical evolution it
is desirable to eliminate purely coordinate-related fluctuations in
$\bar \gamma_{ij}$, which suggests that one may want to construct a
gauge condition that minimizes the time change of the conformally
related metric.  This gauge condition is called {\em minimal
distortion} (see \citet{sy78a,sy78b}, whose derivation we will follow
closely).
 
In Section \ref{sec3.3} we introduced the time derivative $u_{ij}$ of
the conformally related metric,
\begin{equation} 
u_{ij} \equiv \gamma^{1/3} \partial_t (\gamma^{-1/3} \gamma_{ij}),
\end{equation}
(equation (\ref{3_u})).  Since $u_{ij}$ is traceless, we can decompose
it into a transverse-traceless and a longitudinal part
\begin{equation}
u_{ij} = u_{ij}^{TT} + u_{ij}^{L},
\end{equation}
similar to the decomposition of the traceless part of the extrinsic
curvature in Section \ref{sec3.2}.  The divergence of the transverse 
part vanishes
\begin{equation}
D^j u_{ij}^{TT} = 0,
\end{equation}
and the longitudinal part can we written as
\begin{equation} \label{5_uL}
u_{ij}^{L} = D_i X_j + D_j X_i - \frac{2}{3} \gamma_{ij} D^k X_k = 
	(L X)^{ij}
\end{equation}
Since $\bar \gamma_{ij} = \gamma^{-1/3} \gamma_{ij}$ is a vector
density of weight $-2/3$, the right hand side of (\ref{5_uL}) can be
identified with the Lie derivative of $\bar \gamma_{ij}$ along the
vector $X^i$,
\begin{equation}
u_{ij}^{L} = \gamma^{1/3} \Lie_{{\bf X}} \bar \gamma_{ij}.
\end{equation}
Evidently, the longitudinal part can be interpreted as arising from a
change of coordinates, generated by $X^i$.  It represents the
coordinate effects in the time development, which can therefore be
eliminated by choosing $u_{ij}^{L}$ to vanish.  This leaves only the
transverse part $u_{ij}^{TT}$, which implies that the divergence of
$u_{ij}$ itself must vanish
\begin{equation} \label{5_md_0}
D^j u_{ij} = 0.
\end{equation}
Combining this with equation (\ref{3_u_1}) yields
\begin{equation}
D^j (L \beta)_{ij} = 2 D^j (\alpha A_{ij})
\end{equation}
or
\begin{equation} \label{5_md_pm}
(\Delta_L \beta)^i = 2 A^{ij} D_j \alpha 
	+ \frac{4}{3}  \alpha \gamma^{ij} D_j K
	+ 16 \pi \alpha j^i,
\end{equation}
where we have replaced the divergence of $A^{ij}$ with the momentum 
constraint.  This is the minimal distortion condition for the shift
vector $\beta^i$.   This condition can also be derived by minimizing
the action
\begin{equation} \label{5_action}
{\mathcal A} \equiv 
\int u_{ij} u_{kl} \gamma^{ik} \gamma^{jl} \gamma^{1/2} d^3 x
\end{equation}
with respect to $\beta^i$.

It is also useful to express (\ref{5_md_pm}) in terms of conformally
related quantities.  Using $(L \beta)^{ij} = \psi^{-4} (\bar L \beta)^{ij}$
and $D_j S^{ij} = \psi^{-10} \bar D_j (\psi^{10} S^{ij})$ for any
symmetric, traceless tensor, we find
\begin{equation}
(\Delta_L \beta)^i = \psi^{-4} \left ( 
	(\bar \Delta_L \beta)^i + (\bar L \beta)^{ij} \bar D_j \ln \psi^6
	\right)
\end{equation}
and hence
\begin{equation} \label{5_md}
(\bar \Delta_L \beta)^i + (\bar L \beta)^{ij} \bar D_j \ln \psi^6
= 2 \tilde A^{ij} \bar D_j \alpha + \frac{4}{3}  \alpha \bar \gamma^{ij} D_j K
+ 16 \pi \psi^4 \alpha j^i.
\end{equation}
Here we have used the rescaling (\ref{4_A_rescaling}) of Section
\ref{sec4.3} for $ \tilde A^{ij}$.  

Evidently, the minimal distortion condition (\ref{5_md}) is fairly
involved and quite difficult to solve numerically.  \citet{s99c}
therefore suggested simplifying the condition.  Just as was the case
maximal slicing, simplifying is justified because the condition is
only a coordinate condition.  We may hope that minimal distortion has
properties that are advantageous for numerical implementations, but
small modifications of the conditions may still lead to similarly
desirable properties.

\citet{s99c} suggested modifying (\ref{5_md}) in two steps.  In the 
first step, we can express $\tilde A^{ij}$ in terms of 
$(\bar L \beta)^{ij}$ as 
\begin{equation} 
\tilde A^{ij} = \frac{1}{2 \alpha} \left( (\bar L \beta)^{ij} 
- \psi^{-4} u^{ij} \right).
\end{equation}
(compare equation (\ref{3_A_2})).  Assuming $u_{ij} = 0$ in this
expression, and inserting into (\ref{5_md}) we find
\begin{equation} \label{5_md_2}
(\bar \Delta_L \beta)^i 
= 2 \alpha \tilde A^{ij} \bar D_j \ln (\alpha \psi^{-6}) 
+ \frac{4}{3}  \alpha \bar \gamma^{ij} D_j K
+ 16 \pi \psi^4 \alpha j^i.
\end{equation}

Alternatively, this expression can be derived by modifying the
original condition  (\ref{5_md_0}), $D^i u_{ij} = 0$, which is
equivalent to $\bar D^i (\psi^2 u_{ij}) = 0$, to 
\begin{equation} \label{5_md_mod1}
\bar D^i (\psi^{-4} u_{ij}) = 0
\end{equation}
\citep{s99c}.  Note that the condition (\ref{5_md_2}) is identical 
to the shift condition (\ref{3_ts_md}) that we found in the thin
sandwich approach with $u_{ij} = 0$, $K = 0$ and conformal flatness
(Section \ref{sec3.3}).  The assumptions of the thin sandwich approach
hence leads to the minimal distortion condition for the shift.

In the second step \citet{s99c} simplifies the operator $(\bar
\Delta_L \beta)^i$ in (\ref{5_md_2}) by replacing it with the flat
vector Laplacian $(\Delta_L^{\rm flat} \beta)^i$, so that in Cartesian
coordinates the covariant derivatives can be replaced with partial
derivatives.  This condition is called the ``approximate minimal
distortion'' (AMD) condition, and has been used successfully in
various applications \citep{s99a,s99c,su00a,sbs00a,sbs00b}.
Strategies for solving the flat vector Laplacian are discussed
in Appendix \ref{appB}.

In collapse situations, however, \citet{s99c} showed that minimal
distortion leads to shift vectors that point outwards, leading to
coordinate points being shifted outward, and hence to a coarser
resolution at the center of the collapse.  This is clearly not
desirable, because one would probably want an increasingly fine
resolution at the center of the collapse.  \citet{s99c} experimented
with various ways of adding an artificial radial component to the
shift, which compensates this effect and improves the numerical
performance.

A gauge condition that is closely related to minimal distortion is
based on the conformal connection function $\bar \Gamma^i$ of the 
BSSN formulation (Section \ref{sec4.3}).  These functions could
be set to zero, which would result in ``conformal three-harmonic''
coordinates (compare Section \ref{sec5.3}).  Instead, \citet{ab01}
suggested setting their time-derivative to zero, 
\begin{equation} \label{5_GF_1}
\partial_t \bar \Gamma^i = 0.
\end{equation}
Inserting this into (\ref{4_Gammadot3}) leads to the ``Gamma freezing''
condition
\begin{equation} \label{5_GF_2}
\begin{array}{rcl}
& &  	\bar \gamma^{lj} \beta^i_{,lj}
	+ \frac{1}{3} \bar \gamma^{li} \beta^j_{,jl} 
	+ \beta^j \partial_j \bar \Gamma^i 
	- \bar \Gamma^j \partial_j \beta^i
	+ \frac{2}{3} \bar \Gamma^i \partial_j \beta^j \\ 
& &~~~~ = 2 \tilde A^{ij} \partial_j \alpha - 2 \alpha \Big(
	\bar \Gamma^i_{jk} \tilde A^{kj} 
	- \frac{2}{3} \bar \gamma^{ij} \partial_j K
	- \bar \gamma^{ij} S_j + 6 \tilde A^{ij} \partial_j \phi \Big).
\end{array}
\end{equation}
The relation to minimal distortion can be seen by inserting
(\ref{4_ccf}) into the condition (\ref{5_GF_1}), which yields
\begin{equation}
\partial_j \partial_t \bar \gamma^{ij} = \partial_j (\psi^{4} u^{ij}) = 0
\end{equation}
(compare with the divergence conditions (\ref{5_md_0}) and
(\ref{5_md_mod1})).  To simplify the numerical implementation of
condition (\ref{5_GF_2}),
\citet{ab01} converted the elliptic equation into a parabolic one,
\begin{equation}
\partial_t \beta^i = k \, \partial_t \bar \Gamma^i,
\end{equation}
where $k$ is positive and where (\ref{4_Gammadot3}) should be inserted
for $\partial_t \bar \Gamma^i$.  In analogy to the ``K-driver'' of
\citet{bdsstw96} (see Section \ref{sec5.2}), this condition has been 
called the ``Gamma driver'' condition.  \citet{abpst01} also experimented
with a hyperbolic version
\begin{equation}
\partial_t^2 \beta^i = \psi^{-4} k \, \partial_t \bar \Gamma^i 
	- \eta \, \partial_t \beta^i,
\end{equation}
where both $k$ and $\eta$ are positive constants.

Another gauge condition that is closely related to minimal distortion
is the {\em minimal shear} or {\em minimal strain} condition, in which
the time change of the physical metric $\gamma_{ij}$ (as opposed to
the conformal metric) is minimized in the action (\ref{5_action}).
\citet{bct98} suggested to combine this condition with a similar
condition for the lapse to construct ``comoving'' coordinates that are
particularly well suited for simulations of the slow inspiral outside
of the ISCO (see also \citet{gg99} and
\citet{ggio00}).
\section{Locating Black Hole Horizons}
\label{sec6}

A black hole is defined as a region of spacetime out of which no null
geodesics can escape to infinity.  The {\em event horizon}, or surface
of the black hole, is the boundary between those events which can emit
light rays to infinity and those which cannot.  More formally, it is 
defined as the boundary of the causal past of future infinity (see, e.g.,
\citet{he73,w84}).  It is formed by those outward-going, future-directed 
null geodesics which neither escape to infinity nor fall toward the
center of the black hole.  The event horizon is obviously a
gauge-invariant entity, and contains important geometric information
about the spacetime.  Unfortunately, its global properties make it
very difficult to locate, since, in principle, knowledge of the entire
future spacetime is required to decide whether or not any particular
null geodesic will ultimately escape to infinity or not.  An event
horizon can therefore at best be found ``after the fact'', meaning
after an evolution calculation has evolved to some stationary state.

Locating event horizons after the completion of a calculation may be
sufficient for diagnostic purposes, for analyzing the geometrical and
astrophysical results of a black hole simulation, but locating black
holes in numerical simulations is also important for a more technical
reason.  The spacetime singularities at the center of black holes have
to be excluded from the numerical grid, since they would otherwise
spoil the numerical calculation (see also Section \ref{sec8.1}).  As
we have seen in Section~\ref{sec4}, spacetime slicings can be chosen
so that they avoid singularities (for example maximal slicing,
Section~\ref{sec4.2}).  Typically, however, these slices quickly
develop grid pathologies which also cause numerical codes to crash.
Realizing that the interior of a black hole can never influence the
exterior suggests an alternative solution, namely ``excising'' a
spacetime region just inside the event horizon from the numerical
domain~\citep{u84}.  This approach requires at least approximate
knowledge of the location of the horizon at all times during the
evolution, and the construction of the event horizon after the fact is
therefore not sufficient.

In practice one therefore locates {\em apparent horizons} during the
evolution.  The apparent horizon is defined as the outermost smooth
2-surface, embedded in the spatial slices $\Sigma$, whose outgoing
null geodesics have zero expansion.  As we will see, the apparent
horizon can be located on each slice $\Sigma$, and is therefore a
local-in-time concept.  The singularity theorems of general relativity
(see, e.g.,~\citet{he73,w84}) tell us that if an apparent horizon
exists on a given time slice, it must be inside a black hole event
horizon.  This makes it safe to excise the interior of an apparent
horizon from a numerical domain\footnote{This is not necessarily true
in other theories of gravity.  In Brans-Dicke theory, for example,
apparent horizons may exist outside of event horizons, so that only a
part of the region inside the apparent horizon can be excised
(see~\citet{sst95b} for a numerical example).}.  Note, however, that
the absence of an apparent horizon does not necessarily imply that no
black hole is present.  It is possible, for example, to construct
slicings of the Schwarzschild geometry in which no apparent horizon
exists~\citep{wi91}.  The latter clearly demonstrates the
gauge-dependent nature of the apparent horizon.  In numerical
simulations, one simply hopes that the chosen slices are sufficiently
non-pathological, and that the apparent horizon is reasonably close to
the event horizon.  This is the case, for example, for stationary
situations, for which the apparent horizon and the event horizon
coincide.

Recently, the concept of {\em isolated horizons} has also been
introduced (see, e.g., \citet{abf99,abdfklw00}), and the first
implementations in numerical relativity have been reported in
\citet{dkss02}.

\subsection{Locating Apparent Horizons}
\label{sec6.1}

\begin{exmp} \label{6_exmp_1}
In spherical symmetry, the concept of an apparent horizon is quite
transparent, and can be illustrated very easily.  The apparent horizon
is defined as the boundary of the region of trapped surfaces, wherein
the cross-sectional area spanned by a beam of outgoing light rays
immediately evolves to a smaller area.

For the metric~(\ref{ss_metric}), an outgoing light ray satisfies
\begin{equation}
\frac{dr}{dt} = \frac{\alpha}{\psi^2} - \beta,
\end{equation}
and the areal radius is given by $\psi^2r$.  The apparent horizon
condition can therefore be found from the condition that the total
time derivative of $\psi^2r$ along the right ray vanish
\begin{eqnarray} \label{ss_ah1}
\frac{d}{dt} (\psi^2 r) & = & \displaystyle 
	\frac{\partial}{\partial t} (\psi^2 r) 
	+ \frac{dr}{dt} \frac{\partial}{\partial r} (\psi^2 r)  \nonumber \\
& = & 2 r \psi^2 \left( \frac{\dot \psi}{\psi} - \beta \frac{\psi'}{\psi} 
	- \frac{\beta}{2r} \right)  
	+ 2 r \alpha \frac{\psi'}{\psi} + \alpha = 0.
\end{eqnarray}
The expression in brackets can be rewritten in terms of the extrinsic
curvature (\ref{ss_extrinsic_curvature}), which yields 
\begin{equation} \label{ss_ah2}
- \psi^2 r K^{\theta}_{~\theta} + 2 \frac{\psi'}{\psi} r + 1 = 0.
\end{equation}
This condition can be evaluated very easily.  For the Schwarzschild
solution (\ref{ss_psi}), for example, one readily verifies that the
apparent horizon coincides with the event horizon at $r = m/2$.
\end{exmp} 

Consider a closed smooth hypersurface of $\Sigma$, and call it $S$.
By construction, $S$ is spatial and two-dimensional.  Let $s^a$ be its
unit outward pointing normal in $\Sigma$.  Obviously $s^a$ then
satisfies $s_a s^a = 1$ and $s^a n_a = 0$.  Similarly to how the
spacetime metric $g_{ab}$ induces the spatial metric $\gamma_{ab}$ on
$\Sigma$ (see Section~\ref{sec2.1}), the latter now induces a
two-dimensional metric
\begin{equation}
m_{ab} = \gamma_{ab} - s_a s_b = g_{ab} + n_a n_b - s_a s_b
\end{equation}
on $S$.  We now consider the outgoing future-pointing null geodesics 
whose projection on $\Sigma$ is orthogonal to $S$.  Up to an overall
factor, the tangents $k^a$ to these geodesics can be constructed, on 
$S$, from
\begin{equation} \label{6_k}
k^a = s^a + n^a,
\end{equation}
which automatically satisfies $k_a k^a = 0$ and $m_{ab} k^a = 0$. 
We parallel propagate $k^a$ away from $S$ with the geodesic equation $k^a
\nabla_a k^b = 0$.  A {\em marginally trapped surface} (or sometimes
called marginally outer-trapped surface) is now defined as a surface
on which the expansion $\Theta$ of the outgoing null geodesics
orthogonal to $S$ vanishes everywhere\footnote{Note that this
condition is equivalent to $m^{ab}
\nabla_a k_b = 0$.}
\begin{equation} \label{6_ah1}
\Theta = \nabla_a k^a = 0.
\end{equation}
The outermost such surface is called the {\em apparent horizon}.

For the purposes of numerical relativity, it is useful to rewrite this
equation in terms of three-dimensional objects.  To do so, we 
insert~(\ref{6_k}) into~(\ref{6_ah1}), which yields
\begin{equation}
\Theta = \nabla_a (n^a + s^a) = - K + \nabla_a s^b
\end{equation}
where we have used equation~(\ref{2_trace_K}).  The divergence of
$s^a$ can now be rewritten
\begin{eqnarray}
\nabla_a s^b & = & g^a_{~b} \nabla_a s^b 
	= (\gamma^a_{~b} - n^a n_b) \nabla_a s^b
	= D_i s^i - n^a n_b \nabla_a s^b \nonumber \\
	& = & D_i s^i + n^a s^b \nabla_a n_b
	= D_i s^i + s^i s^j K_{ij} + k^a s^b \nabla_a n_b.
\end{eqnarray}
The last term vanishes identically, 
as can be seen by inserting~(\ref{6_k}) again,
\begin{equation}
k^a s^b \nabla_a n_b = s^b k^a \nabla_a k_b - k^a s^b \nabla_a s_b = 0.
\end{equation}
We can now combine results to find the apparent horizon equation
\begin{equation} \label{6_ah2}
\Theta = D_i s^i - K + s^i s^j K_{ij} = 0,
\end{equation}
which can also be written as
\begin{equation} \label{6_ah3}
\Theta = m^{ij} (D_i s_j - K_{ij}) = 0.
\end{equation}
(cf.~\citet{y89} and \citet{g98}).  This condition only depends on
spatial quantities defined within each slice $\Sigma$, which makes it
obvious that it can constructed ``locally in time'' on each slice.

\begin{exmp} \label{6_exmp_2}
We can now verify that condition~(\ref{6_ah3}) leads to~(\ref{ss_ah2})
in spherical symmetry.  In this case, $s^a$ only has a radial
component, $s^r = \psi^{-2}$, and the only non-vanishing components of
$m^{ij}$ are $m^{\theta\theta} = \gamma^{\theta\theta}$ 
and $m^{\phi\phi} = \gamma^{\phi\phi}$.  The term
$m^{ij} K_{ij}$ is therefore
\begin{equation}
m^{ij} K_{ij} = K^{\theta}_{~\theta} + K^{\phi}_{~\phi} = 
2 K^{\theta}_{~\theta},
\end{equation}
and $m^{ij}D_i s_j$ reduces to
\begin{equation}
m^{ij}D_i s_j = m^{ij} (s_{j,i} - \Gamma^k_{ij} s_k) 
  = \frac{2}{\psi^2 r} \left( 2 \frac{\psi'}{\psi} r + 1 \right),
\end{equation}
where we have used the connection coefficients~(\ref{ss_connection}).
Inserting the last two equations into the condition~(\ref{6_ah3})
immediately yields~(\ref{ss_ah2}), as expected.
\end{exmp}

Various different methods have been employed to locate apparent horizons.
Most of these methods have in common that they characterize the horizon
as a level surface of a scalar function, e.g.
\begin{equation}
\tau(x^i) = 0.
\end{equation}
The unit normal $s^a$ can then be expressed as
\begin{equation}
s^i = \lambda \gamma^{ij} D_j \tau
\end{equation}
where $\lambda$ is the normalization factor
\begin{equation}
\lambda \equiv (\gamma^{ij} D_i \tau D_j \tau)^{-1/2}.
\end{equation}
Inserting these into equation~(\ref{6_ah3}) yields
\begin{equation} \label{6_ah4}
\Theta = m^{ij} (\lambda D_i D_j \tau - K_{ij}).
\end{equation}
The expansion $\Theta$ is therefore a second order differential operator
on $\tau$.  The principal part of this operator is the Laplace
operator with respect to the two-dimensional metric $m_{ij}$ 
on $S$ (see also the discussion in~\citet{g98}).

Most authors have chosen $\tau$ to be of the form\footnote{This choice
restricts the topology of $S$ to $S^2$, and $S$ must furthermore be
star-shaped around $C^i$ (compare the discussion in~\citet{g98}).}
\begin{equation} \label{6_tau}
\tau(x^i) = r_C(x^i) - h(\theta,\phi),
\end{equation}
where $r_C$ is the coordinate separation between the point $x^i$ and
some location $C^i$ inside the $\tau = 0$ surface, and where $\theta$
and $\phi$ are polar coordinates centered on $C^i$.  In fact, most
authors assume that $C^i$ is the origin of the coordinate system,
$C^i = 0$.  The function $h$ measures the coordinate distance from
$C^i$ to the $\tau = 0$ surface in the direction $(\theta,\phi)$.

Inserting~(\ref{6_tau}) into~(\ref{6_ah4}) yields a second order
differential operator on $h(\theta,\phi)$ (which, when considered as a
differential operator in two dimensions, is elliptic).  In spherical
symmetry, where $h$ is a constant, the problem reduces to solving an
algebraic equation, as we have seen in Examples~\ref{6_exmp_1}
and~\ref{6_exmp_2}.  In axisymmetry $h$ only depends on $\theta$, and
equation~(\ref{6_ah4}) becomes an ordinary differential equation which
has to be solved with periodic boundary conditions.  Solutions have
been constructed with shooting methods (for example
\citet{c74,b84,st92}), spectral methods \citep{e77} and finite
difference methods \citep{cy90}.  Both spectral and finite difference
methods have also been employed in three-dimensional problems, without
any simplifying symmetry assumptions.

\citet{nko84} (also \citet{nko85}) adopt a spectral method.  They expand 
$h$ in spherical harmonics
\begin{equation} \label{6_ah_expansion}
h(\theta,\phi) = \sum_{l=0}^{l_{\rm max}} \sum_{m=-l}^{l} 
	a_{lm} Y_{lm}(\theta,\phi)
\end{equation}
and construct an iterative algorithm to determine the expansion
coefficients $a_{lm}$, which is most easily explained in the
notation of~\citet{g98}.  Recall that the principal part of the
nonlinear operator $\Theta$ acting on $h$ is a two-dimensional
Laplacian with respect to $m_{ij}$.  The key idea is introduce
a linear elliptic operator, which is easier to invert, and to subtract 
it from the non-linear one.  \citet{nko84} use the flat Laplacian
on a 2-sphere,
\begin{equation} \label{6_nko1}
L^2 h \equiv h_{,\theta\theta} + \cot\theta h_{,\theta} + 
	\sin^{-2}\theta h_{,\phi\phi}
\end{equation}
and rewrite the equation $\Theta = 0$ as
\begin{equation} \label{6_nko2}
L^2 h = \rho \, \Theta + L^2 h.
\end{equation}
Here the scalar function $\rho$ is chosen so that the partial
derivative $h_{,\theta\theta}$ cancels on the right hand side (see
\citet{g98} for a generalization).  Since $L^2 Y_{lm} = - l(l+1) Y_{lm}$, 
we can now multiply both sides
with $Y^*_{lm}$ and integrate over $S$ to find
\begin{equation} \label{6_nko3}
- l(l+1) \, a_{lm} = \int_S  Y^*_{lm} (\rho \, \Theta + L^2 h) \, d\Omega,
\end{equation}
where $d\Omega = \sin \theta d\theta d\phi$, and where we have used the
orthogonality of the spherical harmonics.  A priori, this equation is 
not very helpful, since the right hand side has to be evaluated at $S$, which
depends on the very $a_{lm}$ that we would like to determine.  However,
equation~(\ref{6_nko3}) can now be used to define an iteration procedure,
in which the integral on the right hand side is evaluated using a previous
set of guesses $a_{lm}^n$ to determine a new set $a_{lm}^{n+1}$.  Obviously,
this algorithm works only for $l \geq 1$, and $a_{00}$ has to be determined
independently from the integral on the right hand side alone.  A similar 
scheme has been implemented by \citet{kb91}.

A variant of the spectral method, in which the problem of finding a
root of $\Theta$ is reduced to a multi-dimensional minimization
problem, was proposed by \citet{lmss96} and implemented independently
by \citet{bcsst96} and \citet{aclmss98}.  In this approach, $\Theta^2$
is integrated over $S$
\begin{equation}
{\mathcal S} = \int_S \Theta^2 d\sigma,
\end{equation}
where $d\sigma$ is the proper area element on $S$.  The function $h$
is again expanded as in~(\ref{6_ah_expansion})\footnote{Except that
these authors expand $h$ in terms of symmetric traceless tensors
instead of spherical harmonics, which is completely equivalent, but
more convenient in Cartesian coordinates.}, so that ${\mathcal S}$
becomes a function of the expansion coefficients $a_{lm}$.  A standard
minimization method is then used to vary the $a_{lm}$ until a minimum
of ${\mathcal S}$ has been found.  An apparent horizon has been
located if ${\mathcal S}$, according to some appropriate criterion, is
sufficiently close to zero.

\citet{s97} (also \citet{su00b}) uses the same ansatz~(\ref{6_nko2}) as
\citet{nko84}, but employs a finite difference method instead of a spectral 
method to solve it.  In particular, $S$ is covered with a finite
difference grid $(\theta_i,\phi_j)$, on which $h(\theta,\phi)$ is
represented as $h_{i,j}$.  The operator~(\ref{6_nko1}) is then finite
differenced in a straight-forward, second-order fashion
\begin{eqnarray} \label{6_s1}
(L^2 h)_{i,j} & = & \frac{h_{i+1,j} - 2 h_{i,j} + h_{i-1,j}}{(\Delta \theta)^2}
	+ \cot \theta_i \frac{h_{i+1,j} - h_{i-1,j}}{2 \Delta \theta} + 
	\nonumber \\
	& & \sin^{-2} \theta_i
	\frac{h_{i,j+1} - 2 h_{i,j} + h_{i,j-1}}{(\Delta \phi)^2}. 
\end{eqnarray}
Equation~(\ref{6_nko2}) can again be solved using an iterative
algorithm.  On the right hand side, the operator~(\ref{6_s1}) can be
evaluated for a previous set of values $h^n_{i,j}$.  On the left hand
side, the same operator acts on the new values $h^{n+1}_{i,j}$.
Evaluating~(\ref{6_nko2}) at all gridpoints $(\theta_i,\phi_j)$ then
yields a coupled set of linear equations for the $h^{n+1}_{i,j}$,
which can be solved with standard techniques of matrix inversion
(e.g.~\citet{ptvf92}).  

\citet{t96} and \citet{hcm00} have also implemented finite difference 
methods to locate apparent horizons.  They, however, do not use the
ansatz~(\ref{6_nko2}), and instead finite difference the nonlinear
equation $\Theta = 0$ directly.  The resulting nonlinear system of
equations is then solved with Newton's method.  A similar method 
has been used by \citet{s02}, except that here the tensor fields on
the horizon are represented in Cartesian coordinate components.

Yet another method, a curvature flow method, was proposed by 
\citet{t91}.  This method is related to solving an elliptic
equation by converting it into a parabolic equation.  During the 
evolution in an unphysical ``time'' parameter, the solution of the parabolic
problem settles down into an equilibrium solution, which is the 
solution to the original elliptic problem.  Similarly, \citet{t91}
proposes to deform a trial surface $S$ according to
\begin{equation} \label{6_tod1}
\frac{\partial{x^i}}{\partial \lambda} = - s^a \Theta,
\end{equation}
where $\lambda$ is the unphysical time parameter.  For time-symmetric
data with $K_{ij} = 0$, $\Theta$ reduces to the trace of the extrinsic
curvature of $S$ in $\Sigma$, $D_i s^i$.  The apparent horizon then
satisfies $D_i s^i = 0$ and is therefore a minimal surface (compare
Problem 9.31 in \citet{lppt75}), for which Tod's method is known to
converge.  For general data, the flow~(\ref{6_tod1}) is no longer
guaranteed to converge, but numerical experience shows that in general
it still does.  Implementations of Tod's method have been described in
a number of unpublished reports (see, e.g.~\citet{g98} and
\citet{shm00} for references).

\citet{g98} generalizes the flow prescription~(\ref{6_tod1}) and 
describes a new family of spectral algorithms which include the
methods of \citet{t91} and \citet{nko84} as special cases for
particular values of certain parameters.  It is suggested that other
algorithms in this family may combine the robustness of the former
with the speed of the latter.

\citet{abbgmsw00} have compared the algorithm of \citet{g98} with 
the minimization method of~\citet{aclmss98} for a number of different
test problems, without finding clear advantages of one algorithm over
the other.  The spectral method of \citet{g98} is reported to be
generally much faster than the minimization routine, but the
efficiency of the latter could probably be improved dramatically.
Both \citet{aclmss98} and \citet{bcsst96} employ Powell's method for
the multi-dimensional minimization, since it does not use derivatives
of the function and is hence easy to implement. \citet{ptc00} report
on replacing Powell's method in the code of \citet{bcsst96} with a
Davidson-Fletcher-Powell algorithm (see \citet{ptvf92}) and find a
significant speed-up.

\citet{shm00} implement a finite difference version of a flow description
with is also a slightly generalized version of~(\ref{6_tod1}).  They
also introduce a so-called level flow, in which the flow proceeds to a
surface of constant, but not necessarily zero expansion $\Theta$.  The
apparent horizon is then located by constructing a sequence of
such surfaces, including the case $\Theta = 0$.  This approach may have
certain advantages for handling situations in which multiple apparent
horizons are present.

\subsection{Locating Event Horizons} 
\label{sec6.2} 

Event horizons, formally defined as the boundary of the causal past of
future infinity, are traced out by outgoing light rays that never
reach future null infinity and never hit the singularity.  In
principle, therefore, knowledge of the entire future evolution of a
spacetime is necessary to locate event horizons.  In practice,
however, event horizons can be located fairly accurately after a
finite evolution time, once a spacetime has settled down to an
approximately stationary state.

\citet{hkwwst94} constructed an event horizon finder by evolving
null geodesics
\begin{equation}
\frac{d^2 x^a}{d \lambda^2} 
+ {}^{(4)}\Gamma^a_{bc} \frac{dx^b}{d\lambda} \frac{dx^c}{d\lambda} = 0,
\end{equation}
where $\lambda$ is an affine parameter.  In $3+1$ form, this equation
can be rewritten as
\begin{eqnarray}
\frac{dp_i}{d\lambda} & = & 
- \alpha \alpha_{,i} (p^0)^2 + \beta^k_{~,i}p_k p^0 
- \frac{1}{2} \gamma^{lm}_{~~,i}p_l p_m \\
\frac{dx^i}{d\lambda} & = & \gamma^{ij} p_j - \beta^i p^0,
\end{eqnarray}
where we have used $p^i = dx^i/d\lambda$ and $p^0 = (\gamma^{ij} p_i
p_j)^{1/2}/\alpha$ (which enforces $g^{ab}p_a p_b = 0$).

From each point in a numerically generated spacetime, for which the
lapse $\alpha$, the shift $\beta^i$ and the spatial metric
$\gamma_{ij}$ are known on a numerical grid, light rays can then be
sent out in many different directions $p^i$.  To determine whether or
not this point is inside an event horizon, \citet{hkwwst94} use the
additional knowledge of apparent horizons.  Conceptually, if all light
rays sent out from a point end up inside an apparent horizon (which is
always located inside an event horizon) the point is inside the event
horizon as well.  If, on the other hand, at least one light ray sent
out from the point escapes to large separations, the point is not
inside an event horizon.  In this way, the ejection and propagation of
light rays from various points in spacetime can determine the location
of the event horizon.

An alternative and quite attractive approach was suggested by
\citet{abblmsssw95} and \citet{lmssw96}.  Realizing that the future 
directed light rays diverge away from the event horizon, either toward
the interior of the black hole or toward future null infinity, they
suggest integrating null geodesics {\em backwards} in time, which
converge to the event horizon.  In practice, \citet{hkwwst94} also
employ backwards integration of light rays for the same reason.  This
method is particularly efficient if one can identify a
``horizon-containing'' region in which the event horizon is expected
to reside.  It is then sufficient to integrate light rays from this
limited region, and they will quickly be attracted by the event
horizon.

\citet{abblmsssw95} and \citet{lmssw96} also pointed out that instead
of integrating individual null geodesics, one may integrate entire
null surfaces backward in time.  Defining such a null surface as 
a level surface $f(t,x^i) = 0$, one can rewrite the null condition
\begin{equation}
g^{ab} \partial_a \partial_b f = 0
\end{equation}
to find an evolution equation for $f$
\begin{equation}
\partial_t f = \frac{-g^{ti} \partial_i f + 
\sqrt{
(g^{ti} \partial_i f)^2 - g^{tt} g^{ij}\partial_i f \partial_j f
} }{g^{tt}}.
\end{equation}
An event horizon can then be located by evolving two such surfaces 
defining the inner and outer boundary of the horizon-containing
region backward in time.  The two surfaces will converge very quickly
and will bracket the event horizon.  

Examples of event horizons in numerically generated spacetimes, for 
example for the head-on collision of two black holes, can be found 
in \citet{hkwwst94,abblmsssw95,msssstw95} and \citet{lmssw96}.

\section{Binary Black Hole Initial Data}
\label{sec7}

In this Section we will discuss various approaches to solving the
initial value equations of Section \ref{sec3} for spacetimes that
describe binary black holes in approximately circular orbit.  Such
initial data may be used as initial data for dynamical simulations of
the plunge and merger, as ``snapshots'' of the evolutionary sequence
up to the ISCO, to locate the ISCO, and finally as background models in
quasi-adiabatic evolutionary calculations to simulate the late
inspiral phase.  For a more complete review of initial data for
numerical relativity see \citet{c00}.

Under sufficiently restrictive assumptions, constructing initial data
containing multiple black hole is almost trivial.  Recall from Example
\ref{3_exmp_2} that for conformally flat ($\bar \gamma_{ij} =
\eta_{ij}$), time-symmetric ($K_{ij} = 0$) vacuum spacetimes ($\rho =
j^i = 0$), the momentum constraint (\ref{3_mom_1}) is solved trivially
and the Hamiltonian constraint (\ref{3_ham_1}) reduces to the simple
Laplace equation (\ref{ss_ham_ts}), which, with suitable boundary
conditions, has solutions of the form (\ref{ss_psi}).  Given that the
Laplace equation (\ref{ss_ham_ts}) is linear, we can obviously add
several solutions (\ref{ss_psi}) and find
\begin{equation}
\psi = 1 + \sum_{\alpha} \frac{m_{\alpha}}{2 r_{{\bf C}_{\alpha}}}
\end{equation}
for an arbitrary number of black holes. Here $r_{{\bf C}_{\alpha}} =
\|x^i - C^i_{\alpha}\|$, and $C^i_{\alpha}$ is the coordinate location
of the $\alpha$-th black hole.

However, even for two black holes this construction is not unique.
We could define a coordinate mapping equivalent to that of Example
\ref{3_exmp_2} for each one of the black holes, but would find that 
for each one the presence of the companion would destroy the isometry
that we found for a single black hole \citep{l63}.  In other words, we
can think of the two throats as connecting one sheet or universe with
two black holes with two separate asymptotically flat sheets, each one
only containing one black hole.  This topology is called a
``three-sheeted'' topology.  The isometry can be restored by adding
additional throats inside each one of the already existing throats,
which correspond to mirror images of the companion black hole
\citep{m63,ksy83}.  This ``conformal-imaging'' approach leads to a
two-sheeted topology, in which the two throats connect to identical,
asymptotically flat sheets.  It is also possible to consider a
one-sheeted, but multiply connected topology, in which the two ends of
a ``wormhole'' represent the two black holes \citep{m60}.

The non-uniqueness of these solutions stems from the fact that
Einstein's equations determine the local geometry of a spacetime, but
do not fix its topology.  In more physical terms, solutions with
different topologies differ by their initial gravitational wave
content, which is also not determined by Einstein's equations.

From an astrophysical point of view, the above solutions are not very
interesting because they assume time-symmetry with $K_{ij} = 0$.  For
the construction of binary black holes in binary orbit, we will be
interested in black holes with finite momenta, and hence non-vanishing
extrinsic curvature.  That means that the momentum constraint is no
longer solved as an identity, and that the extrinsic curvature
introduces a non-linear term into the Hamiltonian constraint.  In the
following we will discuss several different approaches to constructing
such binary black hole solutions.  As we have explained in Section
\ref{sec1}, we will be particularly interested in binary black holes
in {\em quasi-circular} orbit.

\subsection{The Bowen-York Approach}
\label{sec7.1}

In the Bowen-York approach \citep{by80,b82,y89}, initial data are
constructed using the conformal transverse-traceless decomposition of
Section \ref{sec3.2}.  Assuming maximal slicing ($K=0$) and
conformal flatness ($\bar \gamma_{ij} = \eta_{ij}$),  the momentum
constraint (\ref{3_mom_1}) reduces to
\begin{equation} \label{7_mom_1}
(\Delta_L^{\rm flat} W)^i = 0 
\end{equation}
(compare Appendix \ref{appB}).  This equation is solved analytically
by
\begin{equation}
W^i_{\bf CP} = - \frac{1}{4r_{\bf C}} (7 P^i + n^i_{\bf C} n^j_{\bf C} P_j)
\end{equation}
where $n^i_{\bf C} = (x^i - C^i)/r$ is the normal vector pointing away
from the center of the black hole at $C^i$, and where $P^i$ is an
arbitrary vector.  Constructing the extrinsic curvature from this
solution then yields
\begin{equation} \label{7_A_1}
\bar A^{ij}_{\bf CP} = \frac{3}{2r^2_{\bf C}} 
	\left(P^i n^j_{\bf C} + P^jn^i_{\bf C} +
	(\eta^{ij} + n^i_{\bf C} n^j_{\bf C}) P_k n^k_{\bf C} \right).  
\end{equation}
Inserting this into (\ref{3_p}) shows that $P^i$ is the linear 
momentum of the black hole.  

Since the momentum constraint (\ref{7_mom_1}) is linear, a binary black
hole solution can now be constructed by superposition of single solutions
\begin{equation} \label{7_A_2}
\bar A^{ij} = \bar A^{ij}_{\bf C_1P_1} + \bar A^{ij}_{\bf C_2P_2}
\end{equation}
The total linear momentum of this solution is $P^i = P_1^i + P_2^i$,
and from (\ref{3_j}) we find that the total angular momentum is given
by
\begin{equation}
J_i =  \epsilon_{ijk} C_1^j P_1^k + \epsilon_{ijk} C_2^j P_2^k.
\end{equation}
The extrinsic curvature can then be inserted into the Hamiltonian
constraint, which for conformal flatness and maximal slicing reduces
to
\begin{equation} \label{7_ham_1}
\Delta^{\rm flat} \psi = - \frac{1}{8} \psi^{-7} \bar A_{ij} \bar A^{ij}.
\end{equation}
Here $\Delta^{\rm flat}$ is the flat Laplace operator.

\subsubsection{The Conformal-Imaging Approach}
\label{sec7.1.1}

At this point, the topology of the solution to be constructed has to
be decided on.  In the conformal-imaging approach a two-sheeted
topology is assumed, so that an isometry holds across the throats.
For this isometry to hold, additional image terms have to be added to
the extrinsic curvature (\ref{7_A_2}) before it is inserted into
the Hamiltonian constraint (\ref{7_ham_1}) \citep{ksy83}. 

\citet{c91} and \citet{ccdkmo93} constructed solutions to the Hamiltonian
constraint for two black holes with arbitrary momenta using the
conformal-imaging approach.  In these simulations, the isometry
conditions on the throats were used as boundary conditions, so that
the singularities inside the throats could be eliminated from the
numerical grid.  The computational disadvantage of this method is that
boundary conditions have to be imposed on fairly complicated surfaces.
In finite difference algorithms, this can be accomplished either with
bispherical or \v{C}ade\v{z} coordinates \citep[see also Appendix C of
\citet{c91}]{c71}, designed such that a constant coordinate surface 
coincides with the throat, or else with fairly complicated algorithms
in Cartesian coordinates.  Both approaches, together with a spectral
method, have been compared in \citet{ccdkmo93}.

To construct equal-mass binary black holes in quasi-circular orbit,
one may assume that $P^i \equiv P_1^i = - P_2^i$, and that $P_i C^i =
0$, where $C^i$ is the separation vector $C^i \equiv C_1^i - C_2^i$.
For a given mass of the binary, the only remaining free parameters are
the separation $C \equiv \| C^i \|$, and the momentum $P \equiv \| P^i
\|$.  It is intuitively clear, however, that for each separation $C$
there is only one momentum $P$ that corresponds to a circular orbit --
namely the one that satisfies the equivalent of Kepler's third law.

General relativity does not admit strictly circular orbits, since the
emission of gravitational radiation will lead to loss of energy and
angular momentum, and hence to a shrinking of the orbit. During most
of the inspiral phase (see Section \ref{sec1}), however, the
separation decreases very slowly, on a timescale much larger than the
orbital period.  It is therefore very reasonable to approximate the
orbit as {\em quasi-circular}.

To determine this quasi-circular orbit, \citet{c94} suggested locating
turning points of the binding energy $E_b$ along a sequence of
constant black hole mass $M_{\rm BH}$ and constant angular momentum $J
= CP$.  Restricting ourselves to non-spinning black holes, the mass of
each individual black hole might be identified with the irreducible
mass\footnote{The mass of a single black hole in the presence of
neighboring black holes is not unambiguously defined in general
relativity, in contrast to the total mass as measured at infinity.}
\begin{equation} \label{7_m_irr}
M_{\rm BH} = M_{\rm irr} \approx \left(\frac{A}{16 \pi}\right)^{1/2},
\end{equation}
where $A$ is the proper area of the black hole's event horizon
\citep{c70}.  In numerical simulations, $A$ is approximated as the 
area of the apparent horizon (see Section \ref{sec6.1}).  The binding
energy can then be defined as 
\begin{equation}
E_b = M - 2 M_{\rm BH},
\end{equation}
where $M$ is the total ADM mass of the system, given by
(\ref{3_mass}).  Lastly, we can define $l$ as the proper separation
between the two horizons, measuring the shortest path from one surface
to the other.

Following \citet{c94}, quasi-circular orbits correspond to turning
points
\begin{equation} \label{7_co}
\left. \frac{\partial E_b}{\partial l} \right|_{M_{\rm BH},J} = 0.
\end{equation}
A minimum corresponds to a stable quasi-circular orbit, while a maximum
corresponds to an unstable orbit.  The transition from stable to
unstable orbits defines the innermost stable circular orbit (ISCO),
which occurs at the saddle point
\begin{equation} \label{7_ISCO}
\left. \frac{\partial^2 E_b}{\partial l^2} \right|_{M_{\rm BH},J} = 0.
\end{equation}
For a quasi-circular orbit, the binary's orbital angular velocity
$\Omega$ as measured at infinity can then be determined
from\footnote{Keeping $l$ fixed in this derivative is equivalent to
taking the derivative along an evolutionary sequence
\citep{c94,ptc00}, since the difference only appears at second
order.  I am grateful to H.~Pfeiffer for pointing this out.}
\begin{equation} \label{7_Omega}
\Omega = \left. \frac{\partial E_b}{\partial J} \right|_{M_{\rm BH},l}
\end{equation}
(compare \citet{fus02,bbd02}).  A motivation and illustration of this
approach can be found in \citet{b01}.

\begin{figure}
\begin{center}
\epsfxsize=3in
\epsffile{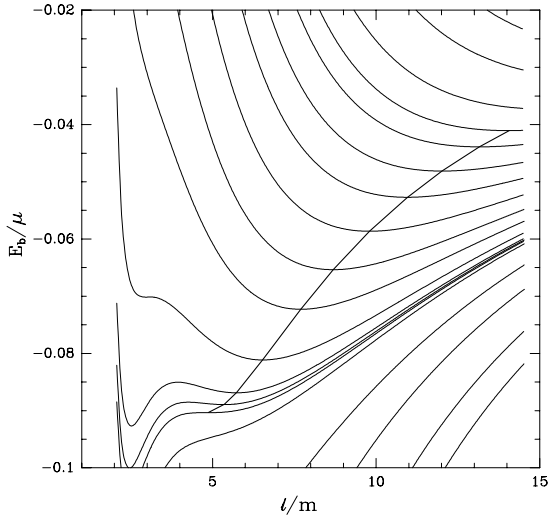}
\end{center}
\caption{The effective potential $E_b/\mu$ as a function of separation
$l$ for various values of the angular momentum $J/\mu m$ (thin lines)
as obtained in the Bowen-York conformal-imaging approach.  Here $\mu$
is the reduced mass $M_{\rm BH}/2$ and $m$ is the sum of the black
hole masses $2 M_{\rm BH}$.  The thick line connects a sequence of
stable quasi-circular orbits, which correspond to minima of the
binding energy (\ref{7_co}).  This sequence ends at the ISCO,
identified by the saddlepoint of the binding energy (\ref{7_ISCO}).
(Figure from \citet{c94}.)}
\label{fig7.1}
\end{figure}

At this point, a word of warning is in order.  As we have discussed
above, relativistic binaries emit gravitational radiation, causing
them to slowly spiral toward each other, and they hence do not follow
strictly circular orbits.  The very concept of an innermost stable
{\em circular} orbit is therefore somewhat ill defined.  Also, the
minimum in the equilibrium energy identifies the onset of a {\em
secular} instability, while the onset of {\em dynamical} instability
may be more relevant for the binary inspiral (see, e.g., the
discussion in \citet{lrs93} and \citet{lrs97}, where it is shown that
the two instabilities coincide in irrotational binaries).  Moreover,
it has been suggested that the passage through the ISCO may proceed
quite gradually \citep{ot00,bd00}, so that a precise definition of the
ISCO may be less meaningful than the above turning method suggests
(see also \citet{dbssu01}).  Ultimately, dynamical evolution
calculations will have to simulate the approach to the ISCO and to
investigate these issues.  For the sake of dealing with a well-defined
problem, we will here identify the ISCO with the saddlepoint of the
equilibrium energy (\ref{7_ISCO}).

Combining the conformal-imaging approach with the turning-point method
\citet{c94} constructed the first models of binary black holes in
quasi-circular orbit (see Figure \ref{fig7.1}, which also illustrates
the construction of circular orbits and the ISCO with the turning
point method).  Numerical values for the non-dimensional binding
energy $\bar E_b \equiv E_b/\mu$, the orbital angular velocity $\bar
\Omega \equiv m \Omega$ and angular momentum momentum 
$\bar J \equiv J/(\mu m)$ at the ISCO as obtained by \citet{c94} and
other authors are listed in Table \ref{7_table1}.  Here $\mu$ is the
reduced mass $\mu = M_{\rm BH}/2$ and $m$ is the sum of the black hole
masses $m = 2 M_{\rm BH}$.  The results of \citet{c94} were
generalized to spinning black holes by \citet{ptc00}.

\begin{table}
\begin{center}
\begin{tabular}{llll}
\hline
\hline
Reference               & $\bar E_b$    & $ \bar J$     &$\bar \Omega$ \\
\hline 
Schwarzschild           & -0.0572       & 3.464         & 0.068 \\ 
\citet{c94}             & -0.09030      & 2.976         & 0.172 \\
\citet{b00}             & -0.092        & 2.95          & 0.18  \\
\citet{ggb01b}		& -0.068	& 3.36		& 0.103 \\
\citet{djs00}		& -0.0668	& 3.27		& 0.0883\\	
\hline
\hline
\end{tabular}
\end{center}
\caption{Values for the binding energy $\bar E_b$, the 
angular velocity $\bar \Omega$ and the angular momentum $\bar J$ 
at the ISCO as obtained in different approaches.}
\label{7_table1}
\end{table}

\subsubsection{The Puncture Approach}
\label{sec7.1.2}

An alternative to the conformal imagining approach is the {\em puncture}
approach suggested by \citet{bb97}, in which the singularities in
the Hamiltonian constraint (\ref{7_ham_1}) are absorbed in an analytic
expression.  To do this, we 
write the conformal factor $\psi$ as a sum
\begin{equation} \label{7_psi}
\psi = u + \frac{1}{a}
\end{equation}
with
\begin{equation}
\frac{1}{a} = \frac{{\mathcal M}_1}{2 r_{{\bf C}_1}} + 
\frac{{\mathcal M}_2}{2 r_{{\bf C}_2}}.
\end{equation}
In the limit of infinite separation, the parameters ${\mathcal M}_1$
and ${\mathcal M}_2$ approach the masses of the individual black
holes, but in general they are simply constants.  Since $1/a$ is a
solution to the (homogeneous) Laplace equation, the Hamiltonian
constraint (\ref{7_ham_1}) now reduces to
\begin{equation} \label{7_ham_2}
\Delta^{\rm flat} u = - b (1 + a u)^{-7},
\end{equation}
where we have abbreviated
\begin{equation} \label{7_b}
b = \frac{1}{8} a^7 \bar A_{ij} \bar A^{ij}.
\end{equation}
The beauty of this approach is that the poles at the center of the
black holes have been absorbed into the analytical terms, and that the
corrections $u$ are regular everywhere \citep{bb97}.  Equation
(\ref{7_ham_2}) can therefore be solved everywhere, and there is no
longer any need to excise the interior of the black holes from the
computational grid.  This eliminates the need for complicated boundary
conditions on the throats, and allows for straight-forward solutions
on very simple computational domains.  Since one no longer needs the
isometry conditions on the throat, one can construct black hole
binaries in a three-sheeted topology, so that no additional mirror
terms need to be added to the extrinsic curvature (\ref{7_A_2}), and
instead (\ref{7_A_2}) can be inserted directly into (\ref{7_b}).  The
only added complication of this approach is that the location of the
throat and horizon is not known a priori, and has to be located after
the fact.  This can be done with the apparent horizon finders
discussed in Section \ref{sec6.1}.

Once a solution $u$ has been found, the ADM energy (\ref{3_mass}) 
can be determined from
\begin{eqnarray}
M & = & - \frac{1}{2 \pi} \oint_{\infty} \bar D^i \psi \,d^2S_i 
= - \frac{1}{2 \pi} \oint_{\infty} \bar D^i 
        \left( \frac{1}{\alpha} \right) d^2S_i 
        - \frac{1}{2 \pi} \int \Delta^{\rm flat} u \,d^3x \nonumber \\
& = & {\mathcal M}_1 + {\mathcal M}_2 
        + \frac{1}{2 \pi} \int b \,(1 + \alpha u)^{-7} d^3x.
\end{eqnarray}

\citet{b00} combined the puncture method of \citet{bb97} with 
the turning-point method of \citet{c94} to construct binary black
holes in quasi-circular orbit.  Numerical values for the ISCO are
included in Table \ref{7_table1}.  \citet{bbd02} recently found very
similar results by keeping a ``bare'' mass fixed instead of the
apparent horizon mass.

\subsection{The Thin-Sandwich Approach}
\label{sec7.2}

As we will see in Section \ref{sec7.3}, the numerical results of
\citet{c94} and \citet{b00} disagree with post-Newtonian values for the ISCO
by disturbingly large factors, which suggests that it would be useful
to explore alternatives to the Bowen-York approach.  

This has been done recently by \citet{ggb01a} and \citet{ggb01b}, who
have constructed binary black holes in the thin-sand\-wich approach
(Section \ref{sec3.3}).  In the thin-sandwich formalism, equations
(\ref{3_ts_ham}) -- (\ref{3_ts_ms}) are solved for the conformal
factor $\psi$, the lapse $\alpha$ and the shift $\beta^i$.  Unlike in
the Bowen-York approach, the extrinsic curvature is not constructed
from the analytical solution (\ref{7_A_2}), but is instead computed
from $\psi$, $\alpha$ and $\beta^i$ (equation (\ref{3_A_2})).  It is
important to note that \citet{ggb01a}, just like \citet{c94} and
\citet{b00}, assume maximal slicing and conformal flatness.

Following \citet{c91}, \citet{ggb01a} and \citet{ggb01b} adopt the
conformal-imaging approach (Section \ref{sec7.1.1}), which means that
suitable boundary conditions have to be imposed on $\psi$, $\alpha$
and $\beta^i$ on the throat.  This leads to a technical problem.  The
isometry condition requires that both $\alpha = 0$ and $\beta^r = 0$
on the throat.  For $\alpha = 0$ (and $\psi$ non-zero) in equation
(\ref{3_A_2}), the extrinsic curvature can only be regular if the
derivative of $\beta^r$ also vanishes.  The shift therefore has to
satisfy more conditions than can be imposed.  \citet{ggb01b}
circumvent this problem by introducing a regularization for the shift
which then slightly violates equation (\ref{3_ts_md}), and proceed by
hoping that this violation is small (see also the discussion in
\citet{c01}, who introduces an alternative set of boundary
conditions).

\begin{figure}
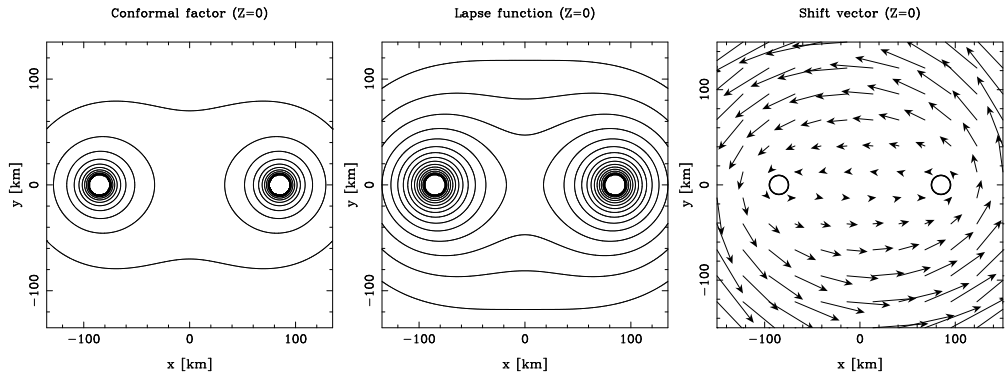

\begin{center}
\epsfxsize=1.7in
\epsffile{Fig_7_2a.ps}
\epsfxsize=1.7in
\epsffile{Fig_7_2b.ps}
\epsfxsize=1.7in
\epsffile{Fig_7_2c.ps}
\end{center}
\caption{Isocontours of the lapse function $\alpha$, the conformal
factor $\psi$ and the shift vector $\beta^i$ in the orbital plane $z =
0$ at the ISCO as obtained with the thin-sandwich approach.  The thick
solid lines denote the surfaces of the throats.  The kilometer scale
of the axis corresponds to an ADM mass of 31.8 $M_{\odot}$.  (Figure
from \citet{ggb01b}.)}
\label{fig7.2}
\end{figure}

The approach of \citet{ggb01b} differs in another way from the
calculations of \citet{c94} and \citet{b00}, namely in the
determination of circular orbits.  Instead of adopting the
turning-point method, they compute the Komar mass \citep{k59}
\begin{equation} \label{7_komar}
M_K = \frac{1}{4 \pi} \oint_{\infty} \gamma^{ij} (D_i \alpha - \beta^k
	K_{ik}) d^2 S_j
\end{equation}
in addition to the ADM mass $M$ (\ref{3_mass}).  In general, the two
mass definitions lead to different masses, but they agree in
stationary spacetimes
\citep{b78}.  Since quasi-equilibrium solutions approximate stationary
spacetimes, \citet{ggb01b} suggest that quasi-circular orbits can
be identified by demanding
\begin{equation}
M = M_K.
\end{equation}
The angular velocity $\Omega$ enters the equations (\ref{3_ts_ham}) --
(\ref{3_ts_ms}) only through the outer boundary condition on the shift
\citep[equation (9)]{ggb01b}.  This value can be varied until
$M = M_K$ has been achieved.  Note that this condition cannot be
applied in the Bowen-York approach, since it does not provide the
lapse and shift that is needed to evaluate the Komar mass
(\ref{7_komar}).

Instead of using equation (\ref{7_Omega}) to determine $\Omega$, as in
the turning-point method, it can now be used to construct an
evolutionary sequence.  The ISCO can be identified by locating a
simultaneous minimum in the ADM mass $M$ and the angular momentum $J$.
Numerical values are included in Table \ref{7_table1}.

\subsection{Comparison and Discussion}
\label{sec7.3}

\begin{figure}
\begin{center}
\epsfxsize=3in
\epsffile{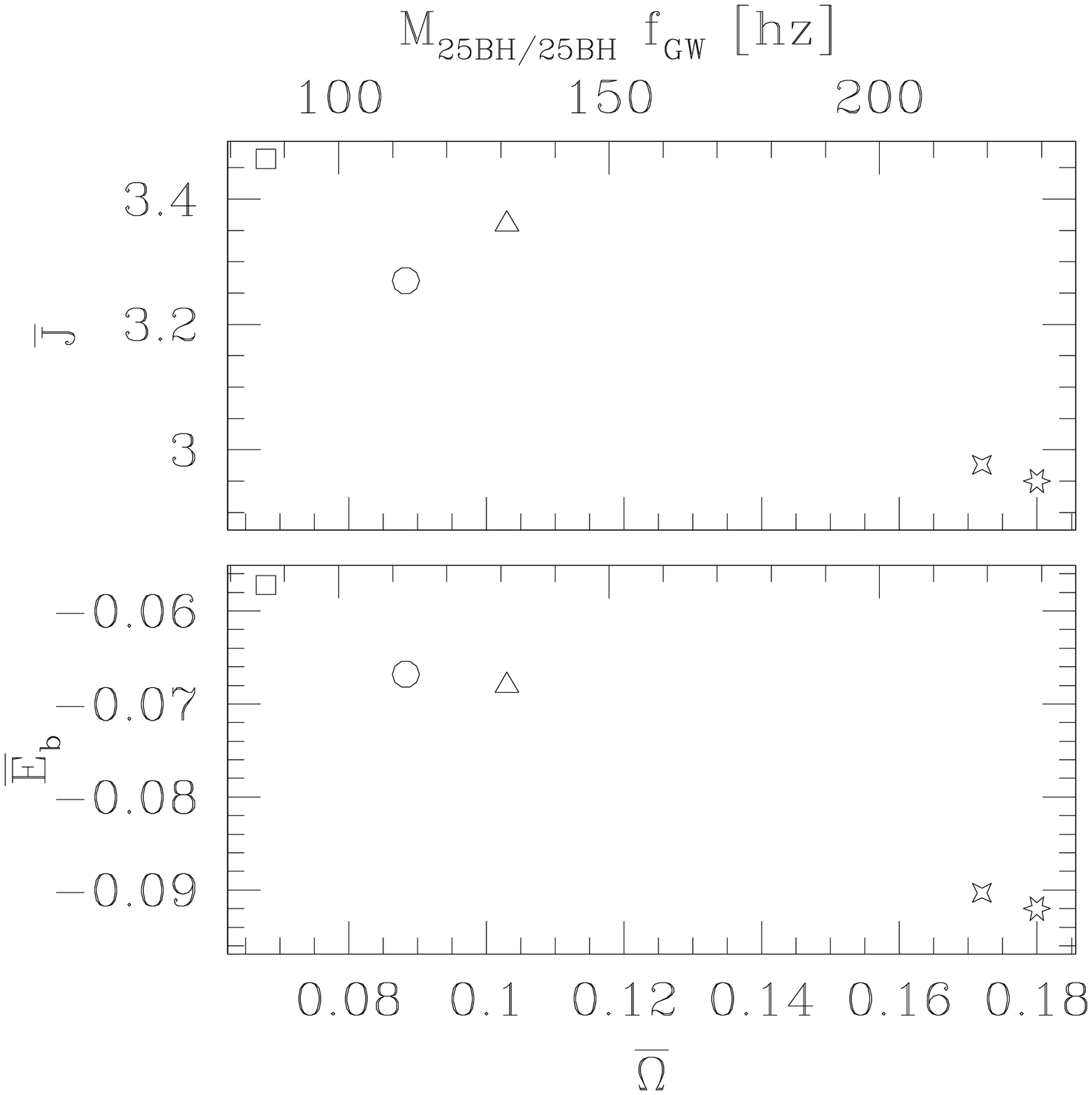}
\end{center}
\caption{Results for the
angular momentum $\bar J$, the binding energy $\bar E_b$, and the
orbital angular velocity $\bar \Omega$ at the ISCO for different
approaches (see also Table \ref{7_table1}).  The square marks the
Schwarzschild test-mass result, the circle the third order
post-Newtonian results of \citet{djs00}, the triangle the numerical
results of \citet{ggb01b}, and the four and six-pointed stars the
results of \citet{c94} and \citet{b00}.  The top label gives the
corresponding gravitational wave frequencies for a binary of two 25
$M_{\odot}$ black holes.  Compare Figures 2 in \citet{b01} and 20 in
\citet{ggb01b}.}
\label{fig7.3}
\end{figure}

In Table \ref{7_table1} we list numerical results from the numerical
calculations of \citet{c94}, \citet{b00} and \citet{ggb01b}.  We also
include the third order post-Newtonian results of \citet[assuming
$\omega_{\rm static} = 0$]{djs00}, as well as the analytical values
for a test particle orbiting a Schwarzschild black hole, $\bar E_b =
\sqrt{8/9} - 1 \sim -0.0572$, $\bar J = 2 \sqrt{3} \sim 3.464$, and
$\bar \Omega = 1/6^{3/2} \sim 0.0680$.  The same numbers are also
plotted in Figure \ref{fig7.3}.

It is very noticable that the results of \citet{ggb01b} agree quite
well with \citet{djs00} (see also \citet{b02,dgg02}), and that
\citet{c94} agrees very well with \citet{b00}.  There is a disturbingly
large discrepancy between these two groups, however, of as much as a
factor of two in the orbital frequency.  It is most likely that these
differences are caused by the different choices in the initial value
decomposition.  In fact, \citet{pct02} recently demonstrated that
``different decompositions generate different physical initial-data
sets for seemingly similar choices for the freely specifiable
pieces'', and that ``the choice of the extrinsic curvature is
critical''.  The different decompositions seem to lead to different
transverse-traceless components of the extrinsic curvature, so that
the resulting data represent physically different slices (see also
\citet{dgg02}).

This conclusion leads to the question which initial value
decomposition leads to more realistic initial data, describing binary
black holes in quasi-circular orbit.  As of now we do not have any
reliable means of knowing.  One way to compare the different initial
data would be to evolve them dynamically and test which one is closer
to being in equilibrium.  While dynamical evolution codes that are
reliable enough to make such comparisons are only now being developed
(see Section \ref{sec8}), preliminary results suggest that evolutions
of the initial data of \citet{b00} do not lead to circular orbits
\citep{as02}.  The fact that the two very different approaches of
\citet{ggb01b} and \citet{djs00} lead to quite similar results also
points to these initial data being more realistic\footnote{Even though
both \citet{ggb01b} and \citet{djs00} make some adhoc assumptions, so
that their agreement could possibly be purely coincidental.}.  This is
in accord with our arguments in Section \ref{sec3.3} that the
thin-sandwich decomposition provides a more natural means of
constructing equilibrium initial data (see also \citet{lp98}).
Together, these findings suggest that the currently most promising
approach to numerically constructing binary black holes in
quasi-circular orbit is the thin-sandwhich formulation.

The good agreement between \citet{c94} and \citet{b00} may not be
surprising, since the two approaches differ, other than in their
numerical approach, mostly in the underlying manifold structure (the
conformal-imaging approach of \citet{c94} adopts a two-sheeted
topology, while the puncture method used by \citet{b00} leads to a
three-sheeted topology).  The small difference in their results
reflects the fact that the strength of the imaged mirror poles in the
conformal-imaging approach is much smaller than the strength of the
poles themselves \citep{m63}.

Several future developments would be very desirable.  As discussed
above, the approach taken by \citet{ggb01b} seems very promising.  The
inconsistency introduced by the regularization of the shift, which may
spuriously affect their results, could be eliminated either by
adopting the boundary conditions of \citet{c01}, or by combining the
thin-sandwich approach with the puncture method and hence completely
eliminating the need for inner boundary conditions.

It would also be desirable to construct binary black holes using yet
other approaches.  \citet[cf.~\citet{pct02}]{mm00} suggested a
construction based on black holes in Kerr-Schild coordinates (see
Section \ref{sec8.1}), which would avoid the assumption of conformal
flatness.  This approach has the additional benefit that the slices
are regular across the horizon, which is very desirable for dynamical
simulations.  Results for quasi-circular orbits, however, are not yet
available.  Another possible approach that avoids the assumption of
conformal flatness is to adopt a post-Newtonian solution as a
background solution, and to solve the Hamiltonian and momentum
constraints for corrections to the post-Newtonian metric
\citep{tbcd02}.

\section{Dynamical Simulations of Binary Black Holes}
\label{sec8}

The dawn of numerical relativity dates back to the calculation in
axisymmetry of the head-on collision from rest at large separation of
two identical black holes by Larry Smarr and his collaborators (see,
e.g., \citet{s79,ahsss93}).  The result was that the emitted
radiation is roughly 0.1 \% $Mc^2$, significantly less than the upper
limit of 29 \% allowed by Hawking's area theorem, but in accord with
expectations from strong-field perturbation theory.

While significant progress has been made recently in the dynamical
evolution of binary black hole scenarios in full 3D, even simulations
of single black holes are still facing stability problems that are
only poorly understood.  A likely candidate for the origin of these
problems is the handling of the black hole singularities, which we
will discuss in Section \ref{sec8.1}.  In spite of these difficulties,
several groups have performed preliminary simulations of binary black
hole mergers.  We will review these efforts in Section \ref{sec8.2}.

\subsection{Singularity Avoidance and Black Hole Excision}
\label{sec8.1}

Simulations of black holes are greatly complicated by the presence of
singularities.  Encountering such a singularity during a computation
would clearly have very unfortunate consequences for the numerical
simulation.  Two approaches have therefore been suggested to avoid
these singularities: using {\em singularity avoiding} coordinates,
and {\em excising} the black hole interior.

\begin{figure}
\begin{center}
\epsfxsize=3in
\epsffile{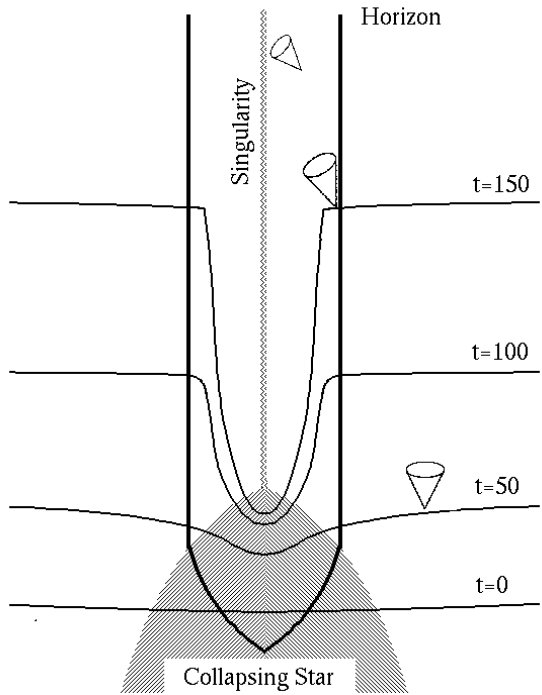}
\end{center}
\caption{A black hole spacetime diagram showing various singularity 
avoiding time slices that wrap up around the singularity inside the
horizon.  Such slicings allow short term success in the numerical
evolution of black holes, while at the same time causing pathological
behavior due to ``grid stretching'' that eventually dooms the
calculation at late times.  (Figure from \citet{admss95}.)}
\label{fig8.1}
\end{figure}

Traditionally, many black hole simulations, in particular in spherical
and axial symmetry, have used {\em singularity avoiding} coordinates
to model black hole spacetimes.  This approach is illustrated in
Figure \ref{fig8.1}, in which a family of singularity avoiding time
slices wrap up around the newly formed singularity inside the horizon.
Among the slicing conditions that avoid singularities is maximal
slicing, which we discussed in Section \ref{sec5.2}.  Another
singularity avoiding slicing condition that proved quite useful in
spherical and axisymmetric simulations is {\em polar slicing}
\citep{e82,bp83,st86,sbp88}.  The properties of maximal and polar slicing
have been analyzed for Oppenheimer-Snyder collapse in
\citet{pst85,pst86}.  In three-dimensional simulations, the ``1+log''
slicing (see Section \ref{sec5.3}) has been used fairly commonly.  It
has been shown to have singularity avoiding properties not unlike
maximal slicing \citep{bmss95,acmsst95}, and as an algebraic condition
it is much easier to implement.

The problem with singularity avoidance is that the time slices become
increasingly pathological, as illustrated in Figure \ref{fig8.1}.  For
late times, they have to ``stretch'' all the way back to an early time
in order to avoid the singularity.  This ``grid stretching'' will lead
to very steep gradients, which will lead to large numerical error, and
which will ultimately cause the code to crash.  One could try to
resolve these steep gradients with an increasingly large number of
gridpoints.  This attempt leads to a ``grid sucking'' effect, and is
similarly undesirable.  Instead of solving the problem it only
postpones the code crash to a slightly later time, and moreover leads
to large amounts of computational resources being spent on
uninteresting regions of the spacetime.

If, unlike in Figure \ref{fig8.1}, the initial time slice already
contains a singularity, it may be possible to absorb the singular part
of the solution into an analytical expression, similar to the puncture
method of \citet{bb97} (see Section \ref{sec7.1.2}).  For a single
black hole, for example, one can factor out the time-independent
conformal factor (\ref{ss_psi}), and evolve only the conformally
related metric, which is regular everywhere.  Functions and their
derivatives can then be evaluated from the numerical functions
together with the analytical factors, as long as the singularity of
(\ref{ss_psi}) does not happen to lie on a grid point.

In the first dynamical simulation of a black hole in three spatial
dimensions, \citet{acmsst95} adopted the traditional ADM formulation
(Section \ref{sec2}) together with geodesic, ``1+log'' and maximal
slicing (Sections \ref{sec5.1}, \ref{sec5.2} and \ref{sec5.3}).  The
simulations in geodesic slicing encounter the central singularity very
early on.  In maximal and ``1+log'' slicing grid-stretching effects
produce very large gradients very rapidly, as expected from
simulations in spherical or axial symmetry, and the code crashes after
a fairly short time (up to about $50 M$ in some cases).
\citet{acmsst95} experimented with different ways of treating the 
singularities at the center of the black hole, including imposing
isometry conditions (see Section \ref{sec7.1.1}) on the black hole
throats and the analytical factoring described above (see also
\citet{b96}).

\citet{u84} pointed out that there is no need to include the interior 
of the black hole in the numerical simulation, because by the very
definition of an event horizon (Section \ref{sec6.2}) the exterior is
causally disconnected from the interior.  If no information can
propagate from the interior to the exterior, there is no need to
numerically simulate the interior\footnote{Unless, of course, one is
interested in the structure of the singularity; see, e.g.,
\citet{bm93,h96,g02}.}.  From Figure \ref{fig8.1} it is evident that 
{\em black hole excision} may avoid both the singularity at the center
of the black hole as well as the grid stretching effects associated
with singularity avoiding timeslices.

Black hole excision is currently considered the most promising
approach to dynamical black hole simulations.  It has been
successfully implemented in spherical symmetry
(e.g.~\citet{ss92,sst95a,sst95b} and \citet{lhabsm00}), but most
implementations in three spatial dimensions are still plagued by
instabilities.

Various design choices have to be made in the implementation of black
hole excision.  As discussed in Section \ref{sec6}, the location of
the event horizon is unknown during the dynamical simulation, so the
excision is instead based on the location of the apparent horizon.
Typically not all gridpoints inside the apparent horizon are excised,
and instead a buffer zone of a few gridpoints is left inside the black
hole.  At each timestep, the grid points on the {\em inner boundary}
have to be filled with numerical data.  If a hyperbolic formulation of
Einstein's equation is used with characteristics that lie either on or
inside the light cone (e.g.~the Einstein-Christoffel system discussed
in Section \ref{sec3.2}), then all these characteristics must be
outgoing on this inner boundary (i.e.~toward the singularity), as is
evident from the tilting of the light cones in Figure \ref{fig8.1}.
In this case, the evolution equations can be used to fill the inner
boundary points (``boundary without a boundary condition'').

The tilting of the light cones points to another computational
problem.  While a normal observer $n^a$ always lies inside the light
cone, a coordinate observer $t^a$ may lie outside the line cone,
resulting in ``super-luminal'' grid velocities.  This happens when the
shift vector $\beta^i$ becomes large (see equation \ref{2_t}).  As a
consequence, a gridpoint $x_{ijk}^{n+1}$ on a timeslice $t^{n+1}$ may
be causally disconnected from the same gridpoint $x_{ijk}^{n}$ on the
previous timeslice $t^{n}$ (see Figure 1 in \citet{sbcst97} for an
illustration).  Some numerical schemes become unconditionally unstable
in such a situation, while others are stable only for very small time
steps $\Delta t$.  To avoid this problem, several groups have used
{\em causal differencing}
\citep{ss92,as94,sbcst97,cetal98,gw99,lhg00}.  Briefly summarized, the
idea behind causal differencing is to evolve the gravitational fields
along $n^a$ instead of $t^a$, and then use an interpolation to shift
the fields along the shift $\beta^i$.

If spatial derivatives have to be evaluated in the updating scheme,
then derivatives in directions orthogonal to the surface of the excised
region have to be computed using extrapolation or one-sided
schemes on the inner boundary.  In general, the coordinates will not be
aligned with this surface, so that an appropriate algorithm has to be
constructed. 

If a black hole moves through the computational grid, then grid points
re-emerge from the excised region on the trailing side of the black
hole.  These grid points have to be filled with data points, presumably
by extrapolation.

\begin{figure}
\begin{center}
\epsfxsize=4in
\epsffile{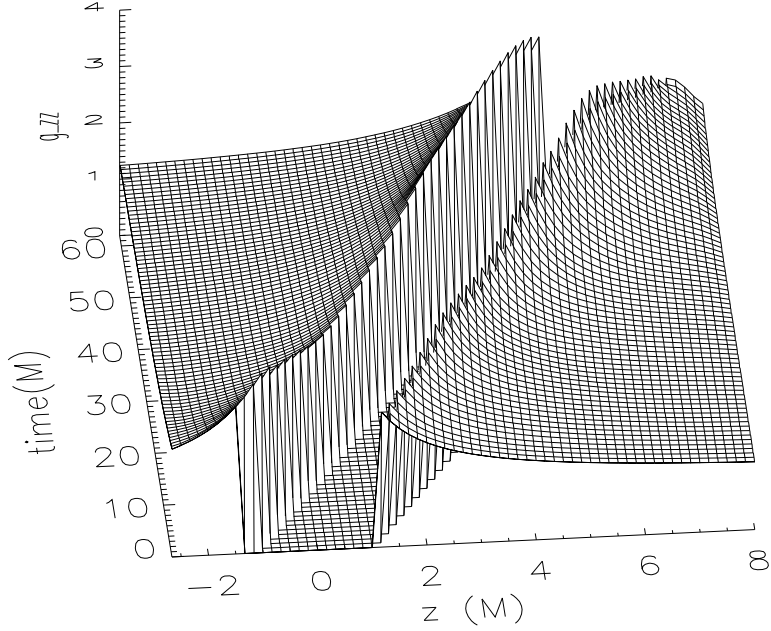}
\end{center}
\caption{Metric component $g_{zz}$ along the $z$-axis as a function 
of time.  The flat region that moves diagonally to the right 
represents the excised region (inside the black hole).  Note that 
points at the trailing edge (left side) are smoothly updated as the
hole moves toward positive $z$.  Coordinate effects are seen to appear
near the inner boundary.  (Figure from \citet{cetal98}.)}
\label{fig8.2}
\end{figure}

\citet{cetal98} implemented black hole excision together with 
causal differencing for both static and boosted black holes, using the
ADM formalism.  As initial data, they adopted Schwarzschild black
holes in {\em Kerr-Schild} form (compare
\citet{c92,mc96,mhs99,mm00,ybs01b}).

In ingoing Kerr-Schild form, the Kerr metric is given by
\begin{equation} \label{8_ks}
ds^2 = g_{ab} dx^{a} dx^{b} =  (\eta_{ab} + 2 H l_{a} l_{b}) dx^{a} dx^{b},
\end{equation}
where $H = H(x^a)$ is a scalar function.  The vector $l_a$ is null both
with respect to $\eta_{ab}$ and $g_{ab}$,
\begin{equation}
\eta^{ab} l_a l_b = g^{ab} l_a l_b = 0.
\end{equation}
Comparing with the ADM metric (\ref{2_final_metric}), we can identify
\begin{eqnarray}
\alpha & = & (1 + 2 H l_t^2)^{-1/2} \\
\beta_i & = & 2 H l_t l_i \\
\gamma_{ij} & = & \eta_{ij} + 2 H l_i l_j.
\end{eqnarray}
For the time-independent Schwarzschild spacetime we have
\begin{eqnarray}
H & = & M/r \\
l_a & = & (1, x_i/r),
\end{eqnarray}
where $M$ is the total mass-energy and $r^2 = x_1^2 + x_2^2 + x_3^2$.
This form is equivalent to the ingoing Eddington-Finkelstein form
(compare \citet{mtw73}).  The Kerr-Schild form (\ref{8_ks}) is
invariant under boosts, and therefore can be used to represent either
a static or boosted Schwarzschild (or Kerr) black hole.  The other
great advantage of Kerr-Schild coordinates is that they are regular on
the horizon and hence extend smoothly into the black hole.  In
Schwarzschild coordinates, for example, the lapse vanishes on the
horizon.  As a consequence, infalling radiation will never cross the
horizon in Schwarzschild coordinates, and will instead ``pile up''
just outside the horizon.  This has unfortunate consequences for
numerical simulations, because this piling up will produce
increasingly small length-scales which cannot be resolved with any
given finite-difference resolution (see \citet{retal98} for an
illustration in three spatial dimensions).

\citet{cetal98} use analytic Kerr-Schild data as initial data,
and adopt the Kerr-Schild lapse and shift to fix the coordinates.  For
static black holes, evolutions of up to $95 M$ are achieved, a slight
improvement over the results of \citet{acmsst95} (compare also with
the run-time of $138 M$ achieved by \citet{d96}\footnote{Note also
that {\em characteristic} evolutions of black holes in three spatial
dimension had already achieved runtimes of about $1400 M$
\citep{getal98}.  However, modeling binary black holes with
characteristic methods requires multiple matching between
characteristic and presumably Cauchy methods, which has not been
established yet.}).  More importantly, \citet{cetal98} demonstrate
that black holes can be advected through a numerical grid.  Figure
\ref{fig8.2} shows the excised region inside the black hole moving
through the grid, and also demonstrates that points at the trailing
edge, which re-emerge from the black hole, are updated smoothly.

\begin{figure}
\begin{center}
\epsfxsize=4in
\epsffile{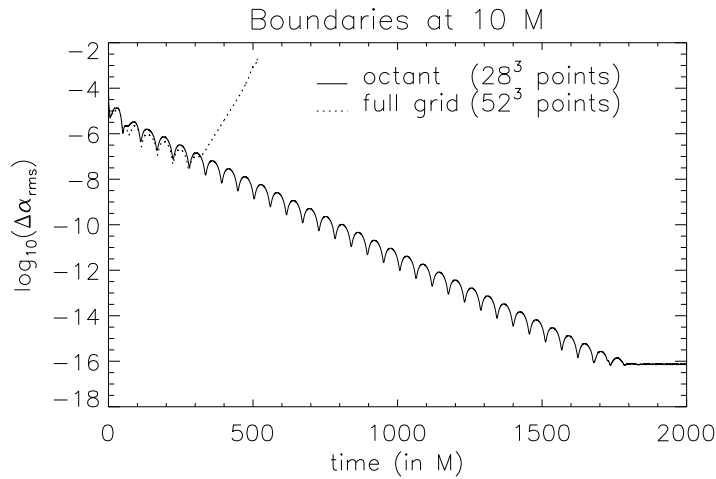}
\end{center}
\caption{The root means square of the change of the lapse function
$\alpha$ from one time step to the next, in the simulations of
\citet{ab01} for ``1+log'' slicing and static shift.  Shown are the
results for two runs, one performed in octant symmetry, and an
identical run except that it does not assume any symmetry.  The former
evolves stably until very late times, while the latter develops 
an instability after a few $100 M$.  (Figure from \citet{ab01}.)} 
\label{fig8.3}
\end{figure}

A new and quite successful approach was suggested by \citet[see also
\citet{abpst01}]{ab01}.  They use the BSSN formalism (Section \ref{sec3.3})
instead of the ADM formalism, and a particular simple black
hole excision scheme.  Instead of excising a (nearly) spherical
region, they excise a cube, which is well adapted to the Cartesian
coordinates.  Instead of trying to construct boundary conditions from
the evolution equations, they use a simple but stable boundary
condition on the inner boundary.  Finally, instead of using causal
differencing, they use centered differences except for advection terms
on the shift (i.e.~terms involving $\beta^i \partial_i$), which are
differenced with an upwind scheme along the shift direction.  A 
modification of this scheme was suggested by \citet{ybs01b}.

\citet{ab01} use a static black hole in Kerr-Schild coordinates
as initial data, and experiment with several slicing and gauge
conditions, including using the analytical lapse and shift of the
Kerr-Schild solution, maximal slicing, ``1+log'' slicing, and
``Gamma-freezing'' (see Section \ref{sec5.4}).  An example of their
results is shown in Figure \ref{fig8.3}, where the root mean square of
the change of the lapse from one time step to the next is plotted as a
function of time for ``1+log'' slicing and static (analytic) shift.
Two findings are quite remarkable.  When evolved in octant symmetry
the evolution settles down exponentially until the changes in the
lapse reach machine precision.  No instability is encountered.
However, when octant symmetry is relaxed, an otherwise identical
evolution develops an instability after a few hundred $M$\footnote{It
is also interesting that when evolved with the ADM formalism instead
of the BSSN formalism, the simulation terminates after about 30 $M$,
even in octant symmetry}.  Similar results were reported by
\citet{sitp00}, who use a completely independent implementation (namely
spectral methods) of a different formulation of Einstein's equations
(namely the Einstein-Christoffel system described in Section
\ref{sec3.2}).  Improvements over these results have been reported
by \citet{ls02,ls02a,ybs02,abdkpst02}.  In the BSSN calculation of
\citet{ybs02}, the octant symmetry has been removed and long-term
stability has been achieved both for static and rotating black
holes.

\subsection{Evolution of Binary Black Holes}
\label{sec8.2}

Disregarding the stability problems in simulations of single black
holes, some groups \citep{b99,bcghlmmnpssw00,abblnst01,bbclt01,bclt02}
have initiated preliminary simulations of binary black holes.

\cite{b99}, \citet{bcghlmmnpssw00} and \citet{abblnst01} simulated 
``grazing'' collisions of black holes.  In these simulations, two
black holes, starting out well within the ISCO, are boosted toward
each other with a certain impact parameter.  These initial data do not
correspond to the ISCO or any other astrophysically motivated
configuration, and are instead chosen so that a common apparent
horizon forms soon after the start of the evolution.

\begin{figure}
\begin{center}
\epsfxsize=4in
\epsffile{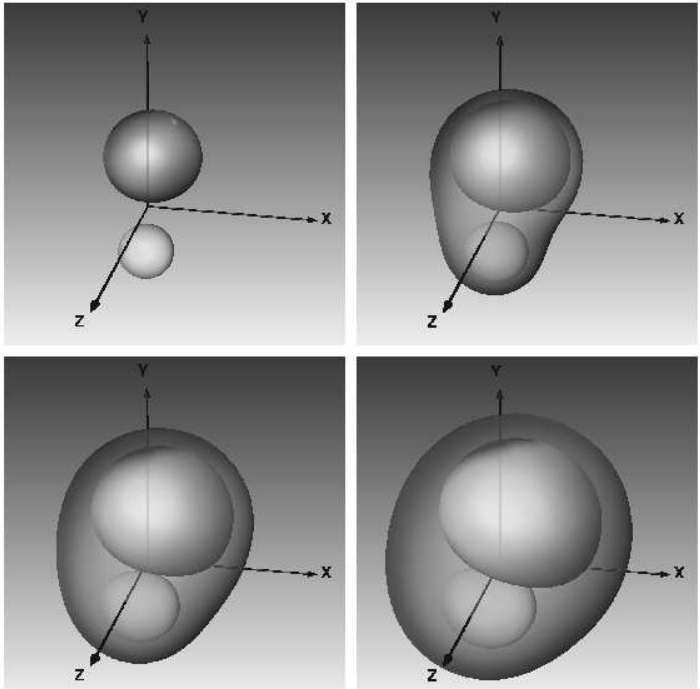}
\end{center}
\caption{The merger of apparent horizons in the simulations of
\citet{abblnst01}.  Shown are marginally trapped surfaces at times
$2.5M$, $3.7M$, $5.0M$, and $6.2M$.  The apparent horizon is the 
outermost of these surfaces.  (Figure from \citet{abblnst01}.)}
\label{fig8.4}
\end{figure}

\citet{b99} uses the puncture method of \citet{bb97} both for the 
construction of initial data (as described in Section
\ref{sec7.1.2}) and to remove the singularities during the evolution.  
In analogy to factoring out the time-independent term (\ref{ss_psi})
for single black holes, \citet{b99} factors out a time-independent
term similar to (\ref{7_psi}) for the binary.  The remaining terms are
regular, and can be integrated everywhere without black hole excision.
In this approach the singularities or ``punctures'', corresponding to
the asymptotically flat region inside the black holes, remain at
constant coordinate locations during the evolution.  \citet{b99}
evolves the remaining regular terms using the ADM formalism with
maximal slicing and zero shift.  The simulation terminates very early
(at about $t=7M$), and the results are too crude to be of any
astrophysical interest.  They nevertheless show some interesting
features, including the formation of a joint apparent horizon, and are
also noteworthy as the first attempt to simulate binary black hole
coalescence in three spatial dimensions.

\citet{bcghlmmnpssw00} adopt as initial data a superposition
of Kerr-Schild solutions, which serves as an approximate solution to
the constraint equations \citep{metal00}.  This approximate solution
also provides a lapse and shift which is used to fix the coordinates
during the evolution.  The gravitational fields are evolved with the
ADM formalism, and the interior of the black holes is excised.  The
results seem to be strongly affected by the outer boundaries, which is
implemented as a so-called ``blending'' to the semi-analytic solution.
The code again crashes too early (after about $t=15M$ on the
``large'' computational domain) to provide any interesting
astrophysical results.

\citet{abblnst01} return to the approach of \citet{b99}, but use
the BSSN formalism instead of the ADM formalism and the ``1+log''
slicing instead of maximal slicing during the evolution.  The initial
data and the ``analytical'' handling of the black hole singularities
are very similar to \citet{b99}.  \citet{abblnst01} take advantage of
large computational resources at the National Center for Supercomputer
Applications (NCSA), allowing them to simultaneously use fairly fine
resolution and far outer boundaries on a uniform Cartesian grid.  With
this ``combination of resolution, outer boundary location and
treatment, coordinate choice, evolution system and puncture method for
the black holes'', \citet{abblnst01} were able to achieve evolutions
well past $t = 30 M$ (see Figure \ref{fig8.4} for a visualization of
the merger of apparent horizons).  This is a significant improvement
over previous results, since this number forms a rough minimum for a
meaningful gravitational wave analysis.  The lowest quasi-normal mode
for a Schwarzschild black hole has a period of about $17M$, which sets
the approximate scale for the expected gravitational wavelengths in
the ringdown phase.  A simulation past $t = 30 M$ therefore allows one
to capture about two periods of this ringdown phase, which
\citet{abblnst01} have been able to identify.  At later times,
however, grid-stretching effects as well as error originating from the
outer boundaries start to degrade the numerical results.
\citet{abdkpst02} have recently reported finding singularity avoiding
lapse and shift conditions which permit long-term stable integrations
in octant symmetry.

Progress in the evolution of single and binary black holes has been
very rapid in the past years (further improvements have already been
reported by \citet{as02}), and it is quite likely that this field will
continue to develop quite quickly in the near future.

As an alternative to a fully self-consistent numerical simulation of
the entire plunge and merger of binary black holes,
\citet{bbclt01,bclt02} recently demonstrated that it may be possible to
match to a perturbative ``close limit'' treatment fairly early (the
``Lazarus'' project; see also \citet{pp94}). \citet{bclt02} adopt
puncture initial data describing binary black hole at the ISCO (see
\citet{b00} and Section \ref{sec7.1.2}), and evolve these with the ADM
formalism, maximal slicing and zero shift (similar to \citet{b99}).
The simulation terminates after about $t = 15 - 20M$.  As an
important self-consistency test, \citet{bclt02} show that the emitted
gravitational wave signal and energy is independent, within a certain
regime, of the time at which the matching is performed.  It is likely
that this or a similar approach will ultimately produce the desired
late-time plunge and merger gravitational waveforms.

\section{Binary Neutron Star Initial Data}
\label{sec9}

In this Section we will discuss initial data describing binary neutron
stars in quasi-equilibrium, quasi-circular orbit.  Like the binary
black hole solutions of Section \ref{sec7} these models may serve as
initial data in dynamical evolution equations (see Section
\ref{sec10}), for the construction of evolutionary sequences up to the
ISCO, and as background data for quasi-adiabatic approximations, as
discussed in Section \ref{sec10.4}.  We first discuss hydrostatic
quasi-equilibrium in general (Section \ref{sec9.1}), and then
specialize to corotational (Section \ref{sec9.2}) and irrotational
binaries (Section \ref{sec9.3}).

\subsection{Hydrostatic Quasi-Equilibrium}
\label{sec9.1}

As we have discussed in Section \ref{sec3.3}, it is natural to adopt
the thin-sandwich formalism for the construction of quasi-equilibrium
initial data, since it allows us to set explicitly the time derivative
of the conformally related metric to zero.  With very few exceptions
\citep{uue00,ue02}, most binary neutron star initial data 
have also been constructed under the assumption of spatial conformal
flatness.  This has the great advantage of dramatically simplifying
the equations, and has been justified naively by arguing that it
approximately minimizes the gravitational wave content in the spatial
slice (see also the discussion in Appendix \ref{appC}).  Under these
assumptions, the metric is written in the conformally flat form
\begin{equation}
ds^2 = - \alpha^2 dt^2 + \psi^4 \eta_{ij} 
	(dx^i + \beta^i dt) (dx^j + \beta^j dt),
\end{equation}
and Einstein's equations reduce to (\ref{3_ts_ham}) through
(\ref{3_ts_ms}) for the lapse $\alpha$, the conformal factor $\psi$
and the shift $\beta^i$\footnote{This formalism was developed
independently by \citet{wm89,wm95} as an approximate approach for
dynamical simulations of binary neutron stars.  We will discuss this
approach in Section \ref{sec10.2}.}.

To determine the matter sources in (\ref{3_ts_ham}) to (\ref{3_ts_ms})
we now have to specify the energy-momentum tensor $T^{ab}$, which,
adopting a perfect fluid, can be written as
\begin{equation} \label{9_se}
T^{ab} = (\rho_0 + \rho_i + P) u^a u^b + P g^{ab}.
\end{equation}
Here $u^a$ is the four-velocity of the fluid, and $\rho_0$, $\rho_i$
and $P$ are the rest-mass density, internal energy density and
pressure as observed by a comoving observer $u^a$.  It is also
useful to introduce the specific enthalpy
\begin{equation} \label{9_enth}
h = \exp \left( \int \frac{dP}{\rho_0 + \rho_i + P} \right).
\end{equation}
For a polytropic equation of state
\begin{equation} \label{9_eos}
P = \kappa \rho_0^{1 + 1/n},
\end{equation}
where $\kappa$ is the polytropic constant and $n$ the polytropic 
index\footnote{It is also common to use the polytropic exponent
$\Gamma = 1 + 1/n$.}, the enthalpy becomes
\begin{equation} \label{9_enth_poly}
h = \frac{\rho_0 + \rho_i + P}{\rho_0}.
\end{equation}
Note that equations (\ref{9_enth}) and (\ref{9_enth_poly}) imply
\begin{equation} \label{9_dh}
dh = \rho_0^{-1} dP.
\end{equation}

The equations of motion
\begin{equation} \label{9_eom}
\nabla_b T^{ab} = 0
\end{equation}
yield the Euler equation, which can be written as
\begin{equation} \label{9_euler}
u^b \nabla_b (h u_a) + \nabla_a h = 0.
\end{equation}
Rest-mass conservation yields the continuity equation,
\begin{equation} \label{9_cont}
\nabla_a (\rho_0 u^a) = 0.
\end{equation}

Following \citet{s98}, we now write the fluid four-velocity $u^a$ as
\begin{equation} \label{9_u}
u^a = u^t(l^a + V^a),
\end{equation}
where $l^a$ is timelike and normalized so that $l^t = 1$, and $V^a$
purely spatial, $n_a V^a = 0$.  We will later assume that $l^a$ is an
approximate Killing vector.  The spatial components of the velocity
vector in a coordinate system comoving with $l^a$, which we will
refer to as a corotating coordinate system, then reduce to $u^t
V^i$.

The four-dimensional equations (\ref{9_euler}) and (\ref{9_cont}) can
now be projected into the slice $\Sigma$.  Expressing the covariant
derivative $l^b \nabla_b u_a$ in terms of a Lie derivative along $l^a$
and using equation (\ref{2_proj_deriv}) yields
\begin{equation} \label{9_euler2}
\gamma_i^{~a} \Lie_{\bf l} (h u_a) 
+ D_i \left(\frac{h}{u^t} + \hat u_j V^j \right)
+ V^j ( D_j \hat u_i - D_i \hat u_j ) = 0
\end{equation}
for the projection of (\ref{9_euler}) and
\begin{equation} \label{9_cont2}
\alpha \left( \Lie_{\bf l} (\rho_0 u^t) + \rho_0 u^t \nabla_a l^a \right)
+ D_i ( \alpha \rho_0 u^t V^i) = 0
\end{equation}
for (\ref{9_cont}) (see \citet{s98} and compare footnote 3 in
\citet{g98a}).  Here we have introduced $\hat u_i$ as the spatial
projection of $h u_i$,
\begin{equation}
\hat u_i = \gamma_i^{~a} h u_a.
\end{equation}

So far we have not made any assumptions about the symmetry of the
spacetime.  Now, however, we assume that $l^a$ is a Killing vector,
$\nabla_a l_b + \nabla_b l_a = 0$, in which case the quantities
$\Lie_{\bf l} (h u_a)$, $\Lie_{\bf l} (\rho_0 u^t)$ and $\nabla_a l^a$
all vanish in the above equations.  The Euler equation
(\ref{9_euler2}) then reduces to the Bernoulli equation
\begin{equation} \label{9_bern}
D_i \left(\frac{h}{u^t} + \hat u_j V^j \right)
+ V^j ( D_j \hat u_i - D_i \hat u_j ) = 0,
\end{equation}
and the continuity equation (\ref{9_cont}) becomes
\begin{equation} \label{9_cont3}
D_i ( \alpha \rho_0 u^t V^i) = 0.
\end{equation}

\begin{exmp} \label{9_exmp_1}
We can now generalize the spherically symmetric, static solutions of
Examples \ref{3_exmp_2} and \ref{5_exmp_2} to include matter.  For a
static star we have $u^a = n^a$ so that $u^t = \alpha^{-1}$.  From
equation (\ref{2_rho}) we find $\rho = \rho_0 + \rho_i$ and from
(\ref{2_S}) $S_{ij} = P \gamma_{ij}$ and hence $S = 3P$.  The
Hamiltonian constraint (\ref{3_ham_1}) becomes
\begin{equation}
\bar D^2 \psi = - 2 \pi \psi^5 \rho,
\end{equation}
which, in spherical symmetry, can be integrated once to yield
\begin{equation} \label{ss_psi_matter}
\frac{\partial \psi}{\partial r} 
= - \frac{2 \pi}{r^2} \int \psi^5 \rho r'^2 dr' 
= -  \frac{m(r)}{2 r^2}
\end{equation}
where $m(r) \equiv 4 \pi \int_0^r \psi^5 \rho r'^2 dr'$.  In the
exterior of the matter source, $m(r)$ is independent of $r$ and
(\ref{ss_psi_matter}) can be integrated once more to yield
(\ref{ss_psi}).  The maximal slicing condition (\ref{3_ms_2}) can be
similarly integrated to yield
\begin{equation} \label{ss_lapse_matter}
\frac{\partial (\alpha \psi)}{\partial r} 
= \frac{2 \pi}{r^2} \int \alpha \psi^5 (\rho + 6 P) r'^2 dr' 
= \frac{\alpha \tilde m(r)}{2 r^2} 
\end{equation}
with $\tilde m(r) \equiv 4 \pi\alpha^{-1}\int_0^r \alpha\psi^5 
(\rho + 6 P) r'^2 dr'$.
For static solutions we have $V^i = 0$, so that the continuity equation
(\ref{9_cont3}) is satisfied identically and the Bernoulli equation
(\ref{9_bern}) reduces to 
\begin{equation} \label{ss_alpha_h}
D_i (\alpha h) = 0.
\end{equation}
This equation can now be rewritten as
\begin{equation}
\alpha D_i h = - h D_i \alpha = - h D_i \, \frac{\alpha \psi}{\psi}
	= - \frac{h}{\psi} (D_i (\alpha \psi) - \alpha D_i \psi)
\end{equation}
or, with (\ref{9_dh}), (\ref{ss_psi_matter}) and (\ref{ss_lapse_matter}),
\begin{equation}
\frac{\partial P}{\partial r} = - 
\frac{\rho_0 h}{\psi} \, \frac{m + \tilde m}{2 r^2}.
\end{equation}
This is the equivalent of the Oppenheimer-Volkoff equation
\citep{ov39} in isotropic coordinates.  In the Newtonian limit this
equation reduces to the Newtonian equation of hydrostatic equilibrium.
We point out, however, that solving the equations of hydrostatic
equilibrium for spherical stars is far more straight-forward in
Schwarzschild areal coordinates then in isotropic coordinates
presented here (e.g.~\citet{mtw73}).
\end{exmp}

For the construction of binaries in quasi-circular\footnote{Like
binary black holes (Section \ref{sec7.1}), binary neutron stars in
general relativity are not in strict equilibrium.  Due to the emission
of gravitational radiation, they loose energy and slowly spiral
inward.  For binaries sufficiently far from the ISCO, however, the
inspiral timescale is much longer than the orbital timescale.  This
justifies the notion of binary neutron stars being in
quasi-equilibrium, quasi-circular orbits, as we are assuming here.}
orbit we assume that the binary has a constant orbital velocity
$\Omega$.  The time vector $l^a$ in a frame corotating with the binary
is therefore a Killing vector and can be constructed from the time
vector in the binary's rest frame $t^a$ as
\begin{equation} \label{9_killing}
l^a = t^a + \Omega \xi^a = \alpha n^a + \beta^a + \Omega \xi^a.
\end{equation}
Here we have used (\ref{2_t}) and have introduced $\xi^a$ as the
generator of rotations about the rotation axis.  In Cartesian
coordinates, $\xi^a = (0,-y,x,0)$ represents a rotation around the
$z$-axis.  The Killing vector $l^a$ is sometimes called the 
{\em helicoidal} Killing vector \citep{bgm97}.

The Bernoulli equation (\ref{9_bern}) and continuity equation
(\ref{9_cont3}) simplify for the two cases $V^a = 0$ and $D_j \hat u_i
- D_i \hat u_j = 0$.  The former corresponds to corotating binaries,
and the latter to the more realistic case of irrotational binaries.
We will discuss both cases separately in Sections \ref{sec9.2} and
\ref{sec9.3}.

\subsection{Corotational Binaries}
\label{sec9.2}

The easier case follows from the assumption that the fluid flow
vanishes in the frame corotating with the binary
\begin{equation} \label{9_v=0}
V^a = 0, 
\end{equation}
which leads to models of {\em corotating} binaries.  With
(\ref{9_v=0}), the continuity equation (\ref{9_cont3}) is solved
identically, and the Bernoulli equation (\ref{9_bern}) reduces to
\begin{equation} \label{9_bern2}
\frac{h}{u^t} = \mbox{const} \equiv C
\end{equation}
(see, e.g., Problem 16.17 in \citet{lppt75}).  

\begin{figure}
\begin{center}
\epsfxsize=3in
\epsffile{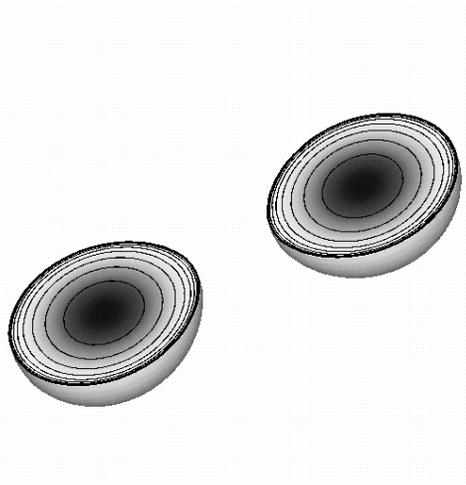}
\end{center}
\caption{Rest density contours in the equatorial plane for a $n=1$ 
neutron star binary close to the ISCO.  Each star has a rest mass of
$\bar M = 0.169$, corresponding to a compaction in isolation of \
$(M/R)_{\infty} = 0.175$.  The contours show isosurfaces of the
rest-density in decreasing factors of 0.556.  (Figure from
\citet{bcsst98b}.)}
\label{fig9.1}
\end{figure}

Using the normalization $u_a u^a = -1$, the $u^t$ component can be
expressed in terms of the spatial components $u_i$ as
\begin{equation} \label{9_ut1}
\alpha u^t = \left( 1 + \gamma^{ij} u_i u_j \right)^{1/2}.
\end{equation}
With $V^a = 0$, and assuming rotation about the $z$-axis,
the four-velocity (\ref{9_u}) reduces to
\begin{equation}
u^a = u^t (1, -\Omega y, \Omega x, 0).
\end{equation}
Since we are also assuming spatial conformal flatness, (\ref{9_ut1})
can then be rewritten as
\begin{equation} \label{9_ut2}
\alpha u^t = \left(1 - \frac{\psi^4}{\alpha^2} 
\left( (\Omega y - \beta^x)^2 + (\Omega x + \beta^y)^2 + (\beta^z)^2\right)
\right)^{- 1/2}.
\end{equation}
Inserting this into (\ref{9_bern2}) shows that the enthalpy $h$ is
given as an algebraic function of $\alpha$, $\psi$ and $\beta^i$,
which are determined by the thin-sandwich equations (\ref{3_ts_ham}) to
(\ref{3_ts_ms}), and the constants $\Omega$ and $C$.  From $h$ the 
fluid variables $\rho_0$, $\rho_i$ and $P$ can be computed.  Finally,
the matter-sources $\rho$, $j^i$ and $S$ in (\ref{3_ts_ham}) to
(\ref{3_ts_ms}) can be computed in terms of the fluid variables and
the velocity from their definitions (\ref{2_rho}), (\ref{2_j}) and
(\ref{2_S}).

\begin{figure}
\epsfxsize=3in
\begin{center}
\leavevmode
\epsffile{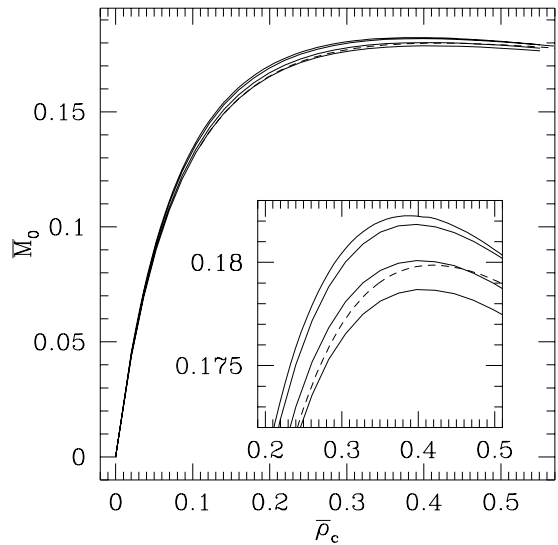}
\end{center}
\caption{Rest mass $\bar M_0$ versus central density $\bar\rho_{\rm c}$
for separations $z_A = 0.3$ (bottom solid line), 0.2, 0.1 and 0.0 (top
line) for corotational, $n=1$, equal mass binary neutron stars.  The
dashed line is the Oppenheimer-Volkoff result.  Due to finite
difference error, the numerical values systematically underestimate
the mass, which explains why some of these values are smaller than the
corresponding Oppenheimer-Volkoff values.  The insert is a blow-up of
the region around the maximum masses.  (Figure from \citet{bcsst97}.)
}
\label{fig9.2}
\end{figure}

Self-consistent solutions to the Bernoulli equation (\ref{9_bern2})
together with the field equations (\ref{3_ts_ham}) to (\ref{3_ts_ms})
can be constructed with iterative algorithms.  The method of
\citet{bcsst97,bcsst98b} is based on a rescaling algorithm that had 
been used earlier for the construction of rotating stars
(e.g. \citet{h86}).  Defining the point on the stellar surface that is
closest to the equal mass companion as $r_A$ and the one that is
furthest from the companion as $r_B$, \citet{bcsst97,bcsst98b} specify
a particular binary model by the maximum density $\rho_0^{\rm max}$
and the {\em relative} separation
\begin{equation} \label{9_z_A}
z_A = r_A/r_B.  
\end{equation}
Starting the iteration with an initial guess for the density profile,
namely that of a spherical star, the field equations (\ref{3_ts_ham})
to (\ref{3_ts_ms}) can be solved.  With a new guess for $\alpha$,
$\psi$ and $\beta^i$, the Bernoulli equation (\ref{9_bern2}) can be
evaluated at $r_A$, $r_B$ and at the point of maximum density, $r_C$.
At all three points the density is known; at $r_A$ and $r_B$ it
vanishes, and at $r_C$ it is pre-determined.  These three relations
can therefore be solved for constants $\Omega$ and $C$ as well as the
{\em absolute} separation $r_B$\footnote{In the Bernoulli equation,
the gravitational fields $\alpha$ and $\psi$ are rescaled in terms of
$r_B$, so that the latter enters the equation implicitly; see
\citet{bcsst98b}.}.  Given these values, the Bernoulli equation can be
solved everywhere, yielding an updated guess for the fluid variables.
The iteration is continued until a certain accuracy has been achieved.
\citet{bcsst97,bcsst98b} implemented this algorithm in Cartesian 
coordinates with a full approximation storage multigrid solver.
Other implementations have been used by \citet{mmw98,ue00,ggtmb01}.
A typical binary configuration close to the ISCO is shown in Figure
\ref{fig9.1}.

\begin{table}
\begin{center}
\begin{tabular}{llllll}
\hline
\hline
$\bar M_0$ & $\bar M_{\infty}$ & $(M/R)_{\infty}$ & $M_0
\Omega_{\rm ISCO}$ & $(J_{\rm tot}/M_{\rm tot}^2)_{\rm ISCO}$ & 
$\bar M_{\rm ISCO}$\\
\hline
0.0597	& 0.0582	& 0.05	& 0.003 & 1.69 & 0.0578 \\
0.1122	& 0.1066	& 0.1	& 0.01	& 1.22 & 0.1055 \\
0.1341	& 0.1259	& 0.125	& 0.015	& 1.12 & 0.1248	\\
0.153	& 0.1423	& 0.15	& 0.02	& 1.05 & 0.1408 \\
0.169	& 0.1547	& 0.175	& 0.025	& 1.00 & 0.1529 \\
0.178 	& 0.1625 	& 0.2	& 0.03	& 0.97 & 0.1601 \\
\hline
\hline
\end{tabular}
\end{center}
\caption{Numerical results for the orbital angular velocity $\Omega$,
angular momentum $J$ and the ADM mass $\bar M$ at the ISCO, for
corotating, equal mass binary neutron stars with polytropic index
$n=1$.  We tabulate the individual rest mass $\bar M_0$, the
mass-energy $\bar M_{\infty}$ and the compaction $(M/R)_{\infty}$ each
star would have in isolation (where $R$ is the Schwarzschild radius),
as well as the angular velocity $M_0 \Omega$, the angular momentum
$J_{\rm tot}/M_{\rm tot}^2$ and the ADM mass $\bar M$ at the ISCO. For
$n=1$, the maximum rest-mass in isolation is $\bar M_0^{\rm max} =
0.180$. (Table adapted from \citet{bcsst98b}.)}
\label{9_table1}
\end{table}

Physical units enter the problem only through the polytropic constant
$\kappa$ in the polytropic equation of state (\ref{9_eos}).  It is
therefore convenient to nondimensionalize all equations, and present
results in terms of dimensionless quantities.  Since $\kappa^{n/2}$ has
units of length, the non-dimensional mass $\bar M$, for example, is
defined as
\begin{equation}
\bar M = \kappa^{-n/2} M,
\end{equation}
and similar for other quantities.

One question of interest is how the maximum allowed rest mass changes
with separation.  In Figure \ref{fig9.2}, the rest mass $\bar M_0$ of
$n=1$ polytropes is plotted as a function of the central density $\bar
\rho_c = \kappa^n \rho_c$ for different relative separations $z_A$.  
Clearly, the maximum allowed mass increases with decreasing
separation, suggesting that neutron stars in corotating binaries are
more stable than in isolation.  Evolutionary sequences can be
constructed by keeping the rest mass $\bar M_0$ constant.  In Figure
\ref{fig9.2}, horizontal lines therefore correspond to evolutionary
sequences.  Evolving from a large separation to a smaller separation,
the central density decreases (see also Figure \ref{fig9.4} below).
Similar results hold for other polytropic indices
\citep{bcsst98b}.

The ISCO can be located by finding, for example, the minimum of the
total energy $\bar M$ along an evolutionary sequence $\bar M_0 =
\mbox{const}$.  According to the relation
\begin{equation}
dM = \Omega dJ
\end{equation}
(equation (33.61) in \citet{mtw73}, compare equation (\ref{7_Omega})),
a minimum in the energy should coincide with a minimum in the angular
momentum.  Values of the orbital angular velocity $M_0 \Omega$ and
angular momentum $J/M^2$ at the ISCO for different values of the
stellar mass $\bar M_0$ are tabulated in Table \ref{9_table1} for
$n=1$ polytropes.  \citet{bcsst98b} found an ISCO only for
sufficiently stiff equations of state, $n \lesssim 1$.  This result is
in qualitative agreement with various Newtonian results (see, e.g.,
the review by \citet{rs99}).  Stars with softer equations of state are
more centrally condensed and have more extended envelopes.  Such stars
come into contact and merge before encountering an ISCO.

The above arguments -- the increase in the maximum allowed mass and
the decrease in the central density -- suggest that neutron stars in
corotating binaries are more stable than in isolation.  A
rigorous stability analysis, based on turning point methods, comes to
the same conclusion and shows that corotating binaries do not
encounter any instabilities until they reach the ISCO
\citep{bcsst98a}.  These results are in contrast to the claims
of \citet{wm95}, who found in dynamical simulations that neutron stars
collapsed to black holes individually before they reached the ISCO
(``binary-induced collapse''; compare Section \ref{sec10.2}).  It was
later found that at least the size these results was erroneous and
caused by a mistake in the derivation of the equations
\citep{f99,mw00}.  Binary-induced collapse can occur for binaries made of 
collisionless matter \citep{sls98,defhs99}).  As seen in an inertial
frame, the neutron stars in corotating binaries are rapidly spinning,
and in fact the mass increases found in corotating binaries are quite
similar to those found in individual stars spinning with the same
angular velocity (see footnote [19] in \citet{bcsst97}).  This adds
one more motivation to studying irrotational binaries, which are more
realistic and where stabilizing effects of rotation are potentially
much smaller.

\subsection{Irrotational Binaries}
\label{sec9.3}

The presence of viscosity is necessary to maintain binaries in
corotation.  Estimates of the viscosity in neutron stars
\citep{k92,bc92} show that the viscous timescale is much longer than
the inspiral timescale, suggesting that the assumption of corotation
is quite unrealistic.  Instead, it may be more reasonable to assume
that the typical spin frequency of neutron stars is much less than the
orbital frequency of close neutron star binaries.  Since the emission
of gravitational radiation, which drives the inspiral of compact
binaries, conserves circulation, it is then more realistic to assume
neutron stars to be {\em irrotational} (zero circulation), even in
close binaries.

The first relativistic formulation for irrotational binaries in
quasi-equilibrium was given by \citet{bgm97}.  Inspired by this
approach, less complicated formalisms were developed independently by
\citet{a98}, \citet{t98} and \citet{s98}.  \citet{g98a} demonstrated 
that all of these formulations are equivalent.  The derivation here
follows most closely that of \citet{s98}.

In general relativity, the vorticity tensor $\omega_{ab}$ is defined
as
\begin{equation}
\omega_{ab} = P_a^{~c} P_b^{~d} 
	\left( \nabla_d (h u_c) - \nabla_c (h u_d) \right),
\end{equation}
where $P_a^{~b} = g_a^{~b} + u_a u^b$ is the projection operator with
respect to the fluid's four-velocity $u^a$.  For irrotational
binaries, the vorticity vanishes, and the quantity $hu_a$ can be
expressed as the gradient of a potential $\Phi$
\begin{equation}
h u_a = \nabla_a \Phi.
\end{equation}
It is easy to see that with this assumption the Euler equation
(\ref{9_euler}) is satisfied identically, while the continuity
equation becomes
\begin{equation}
\nabla^a \left( (n/h) \nabla_a \Phi \right) = 0
\end{equation}
(see, e.g., \citet{t98}).

Projecting these relations into $\Sigma$ shows that for irrotational
stars, the three-dimensional vorticity vanishes
\begin{equation}
D_j \hat u_i - D_i \hat u_j = 0,
\end{equation}
so that 
\begin{equation}
\hat u_i = D_i \Phi.
\end{equation}
The Bernoulli equation (\ref{9_bern}) then reduces to
\begin{equation} \label{9_bern3}
\frac{h}{u^t} + \hat u_i V^i = C,
\end{equation}
where $C$ is a constant.

\begin{figure}
\epsfxsize=3in
\begin{center}
\leavevmode
\epsffile{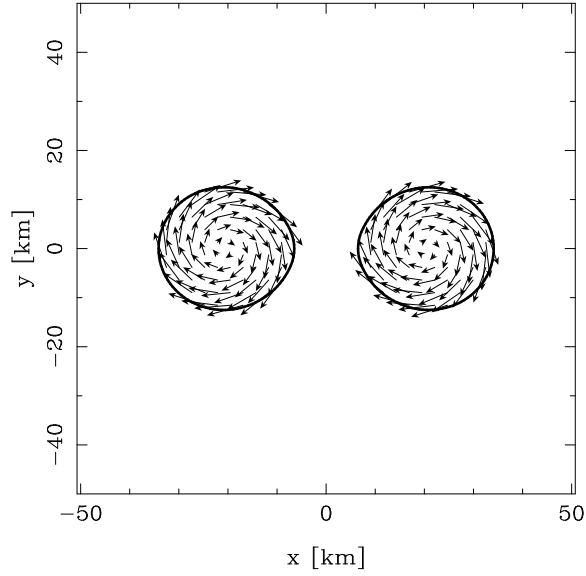}
\end{center}
\caption{The internal velocity field with respect to the co-orbiting 
frame in the orbital plane for stars of rest-mass $M_0 = 1.625
M_{\odot}$ at a coordinate separation of 41 km, for an $n=1$ polytrope
with $\kappa = 1.8 \times 10^{-2}
\mbox{Jm}^3\mbox{kg}^{-2}$.  The thick lines denote the surfaces of
the stars.  (Figure from \citet{ggtmb01}.)}
\label{fig9.3}
\end{figure}

It is now convenient to introduce the rotational shift vector
\begin{equation}
B^i = \beta^i + \Omega \xi^i = l^a - \alpha n^a.
\end{equation}
From (\ref{9_u}), we can express $V^a$ as
\begin{equation} \label{9_V}
V^i = \frac{1}{u^t h} D^i \Phi - B^i,
\end{equation}
and the normalization $u^a u_a = -1$ yields
\begin{equation} \label{9_ut3}
\alpha u^t = \left( 1 + h^{-2} D_i \Phi D^i \Phi \right)^{1/2}
\end{equation}
(see (\ref{9_ut1})).  Inserting (\ref{9_V}) and (\ref{9_ut3}) into
the Bernoulli equation (\ref{9_bern3}) also yields the following
expression for $\alpha u^t$\footnote{This relation can also be 
derived by observing that $\Lie_{\bf l} (h {\bf u}) = 
\Lie_{\bf l} ({\bf d} \Phi) = {\bf d} \Lie_{\bf l} \Phi =
0$ implies that $l^a \nabla_a \Phi = C$, where $C$ is a constant.
Expressing $n^a$ in $h\alpha u^t = h n^a u_a = n^a \nabla_a \Phi$
in terms of $l^a$ and $B^i$ then yields (\ref{9_ut4}); see
\citet{t98}.}
\begin{equation} \label{9_ut4}
\alpha u^t = \frac{1}{\alpha h} \left(C + B^i D_i \Phi \right).
\end{equation}
The last two equations can be solved for the enthalpy $h$
\begin{equation} \label{9_h_irr}
h^2 = \alpha^{-2} (C + B^i D_i \Phi) - D_i \Phi D^i \Phi.
\end{equation} 
An equation for the velocity potential $\Phi$ can be found by
inserting (\ref{9_V}) and (\ref{9_ut4}) into the continuity equation
(\ref{9_cont3})
\begin{equation} \label{9_phi_irr}
\begin{array}{rcl}
& & \displaystyle
D_i D^i \Phi 
- B^i D_i \left(\frac{C + B^j D_j \Phi}{\alpha^2} \right)
- \frac{C + B^j D_j \Phi}{\alpha} K \\
& &  \displaystyle \mbox{~~~~~~~} = 
\left( \frac{C + B^j D_j \Phi}{\alpha^2} B^i - D^i \Phi \right)
D_i \frac{\alpha \rho_0}{h} .
\end{array}
\end{equation}
At the surface of the stars the density vanishes $\rho_0 = 0$, so that
regularity of the right hand side implies the boundary condition
\begin{equation} \label{9_bc_irr}
\left. \left(  \frac{C + B^j D_j \Phi}{\alpha^2} B^i - D^i \Phi \right) 
	D_i \rho_0 \right|_{\rm surface} = 0.
\end{equation}
This boundary condition can also be derived by requiring that at the
surface the fluid flow be tangent to the surface, $u^a \nabla_a
\rho_0 = 0$.

Evidently, constructing irrotational binaries is much more involved
than constructing corotational binaries.  Unlike for corotational
binaries, where only one algebraic equation (\ref{9_bern2}) has to be
solved for the enthalpy, one has to solve the elliptic equation
(\ref{9_phi_irr}) for the velocity potential $\Phi$ together with the
enthalpy equation (\ref{9_h_irr}).  The boundary condition
(\ref{9_bc_irr}) has to be imposed on the surface of the star, which
adds another complication since the location of the surface is not
known a priori.

\begin{figure}
\epsfxsize=3in
\begin{center}
\leavevmode
\epsffile{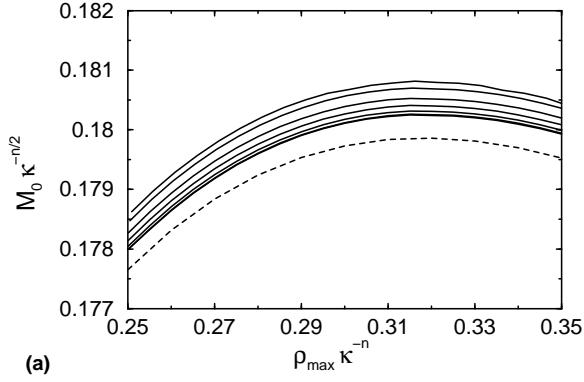}
\end{center}
\caption{The rest mass $\bar M_0$ as a function of central density
$\bar \rho_0$ for separations $\bar d =$ 1.3125, 1.375, 1.5 1.625,
1.75, 1.875 and 2 (thick lines from top to bottom) for irrotational
$n=1$ binary neutron stars.  The dashed line is the
Oppenheimer-Volkoff result.  (Figure from \citet{use00}.)}
\label{fig9.4}
\end{figure}

\citet{bgm99a} and \citet{ggtmb01} solved this problem with the help of a 
multi-domain, spectral method.  In this method, space is covered with
several patches of coordinate systems.  In particular, the interiors
of the stars are covered with spherical-type coordinate systems, which
are constructed so that the surface of the star lies at a constant
value of the radial coordinate.  Such coordinate systems are called
``surface-fitting'' coordinates, and are very well suited for imposing
the boundary condition (\ref{9_bc_irr}).  A similar algorithm, based
on Newtonian simulations of irrotational neutron star binaries
\citep{ue98a,ue98b}, was used by \citet{ue00} and \citet{use00}.
\citet{mmw99} constructed irrotational models of neutron star binaries in 
Cartesian coordinates.  They simplified the boundary condition
(\ref{9_bc_irr}) by assuming that the stars are spherical, so that the
gradient of the density in (\ref{9_bc_irr}) is aligned with a radial
vector.  While this may be adequate as long as the separation between
the stars are large, this approximation seems problematic for small
separation, and may explain why the results of \citet{mmw99} differ
from those of
\citet{bgm99a,ggtmb01,ue00} and \citet{use00} for close binaries.

\begin{figure}
\epsfxsize=3in
\begin{center}
\leavevmode
\epsffile{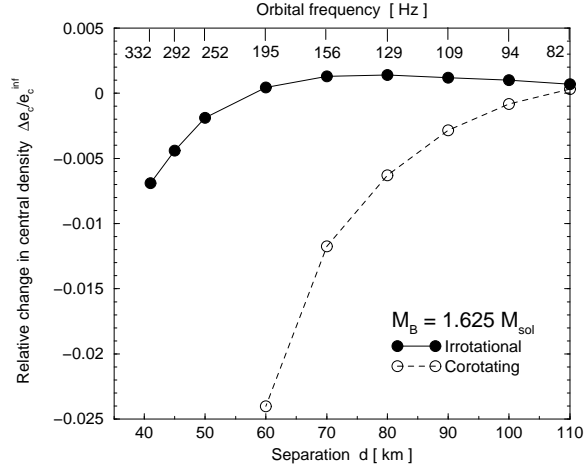}
\end{center}
\caption{Relative change of the central energy density $e = \rho_0 + \rho_i$
with respect to its value at infinite separation $e_c^{\rm inf}$ as a
function of the coordinate separation $d$ (or the orbital frequency
$\Omega/(2 \pi)$) for corotational and irrotational constant rest mass
sequences with $M_0 = 1.625 M_{\odot}$.  The sequences were computed
for a $n =1$ polytrope with $\kappa = 1.8 \times 10^{-2}
\mbox{Jm}^3\mbox{kg}^{-2}$.  (Figure from \citet{bgm99a}.)}
\label{fig9.5}
\end{figure}

A typical binary configuration and its internal velocity field is
shown in Figure \ref{fig9.3}.  The maximum allowed mass of neutron
stars in irrotational binaries can be found by finding the mass as a
function of central density for fixed separation.  Results of
\citet{use00} are shown in Figure \ref{fig9.4}.  This demonstrates that,
as in corotational binaries, the maximum mass increases with
decreasing separation.  However, comparing with Figure \ref{fig9.2},
we find that the increase in maximum mass is smaller for irrotational
binaries than for corotational binaries (especially taking into
account that the fairly coarse resolution results of \citet{bcsst98b}
underestimate the masses in Figure \ref{fig9.2}).  This result is not
surprising, since neutron stars in corotational binaries are rotating
with respect to the rest frame of the binary, which by itself
increases their maximum mass (e.g. \citet{cst94}).  It is also evident
from Figures \ref{fig9.4} and \ref{fig9.2} that while the density
along evolutionary sequences $\bar M_0 = \mbox{const}$ of irrotational
binaries decreases with decreasing separation, the decrease is less
than for corotational binaries.  This can be seen more clearly in
Figure \ref{fig9.5}, which demonstrates that while it is possible that
the central density increases with decreasing separation by a very
small amount for intermediate separations in irrotational binaries, it
certainly decreases for small separations where tidal deformation
dominates over any other effects.  Similar results were found by
\citet{ue00} and \citet{use00}.  While these findings have no immediate 
implications for the stability of neutron stars in irrotational
binaries, they offer no evidence for an instability as initially
reported by \citet{wm95}.

\begin{table}
\begin{center}
\begin{tabular}{llllll}
\hline
\hline
$\bar M_0$ & $\bar M_{\infty}$ & $(M/R)_{\infty}$ & $M_0
\Omega_{\rm CUSP}$ & $(J_{\rm tot}/M_{\rm tot}^2)_{\rm CUSP}$ & 
$\bar M_{\rm CUSP}$ \\
\hline
0.112	& 0.107	& 0.1	& 0.0106	& 1.09 	& 0.105 \\
0.130	& 0.122 & 0.12  & 0.0144	& 1.02	& 0.121 \\
0.146	& 0.136	& 0.14	& 0.0187	& 0.971 & 0.135 \\
0.166	& 0.153	& 0.17	& 0.0263	& 0.919	& 0.150 \\
0.175	& 0.160	& 0.19	& 0.0320	& 0.895 & 0.157 \\
\hline
\hline
\end{tabular}
\end{center}
\caption{Numerical results for the orbital angular velocity $\Omega$,
the angular momentum $J$ and the ADM mass $\bar M$ at cusp formation,
for irrotational binary neutron stars with polytropic index $n=1$.  We
tabulate the individual rest mass $\bar M_0$, the mass-energy $\bar
M_{\infty}$, the compaction $(M/R)_{\infty}$ each star would have in
isolation, as well as the angular velocity $M_0 \Omega$, the angular
momentum $J_{\rm tot}/M_{\rm tot}^2$ and ADM mass $\bar M$ at cusp
formation. For $n=1$, the maximum rest-mass in isolation is $\bar
M_0^{\rm max} = 0.180$. (Table adapted from \citet{use00}.)}
\label{9_table2}
\end{table}

While evolutionary sequences of corotational binaries end either at
the ISCO or at contact, irrotational sequences typically end with the
formation of a cusp before they reach the ISCO
\citep{bgm99a,ue00,use00}.  As analyzed by \citet{use00}, this cusp
corresponds to an inner Lagrange point, across which neutron stars
will transfer mass.  Such configurations are likely to form
dumbbell-like structures, with mass flowing between the two stars.
\citet{use00} find that only binaries with very stiff equations
of state ($n \lesssim 2/3$) reach an ISCO before they form a cusp,
while binaries with softer equations of state form a cusp first.
Numerical values for irrotational $n=1$ binaries at cusp formation are
listed in Table \ref{9_table2}.  Comparing with the ISCO values of
corotational binaries in Table \ref{9_table1}, one finds that the cusp
and ISCO occur at very similar frequencies.  The corotational binaries
have more angular momentum, because the individual stars carry a spin
angular momentum in addition to the orbital angular momentum of the
binary.  We also find that the binding energy $(\bar M - \bar
M_{\infty})/\bar M_0$ of corotational binaries is slightly larger than
for irrotational binaries, because the ADM mass $\bar M$ of the former
include the additional spin kinetic energy of the individual stars
(see also the discussions in \citet{dbssu01} and Section
\ref{sec10.4}).  More numerical results and a detailed discussion of
the ISCO and cusp formation in irrotational binaries can be found in
\citet{use00}.

\section{Dynamical Simulations of Binary Neutron Stars}
\label{sec10}

In this Section we present approaches and results for dynamical
simulations of binary neutron stars.  We discuss formulations of
relativistic hydrodynamics in Section \ref{sec10.1}, the conformal
flatness approximation in Section \ref{sec10.2}, fully
self-consistent simulations in Section \ref{sec10.3}, and a
quasi-adiabatic approximation of the slow inspiral just outside of the
ISCO in Section \ref{sec10.4}.

\subsection{Relativistic Hydrodynamics}
\label{sec10.1}

Since numerical hydrodynamics in general relativity has been the
subject of a recent review \citep{f00} we will provide only a very
brief overview.

The equations of relativistic hydrodynamics can be derived from the
local conservation laws of the stress-energy tensor $T^{ab}$
(\ref{9_eom})
\begin{equation} \label{10_se_cons}
\nabla_a T^{ab} = 0
\end{equation}
and the continuity equation (\ref{9_cont})
\begin{equation} \label{10_cont1}
\nabla_a (\rho_0 u^a) = 0,
\end{equation}
where $\rho_0$ is the rest mass density and $u^a$ the fluid
four-velocity.  For a perfect fluid, the stress-energy tensor is
(\ref{9_se})
\begin{equation} \label{10_se}
T^{ab} = (\rho_0 + \rho_i + P) u^a u^b + P g^{ab}
\end{equation}
where $\rho_i$ is the internal energy density and $P$ the pressure.

The equations of motion can be derived from (\ref{10_se_cons}) and
(\ref{10_cont1}) in various different ways, depending on how the basic
dynamical fluid variables are defined.  One might be tempted to adopt
a projection of $T^{ab}$ with respect to the normal vector $n^a$,
which yields the matter variables $\rho$ (\ref{2_rho}) and $j^i$
(\ref{2_j}) as they appear in the ADM equations.  It turns out,
however, that projections with respect to the fluid four-velocity
$u^a$ yields equations that are simpler and closer to the Newtonian
equations of hydrodynamics.

\citet[see also \citet{hsw84a,hsw84b,wm89,wmm96}]{w72}, 
who pioneered the field of relativistic hydrodynamics in
multi-dimensions, used the density definition $D = W \rho_0$, where
$W$ is the Lorentz factor between the fluid four-velocity $u^a$ and
normal observers $n^a$
\begin{equation}
W \equiv n_a u^a = \alpha u^t.
\end{equation}
The equation of continuity (\ref{10_cont1}) then becomes
\begin{equation}   \label{10_density1}
\partial_t (\sqrt{\gamma} D) 
+ 
\partial_j (\sqrt{\gamma} D v^j) = 0.
\end{equation}
Here $v^i = u^i/u^t$ is the transport velocity with respect to a
coordinate observer, and $\gamma$ is the determinant of the spatial
metric $\gamma_{ij}$.  Defining the momentum as $S_a = \rho_0 h W
u_a$, the spatial components of (\ref{10_se_cons}) reduce to the Euler
equation
\begin{equation} \label{10_euler1} 
\frac{1}{\sqrt{\gamma}} 
\partial_t (\sqrt{\gamma} S_i) 
+ 
\frac{1}{\sqrt{\gamma}} 
\partial_j (\sqrt{\gamma} S_i v^j)
= - \alpha \partial_i P - \frac{S_a S_b}{2 S^0} \partial_i g^{ab}. 
\end{equation}
The time component $S^0$ is found from the spatial components through
the normalization $u^a u_a = -1$.  With the energy density $E = W
\rho_i$, the timelike component of (\ref{10_se_cons}) yields the energy
equation
\begin{equation}  \label{10_energy1}
\partial_t (\sqrt{\gamma} E) 
+ 
\partial_j (\sqrt{\gamma} E v^j) 
= - P \left( \partial_t (\sqrt{\gamma} W) + 
\partial_i (\sqrt{\gamma} W v^i) \right).
\end{equation}

\citet[see also \citet{sbs98}]{s99a} modified this scheme by absorbing
the determinant into the definition of the density $\rho_* = \sqrt{\gamma} W
\rho_0$, for which the equation of continuity becomes
\begin{equation} \label{10_density2}
\partial_t \rho_{*} + \partial_j (\rho_* v^j) = 0.
\end{equation}
Also, for gamma-law equations of state
\begin{equation} \label{10_gleos}
P = (\Gamma - 1) \rho_i
\end{equation}
the source term on the right hand side of the energy equation
(\ref{10_energy1}) can be eliminated
\begin{equation}
\partial_t e_{*} + \partial_j (e_* v^j) = 0
\end{equation}
if the energy variable is defined as $e_* = \sqrt{\gamma} W
\rho_i^{1/\Gamma}$.  Defining $\tilde u_i = h u_i$, the Euler equation
(\ref{10_euler1}) becomes
\begin{eqnarray} \label{10_euler2}
\partial_t (\rho_* \tilde u_i) + \partial_j (\rho_* \tilde u_i v^j) & = &
        - \alpha e^{6 \phi} \partial_i P - \rho_* \Big( W h \partial_i \alpha
	- \tilde u_j \partial_i \beta^j \nonumber \\
& & 	+ \frac{\alpha e^{-4 \phi} \tilde u_j \tilde u_k}{2 W h}
	\partial_i \bar \gamma^{jk} -
	\frac{2 \alpha h (W^2 - 1)}{W} \partial_i \phi \Big)
\end{eqnarray}
where we have expressed the right hand side of (\ref{10_euler1}) in
terms of the metric quantities introduced in the BSSN formalism
(Section \ref{sec4.3}; in particular $\sqrt{\gamma} = e^{6 \phi}$).
The transport velocity $v^i$ can be found from
\begin{equation}
v^i = - \beta^i + \frac{\alpha \bar \gamma^{ij} \tilde u_j}{W h e^{4 \phi}},
\end{equation}
and $W$ can be found from the normalization (\ref{9_ut1}), which can
be expressed as
\begin{equation}
W^2 = 1 + \frac{\bar \gamma^{ij} \tilde u_i \tilde u_j}{e^{4 \phi}}
\left(1 + \frac{\Gamma e_*^{\Gamma}}{\rho_* (W e^{6 \phi})^{\Gamma - 1}}
\right)^{-2}.
\end{equation}

\begin{exmp} \label{10_exmp_1}
For static configurations we have $\tilde u_i = 0$ and hence $W=1$
and, in the BSSN formulation, $\rho_* = e^{6 \phi} \rho_0$.  The Euler
equation (\ref{10_euler2}) then reduces to
\begin{equation}
\alpha \partial_i P + \rho_0 h \partial_i \alpha = 0, 
\end{equation} 
which can be combined with (\ref{9_dh}) to yield equation
(\ref{ss_alpha_h}) of Example \ref{9_exmp_1} for hydrostatic
equilibrium.
\end{exmp}

Other groups have cast the equations of relativistic hydrodynamics
into the flux-conservative form
\begin{equation}
\partial_t {\mathcal U} + \partial_i {\mathcal F}^i = {\mathcal S}
\end{equation}
where ${\mathcal U}$ is the state vector containing the so-called
primitive fluid variables, ${\mathcal F}$ is the flux vector, and the
source vector ${\mathcal S}$ does not contain any derivatives of the
fluid variables (see, e.g., \citet{mim91}, \citet{f00} and references
therein).  \citet{fmst00,fgimrssst01} have implemented such a scheme
with
\begin{equation}
{\mathcal U} = 
\left( \begin{array}{c}
\tilde D \\
\tilde S_j \\
\tilde \tau 
\end{array} \right) 
=
\left( \begin{array}{c}
\sqrt{\gamma} W \rho_0 \\
\sqrt{\gamma} \rho_0 h W^2 {\tilde v}_j \\
\sqrt{\gamma}(\rho_0 h W^2 - P - W\rho_0)
\end{array} \right),
\end{equation}
where ${\tilde v}^i = u^i/W + \beta^i/\alpha$ is the velocity of the
fluid with respect to a normal observer, and where $\gamma$ is the
determinant of the spatial metric $\gamma_{ij}$.  The flux vector
${\mathcal F}$ is then given by
\begin{equation}
{\mathcal F}^i =
\left( \begin{array}{c}
(\alpha {\tilde v}^i - \beta^i) \tilde D \\
(\alpha {\tilde v}^i - \beta^i) \tilde S_j + \alpha \sqrt{\gamma} P \delta^i_{~j} \\
(\alpha {\tilde v}^i - \beta^i) \tilde \tau + \alpha \sqrt{\gamma} {\tilde v}^i P
\end{array} \right),
\end{equation}
and the source vector ${\mathcal S}$ by
\begin{equation}
{\mathcal S} = 
\left( \begin{array}{c}
0 \\
\alpha \sqrt{\gamma} T^{ab} g_{bc} {}^{(4)}\Gamma^c_{aj} \\
\alpha \sqrt{\gamma} (T^{a0} \partial_a \alpha - 
\alpha T^{ab} {}^{(4)}\Gamma^0_{ab})
\end{array} \right).
\end{equation}
The virtue of these schemes is that the local characteristic structure
can be determined, which is crucial for the implementation of
high-resolution shock-capturing schemes (see below).

Once the equations have been brought into a particular form, a
numerical strategy for their numerical implementation has to be chosen
(see \citet{f00}).  This is true for all equations in this article,
but hydrodynamics poses the additional challenge that shocks may
appear.  In a shock, hydrodynamic quantities develop discontinuities
when perfect fluids are assumed, and macroscopic fluid flow is
converted into internal energy.  Neither one of these phenomena can be
captured by traditional finite difference schemes, which therefore
have to be modified.

Von \citet{vr50} suggested an {\em artificial viscosity} scheme, in which 
an artificial term $Q$ is added to the pressure $P$.  They showed that
the jump conditions across shocks are well satisfied for a Newtonian,
one-dimensional fluid flow if $Q$ is defined as
\begin{equation}
Q = \left\{ \begin{array}{lcl}
- \rho (k \Delta x)^2 (\partial_x v)^2  & ~~ 
	&\mbox{if~~} \partial_x v < 0  \\
0					& ~~ 
	&\mbox{otherwise},
\end{array} \right.
\end{equation}
where $\Delta x$ is the grid resolution and $k$ an adjustable constant
of order unity.  The effect of the artificial viscosity is to spread
out the shock over approximately $k$ gridzones.  It appears in the
source terms of the Euler equation (\ref{10_euler1}) and the energy
equation (\ref{10_energy1}), and, as desired, converts the kinetic
energy of bulk fluid flow into internal energy in accord with the
Rankine-Hugoniot jump conditions.

Artificial viscosity schemes have been generalized for applications
in general relativistic hydrodynamics by numerous authors (including
\citet{mw66,w72,v79,st80,bst95,s99a}).  These schemes have the virtue of 
being quite robust and very easy to implement.  However, it has been
shown that artificial viscosity schemes do not work well for
ultra-relativistic fluid flow \citep{nw86}.

Alternatively, {\em high-resolution shock-capturing} (HRSC) schemes
can be employed.  These schemes are based on the idea of treating all
fluid variables as constant in each grid cell.  The discontinuity at
each grid interface can then be viewed as the initial data for a
Riemann shock tube problem, which can be solved either exactly or
approximately.  These schemes rely on the local characteristic
structure of the equations for the solution of the Riemann problem,
and are therefore used with equations in flux-conservative form,
\citep{fmst00,fgimrssst01}.  By construction
HRSC are capable of handling shocks and do not need additional
artificial viscosity terms.  In various applications, HRSC have been
found to be more accurate than artificial viscosity schemes, in
particular for highly relativistic flows.  For more information on
HRSC schemes in relativistic hydrodynamics see the review articles by
\citet{mm99} and \citet{f00}.

It is also possible to avoid finite difference methods altogether, and
to solve the equations of relativistic hydrodynamics with either {\em
smoothed particle hydrodynamics} (SPH) or {\em spectral} methods.  SPH
methods are quite common in Newtonian hydrodynamics, and have also
been applied in relativistic hydrodynamics in stationary spacetimes
(see \citet{f00} for references).  Quite recently, SPH methods have
also been used for the modeling of binary neutron star coalescence in
post-Newtonian theory \citep{fr02} as well as in full general
relativity \citep{ort01}.  Spectral methods (see \citet{bgm99b} for a
review) have been used quite commonly for the solution of the elliptic
equations in initial value problems
(e.g.~\citet{bgm99a,ggb01b,ggtmb01}) and recently also for the
dynamical evolution of gravitational (vacuum) fields \citep{kst01}.
For hydrodynamic simulations, in particular in three dimensions,
spectral methods are less common because of difficulties in treating
the discontinuities in shocks, which lead to spurious oscillations
(Gibbs phenomenon).

\subsection{The Wilson-Mathews approximation}
\label{sec10.2} 

\citet[also \citet{wmm96}]{wm89,wm95} pioneered the conformal
flatness approximation to the simulation of neutron
star binaries.  In this approach, one assumes that the dynamical
degrees of freedom of the gravitational fields, i.e.~the gravitational
radiation, play a negligible role for the structure of binary neutron
stars.  To simplify the metric and the field equations,
\citet{wm89,wm95} therefore suggest assuming that the spatial metric is
conformally flat, so that the spacetime metric takes the simplified
form
\begin{equation} \label{10_metric1}
ds^2 = - \alpha^2 dt^2 +  \psi^4 \eta_{ij}(dx^i + \beta^i dt)
	(dx^j + \beta^j dt).
\end{equation}
\citet{wm89,wm95} further assume that the spatial metric 
remains conformally flat at all times.  The traceless part of
equation (\ref{2_gdot_2}) then has to vanish, which yields
\begin{equation} \label{10_K1}
A^{ij} = \frac{1}{2\alpha} (L \beta)^{ij}.
\end{equation}
Here $A^{ij}$ is the traceless part of the extrinsic curvature, and
the vector gradient $L$ is defined in (\ref{3_LO}).  

Equation (\ref{10_K1}) can be inserted into the momentum constraint
(\ref{2_mom_2}), resulting in an equation for the shift $\beta^i$.
The conformal factor $\psi$ can be found from the Hamiltonian
constraint (\ref{2_ham_2}), and to determine the lapse $\alpha$
\citet{wm95} adopt maximal slicing (Section \ref{sec5.2}).  If one
also adopts the conformal rescaling (\ref{3_A_1}), these equations
reduce to the thin-sandwich equations (\ref{3_ts_ham}) through
(\ref{3_ts_ms}), which now completely determine the metric
(\ref{10_metric1}).

All unknowns in the metric (\ref{10_metric1}) are determined by
elliptic equations, and in this sense dynamical degrees of freedom
have been removed from the gravitational fields.  In this approach,
one solves an initial value problem at each instant of time, as
opposed to dynamically evolving the gravitational fields.  While one
may worry about the accuracy of this approximation (see also Appendix
\ref{appC}), it greatly simplifies the field equations and allowed
\cite{wm95} to perform the first fairly detailed simulations of binary
neutron stars.  In this approach, the time step is limited by the
matter sound speed and not the light speed, so it can be much larger
than in fully self-consistent algorithms.

In these simulations, \cite{wm95} combined equations (\ref{3_ts_ham})
through (\ref{3_ts_ms}) for the metric (\ref{10_metric1}) with
Wilson's formulation of the relativistic equations of hydrodynamics
(\ref{10_density1}) through (\ref{10_energy1}).  In each timestep, one
first evolves the matter variables, and then solves the field
equations for the metric (\ref{10_metric1}) with the new density
distribution as sources.  If desired, the gravitational wave emission
can be estimated with the quadrupole formula.

The initial results of \cite{wm95} showed a so-called crushing effect,
in which the neutron stars were compressed to very high densities even
far outside of the ISCO, and ultimately collapsed to two individual
black holes (``binary-induced collapse'').  Since this effect
contradicted expectations from Newtonian and post-Newtonian theory as
well as relativistic quasi-equilibrium results (which predict that the
density should be reduced by tidal deformations, see Sections
\ref{sec9.2} and \ref{sec9.3}), their findings spurned a fairly 
intense debate until it was discovered that at the very least the size
of the effect was caused by an inconsistency in the derivation of the
equations \citep{f99,mw00}.

The unfortunate consequence of this debate is that the assumption of
conformal flatness itself was often incorrecty blamed for the spurious
results.  While this approximation is obviously not suitable for every
situation, it may be very good for the modeling of compact objects in
some regimes, where its errors are probably much smaller than other
sources of error (see also the discussion in Appendix \ref{appC}).  It
has recently been adopted by \citet{ort01}, who combined it with an
SPH method to solve the equations of relativistic hydrodynamics.

\subsection{Fully Self-consistent Simulations}
\label{sec10.3} 

Several groups have launched efforts to self-consistently solve the
equations of relativistic hydrodynamics together with Einstein's
equations, and model the coalescence and merger of binary neutron
stars \citep{on97,bhrss99,fmst00,fgimrssst01}.  The first successful
simulations of binary neutron star mergers are those of \citet[see also
\citet{s99a,su02}]{su00a}, which we will focus on in this Section.

\begin{table}
\begin{center}
\begin{tabular}{llllll}
\hline
\hline
Model &
$\bar \rho_{\rm max}$ &
$\bar M_0$ & $\bar M$ & $J/M^2$ & remnant \\
\hline
I1	& 0.0726 & 0.261 & 0.242 & 0.98 & neutron star \\
I2	& 0.120  & 0.294 & 0.270 & 0.93 & black hole \\
I3	& 0.178  & 0.332 & 0.301 & 0.88 & black hole \\
\hline
\hline
\end{tabular}
\end{center}
\caption{Summary of the initial data for the coalescence simulations
of $n=1$ irrotational binary neutron stars by \citet{su00a}.  Here
$\bar M_0$, $\bar M$ and $J$ are the total rest mass, mass and angular
momentum of the binary.  In these dimensionless units, the maximum
allowed rest mass of an isolated, non-rotating star is $\bar M_0^{\rm
max} = 0.180$.  (Table adapted from \citet{su00a}.)}
\label{10_table1}
\end{table}

\begin{figure}[t]
\begin{center}
\epsfxsize=2.2in
\leavevmode
\epsffile{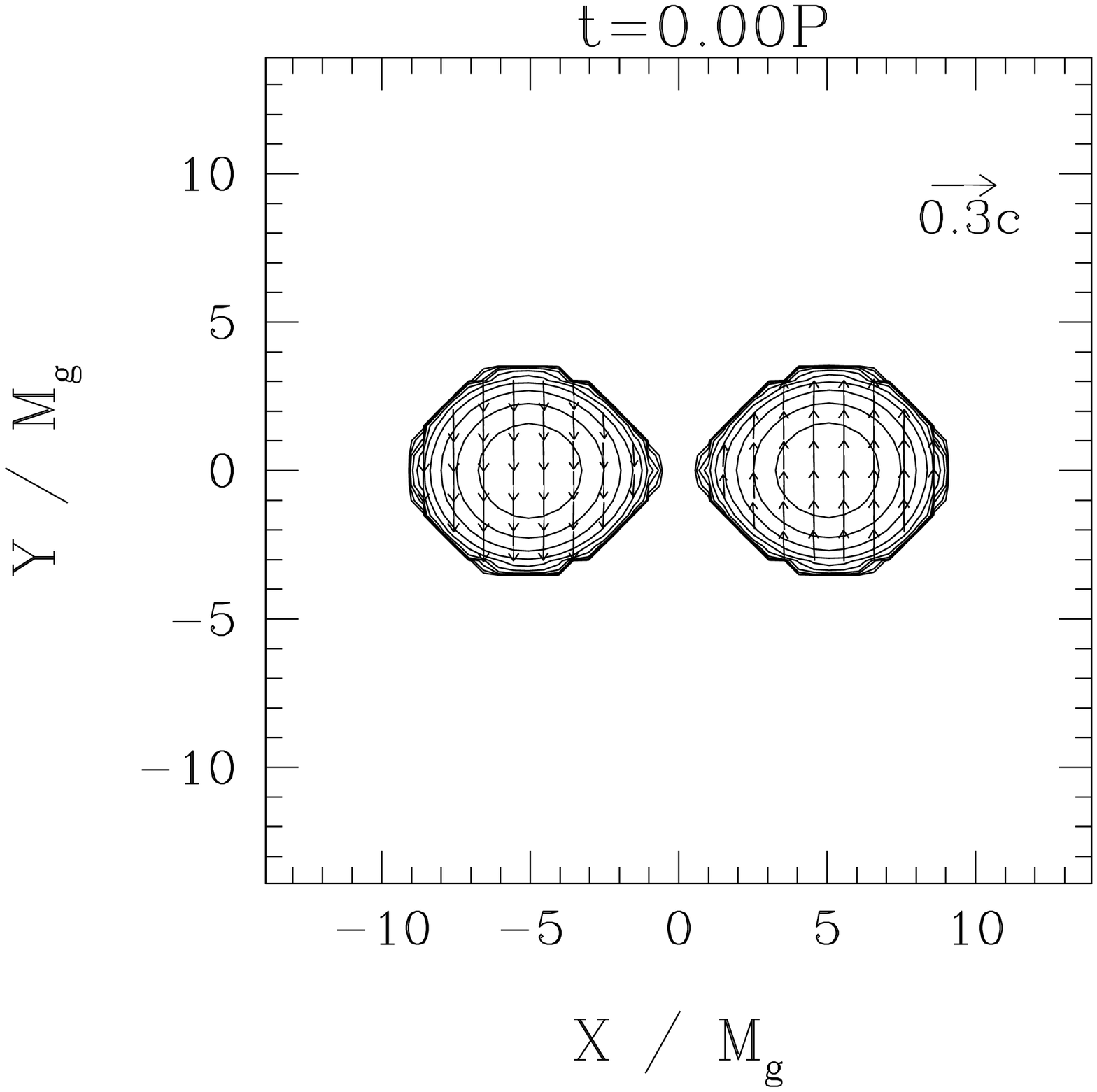}
\epsfxsize=2.2in
\leavevmode
\epsffile{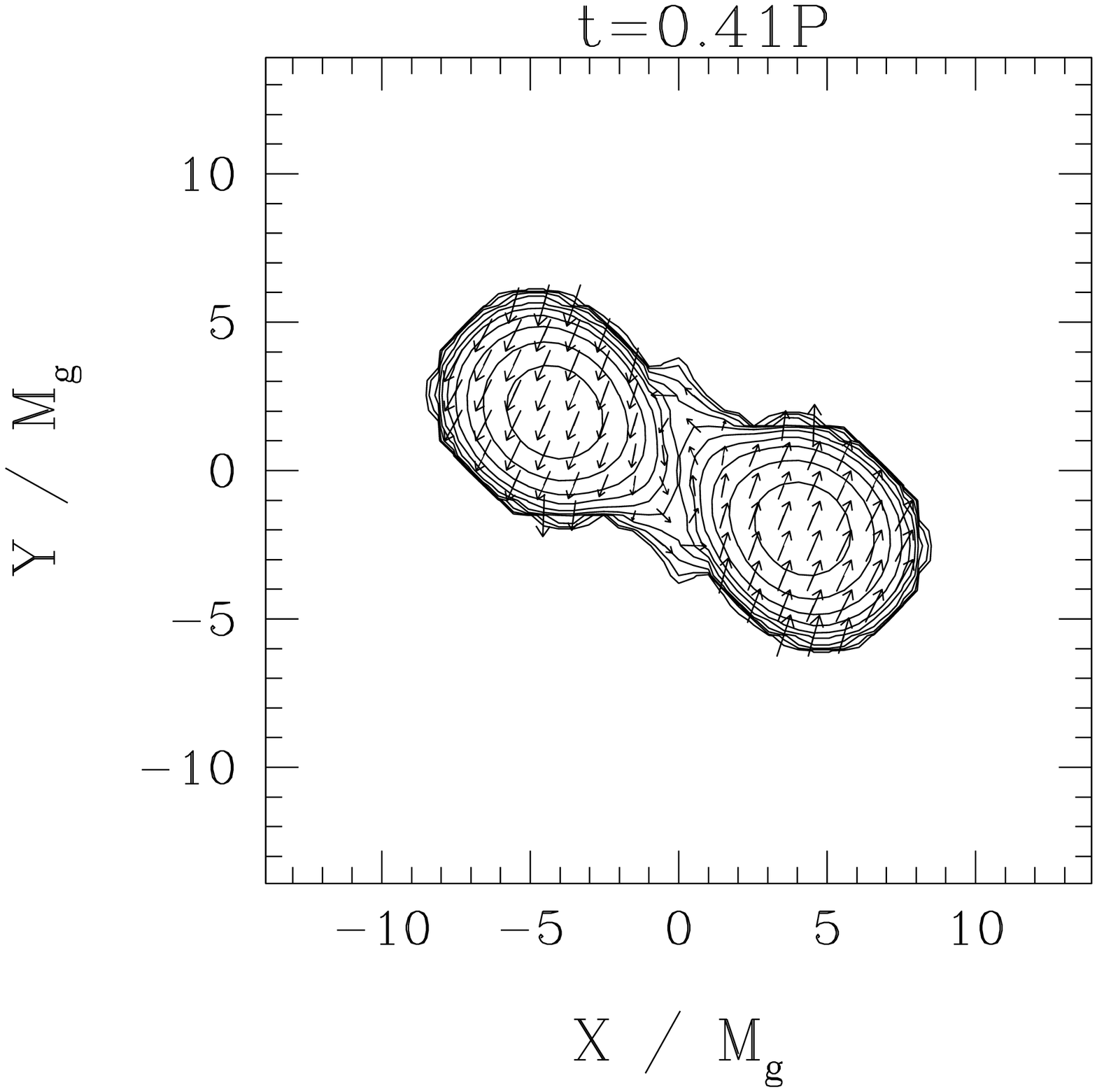}\\
\epsfxsize=2.2in
\leavevmode
\epsffile{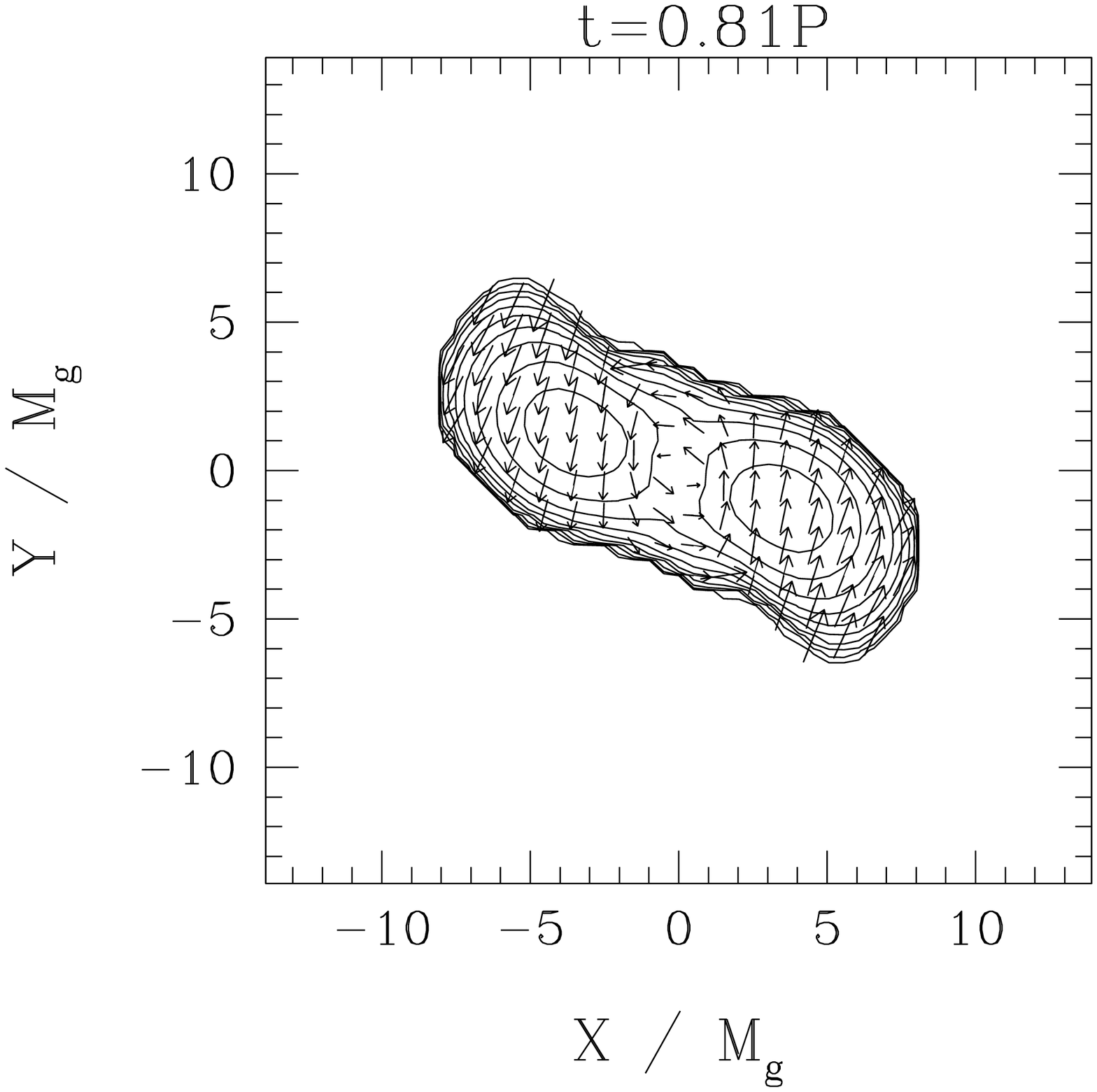}
\epsfxsize=2.2in
\leavevmode
\epsffile{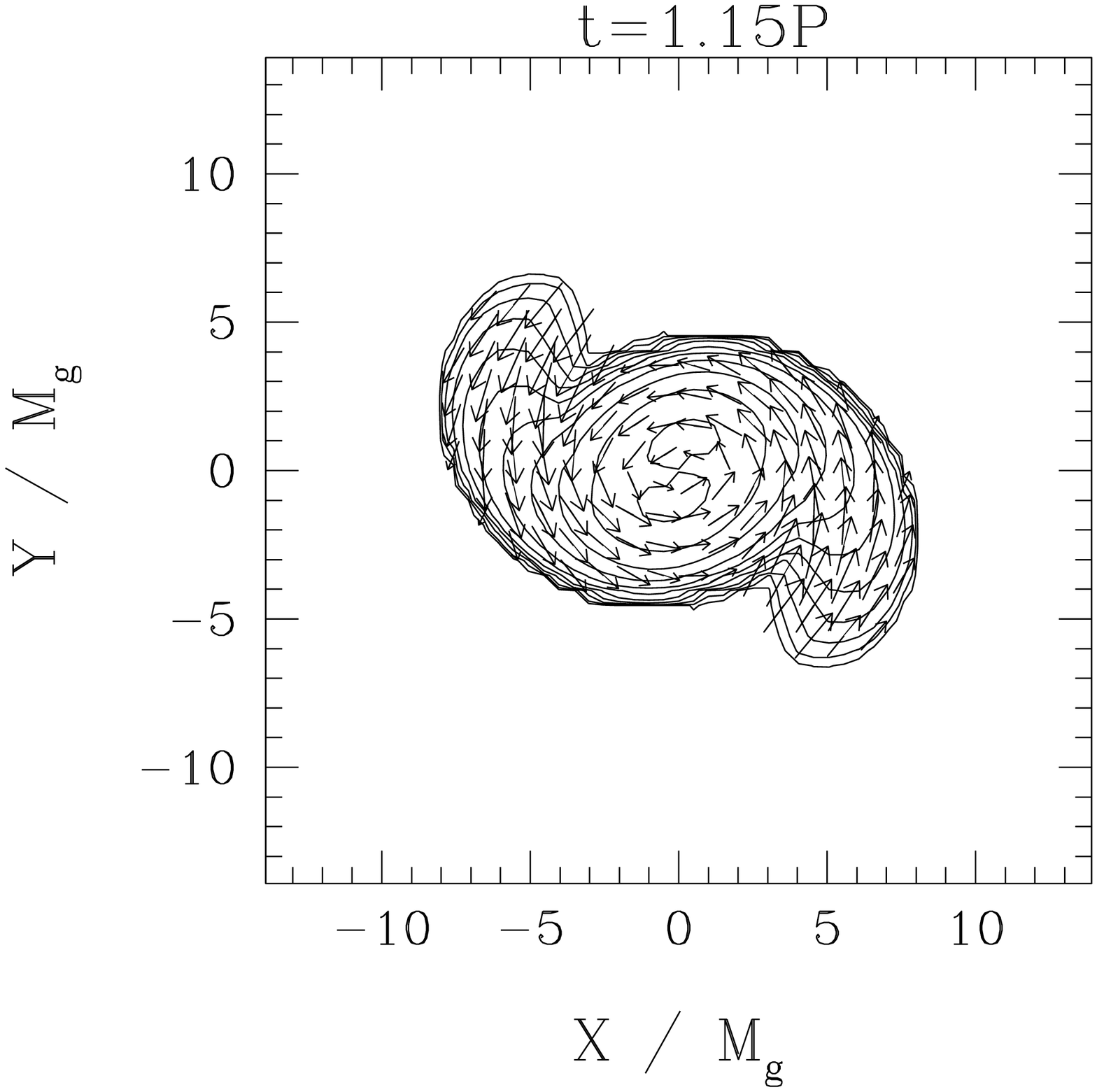} \\
\epsfxsize=2.2in
\leavevmode
\epsffile{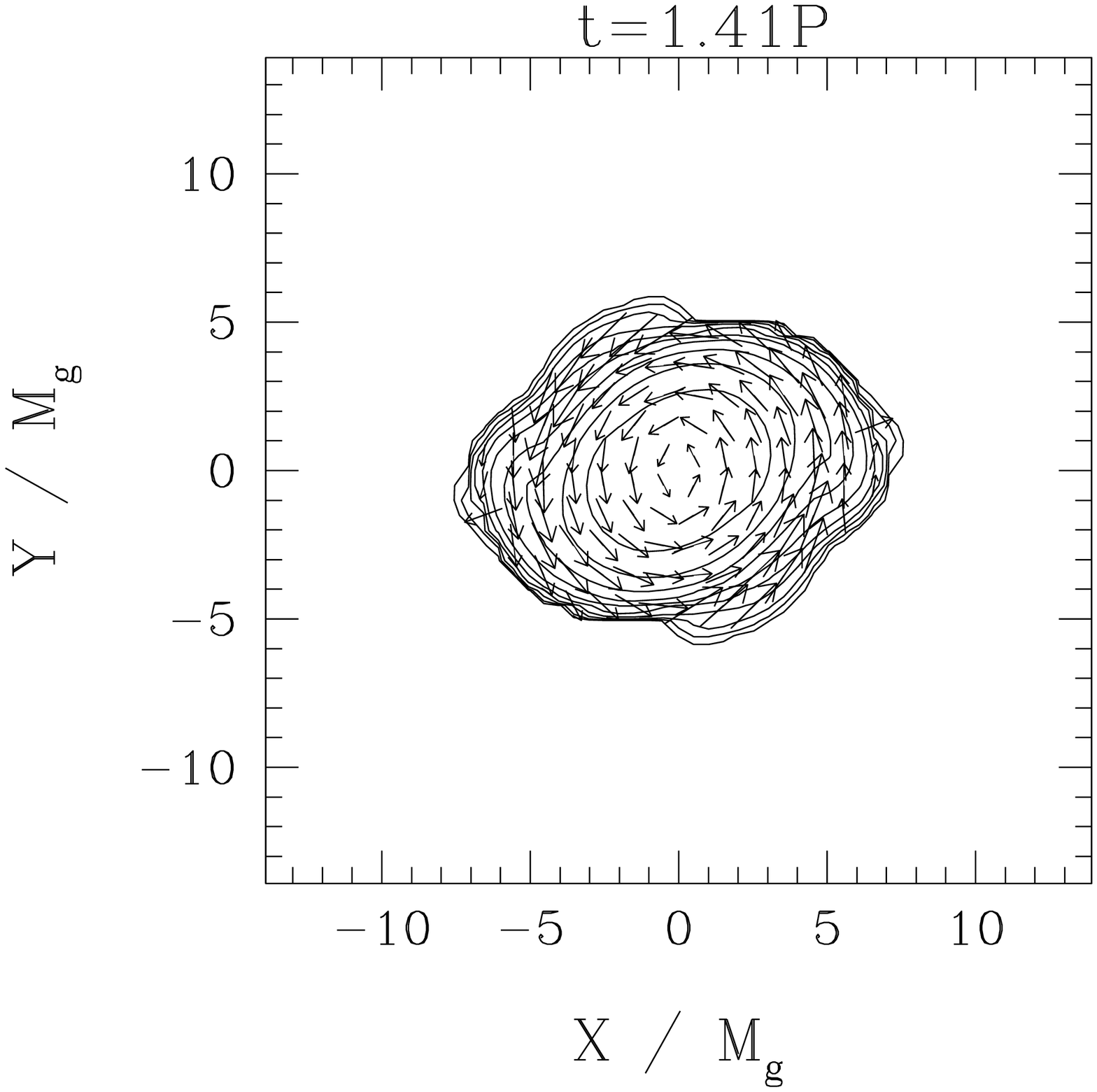}
\epsfxsize=2.2in
\leavevmode
\epsffile{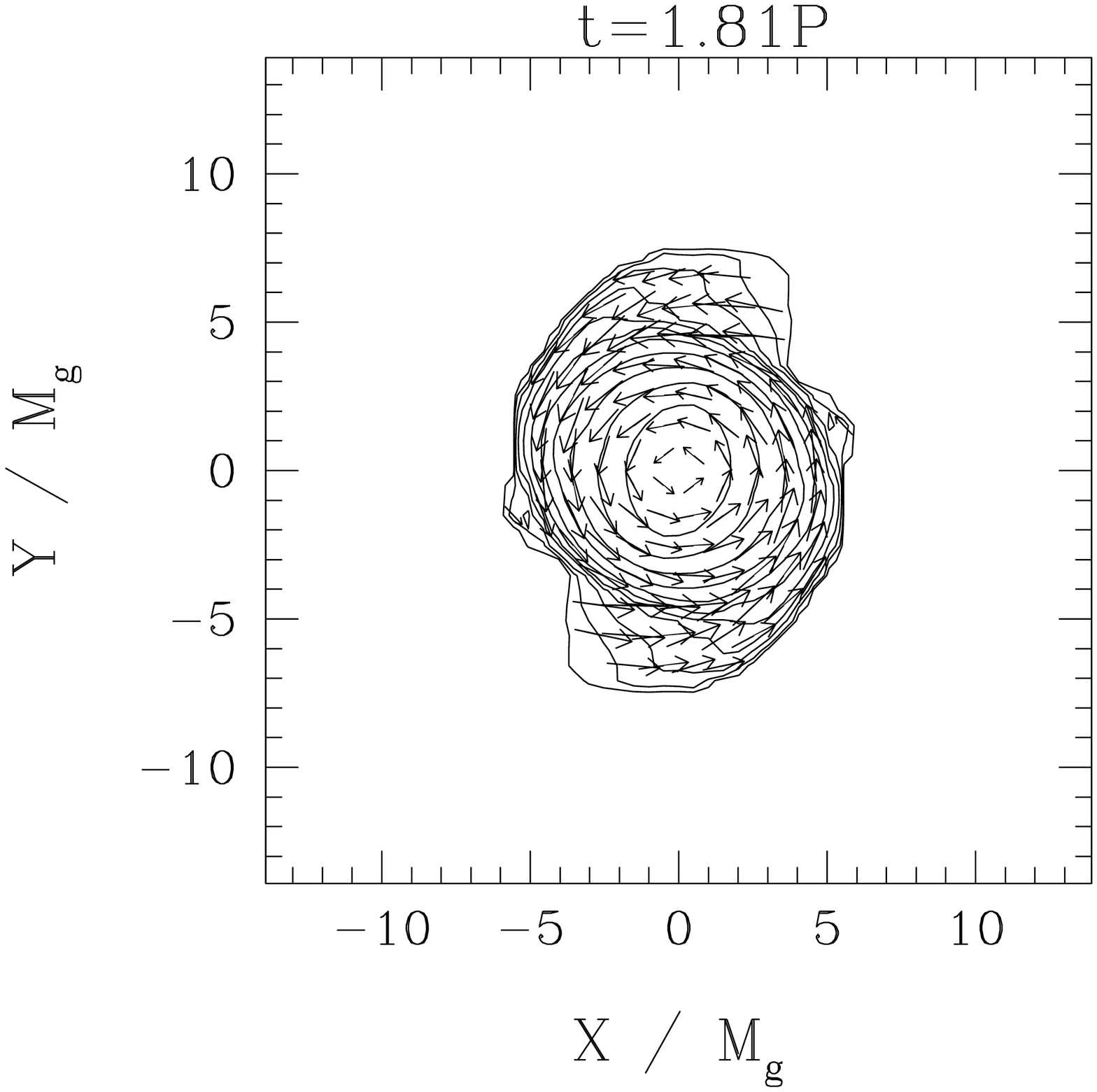} 
\caption{
Snapshots of density contours for $\rho_*$ (see equation
(\ref{10_density2})) and the velocity field $(v^x,v^y)$ in the
equatorial plane for the coalescence of an irrotational binary of
total rest mass $\bar M_0 = 0.261$ (Model (I1) in \citet{su00a}).  Time is
measured in terms of the orbital period $P$.  The contour lines denote
densities $\rho_*/\rho_{*~{\rm max}}=10^{-0.3j}$ with $\rho_{*~{\rm
max}}=0.255$ and $j=0,1,2,\cdots,10$.  (Figure from
\citet{su00a}.)}
\label{fig10.1}
\end{center}
\end{figure}

\citet{su00a} adopt a gamma-law equation of state (\ref{10_gleos}) with 
index $\Gamma = 2$ ($n=1$).  They solve the equations of relativistic
hydrodynamics in the form (\ref{10_density2}) to (\ref{10_euler2}),
and Einstein's equations in the original form \citep{sn95} of what is
now often referred to as the BSSN equations (Section \ref{sec4.3}).
They use ``approximate maximal slicing'' (Section \ref{sec5.2}) to
specify the lapse $\alpha$ and ``approximate minimal distortion''
(Section \ref{sec5.4}) to determine the shift $\beta^i$.  They also add 
a radial component to the shift vector to avoid grid-stretching in
collapse situations. 

As initial data, \citet{su00a} prepare equal-mass polytropic ($n=1$)
models of binary neutron stars in quasi-equilibrium with both
corotational (see Section
\ref{sec9.2}) and irrotational (Section \ref{sec9.3}) velocity
profiles.  For both velocity profiles they prepare three different
models with individual stellar masses ranging from about 75\% to 100\%
of the maximum allowed mass of non-rotating stars in isolation.  For
corotating models, \citet{su00a} adopt the contact models ($z_A = 0$
in the parametrization (\ref{9_z_A})) as initial data, which are
fairly close to the ISCO (see Section \ref{sec9.2}).  Irrotational
sequences terminate at cusp formation (see Section \ref{sec9.3}),
which is still outside of the ISCO.
\citet{su00a} therefore adopt the cusp model as initial data, and
induce collapse by artificially reducing the angular momentum by about
2.5\%.  Since irrotational velocity profiles are probably more
realistic, we will discuss their models (I1), (I2) and (I3) for
irrotational binaries.  The initial data for these three models are
summarized in Table \ref{10_table1} (more information can be found in
Table 1 of \citet{su00a}).

\begin{figure}
\begin{center}
\epsfxsize=3in
\epsffile{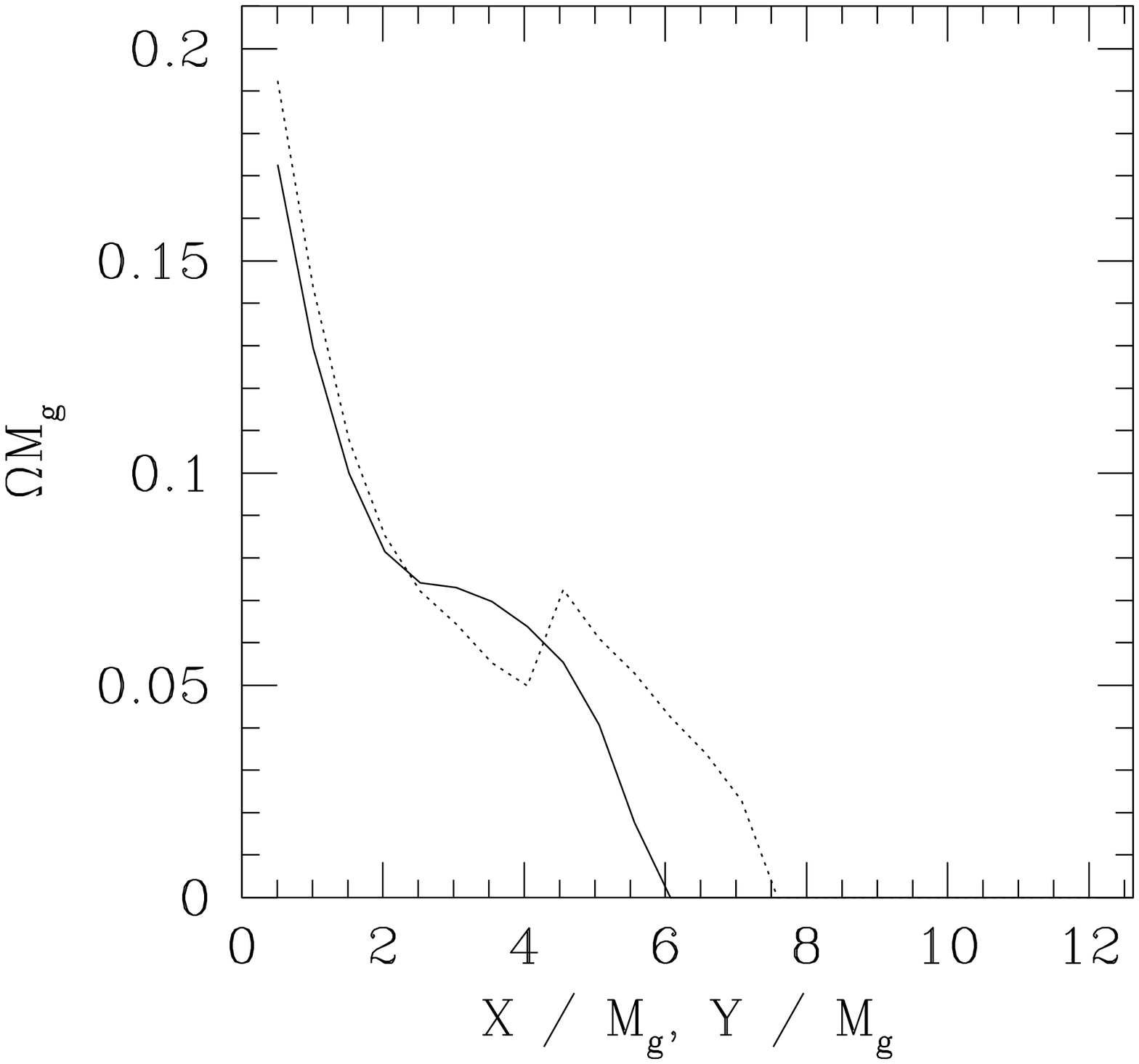}
\end{center}
\caption{The angular velocity $\Omega$ along the $x$-axis (solid line)
and the $y$-axis (dotted line) at $t = 1.81 P_{\rm orb}$ for model
(I1).  Here $M_g$ is the gravitational mass $M$.  (Figure from
\citet{su00a}.)}
\label{fig10.2}
\end{figure}

The first simulation, model (I1) in \citet{su00a}, is for a binary of
total rest mass $\bar M_0 = 0.261$ \footnote{We have converted the
units of \citet{su00a} into the same dimensionless units adopted in
Section \ref{sec9} for easier comparison.}.  The angular momentum of
the initial data is $J/M^2 = 0.98$, where $M$ is the total
gravitational mass, and is hence smaller than the Kerr limit $J/M^2 =
1$.  We show snapshots of density contours in Figure \ref{fig10.1}.

\begin{figure}[t]
\begin{center}
\epsfxsize=2.2in
\leavevmode
\epsffile{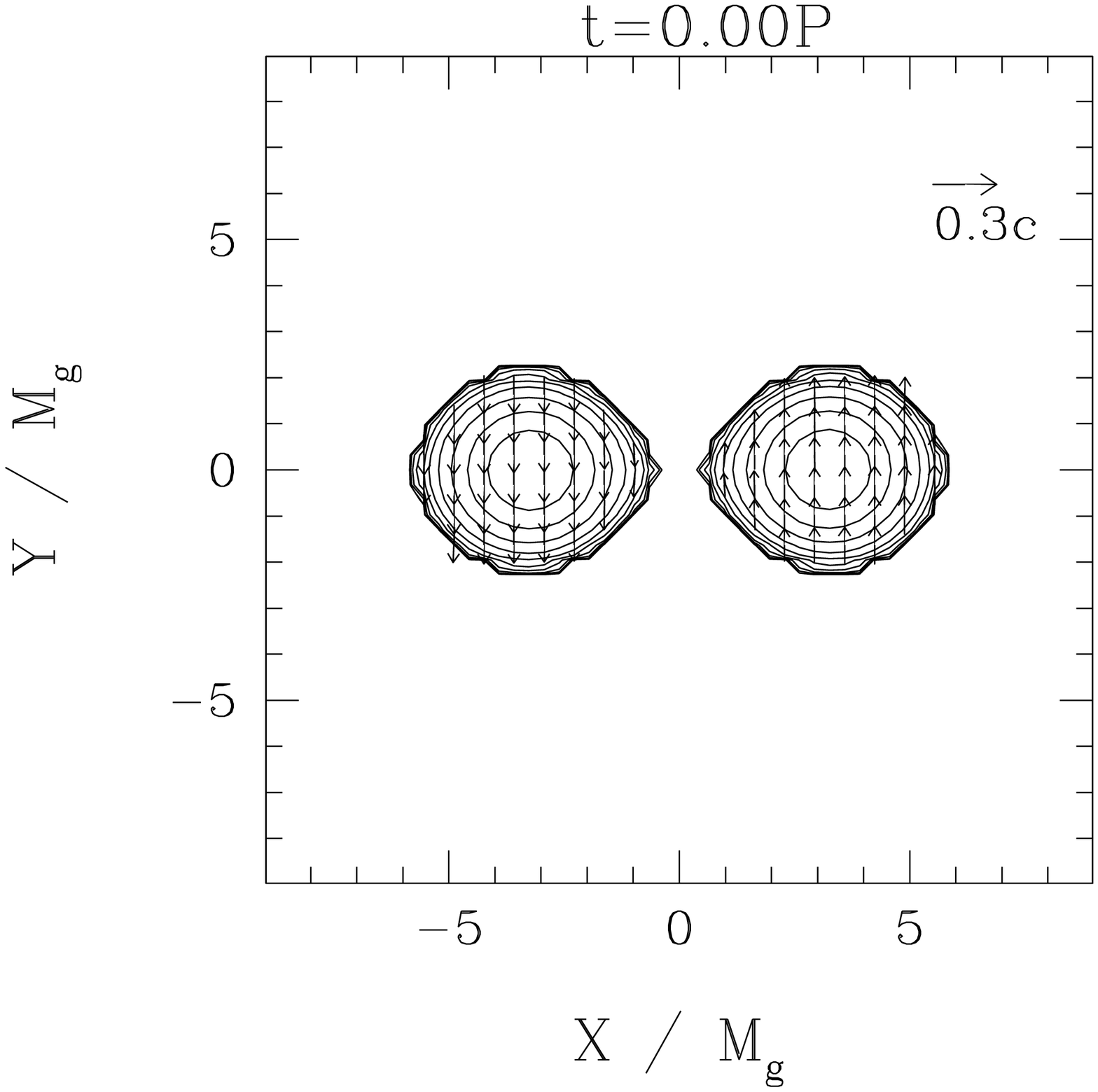}
\epsfxsize=2.2in
\leavevmode
\epsffile{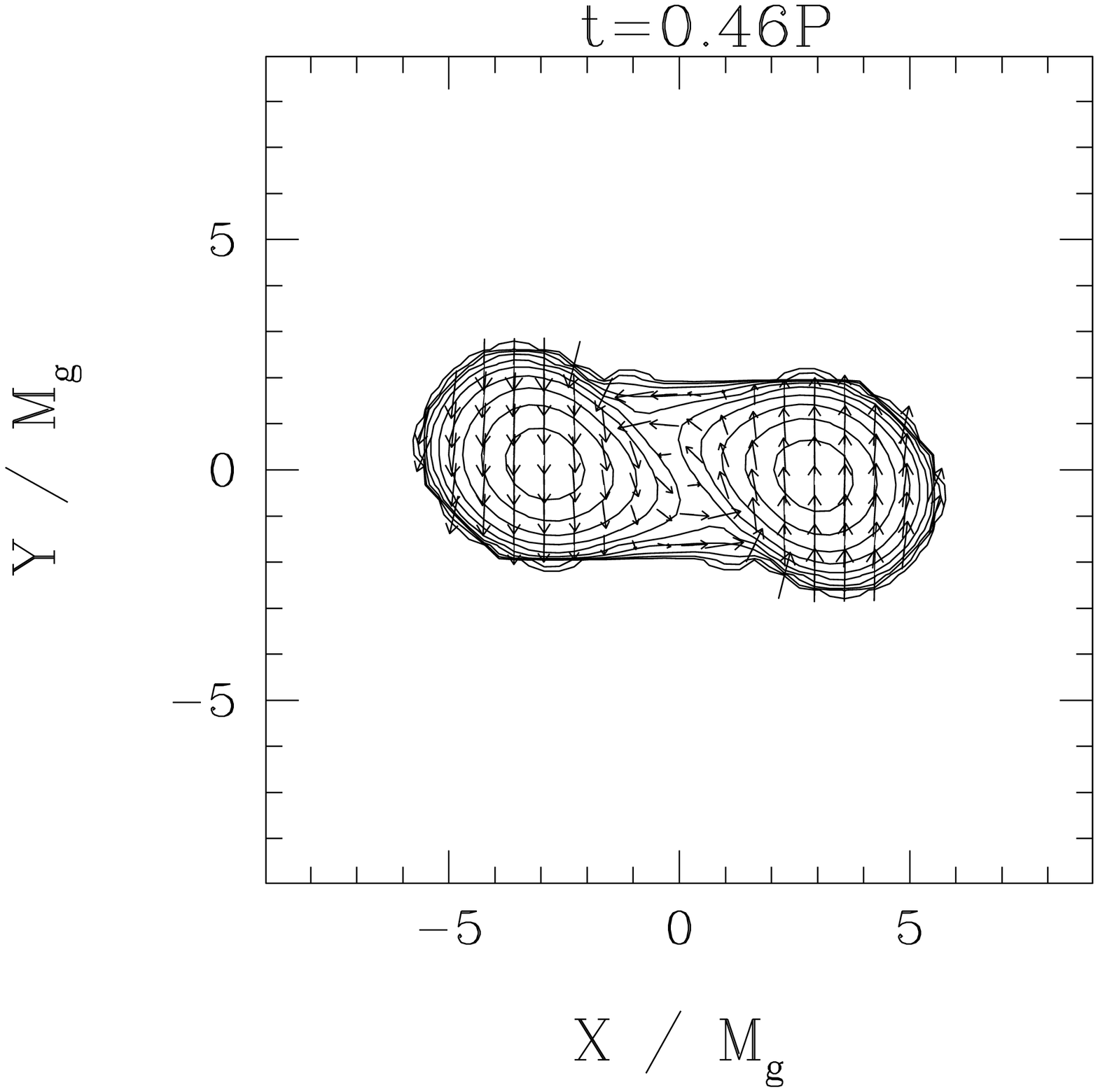}\\
\epsfxsize=2.2in
\leavevmode
\epsffile{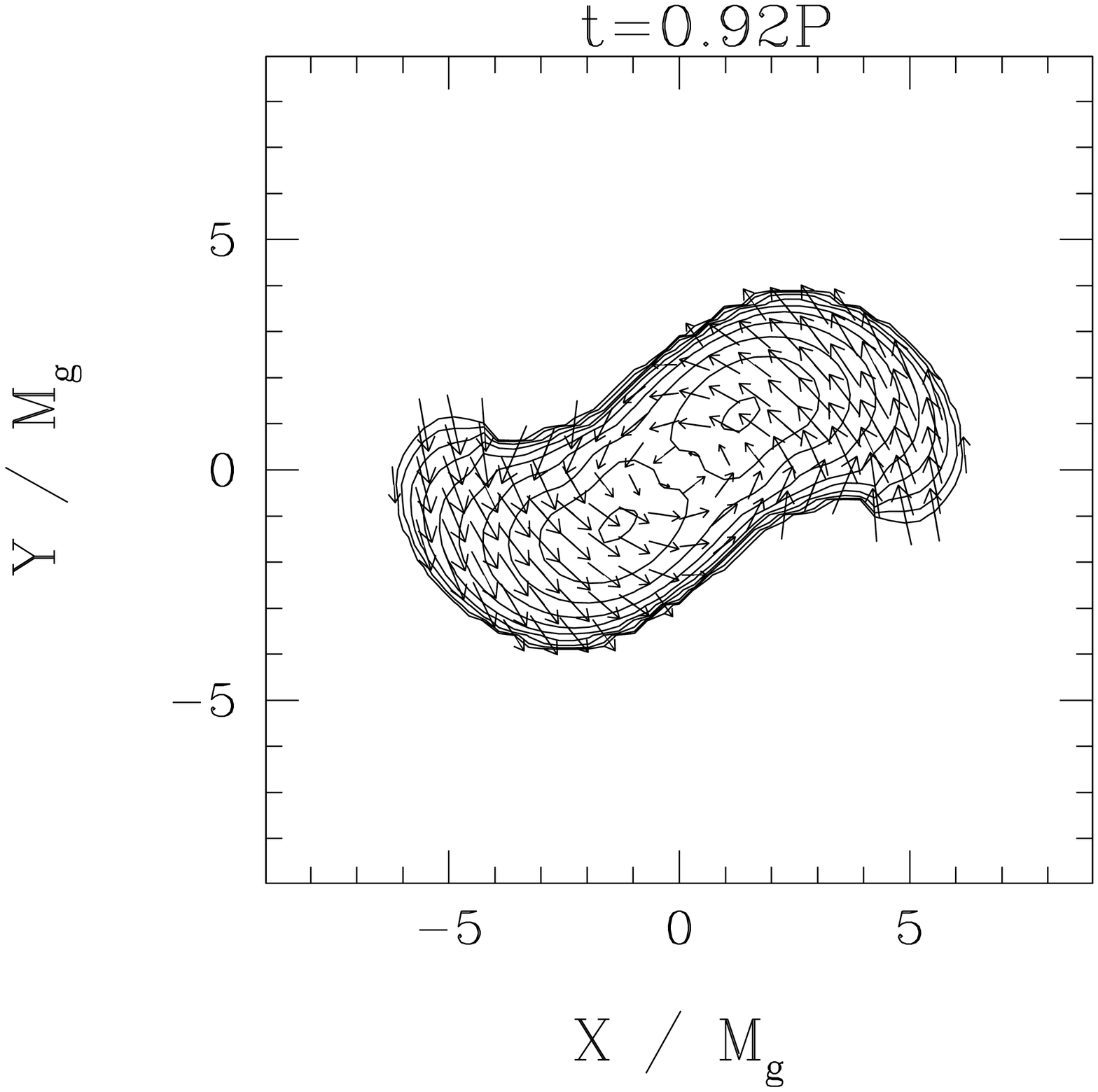}
\epsfxsize=2.2in
\leavevmode
\epsffile{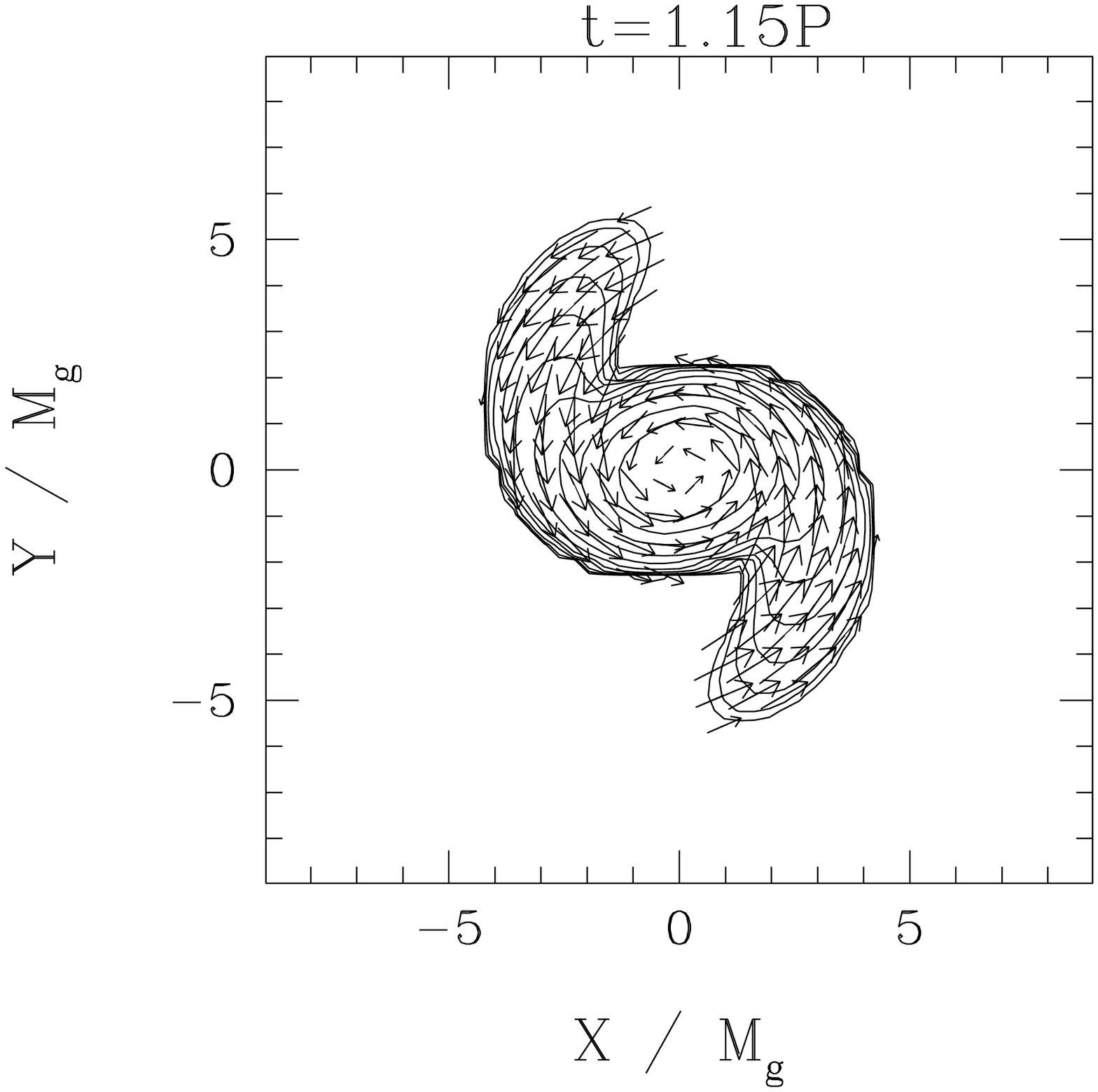} \\
\epsfxsize=2.2in
\leavevmode
\epsffile{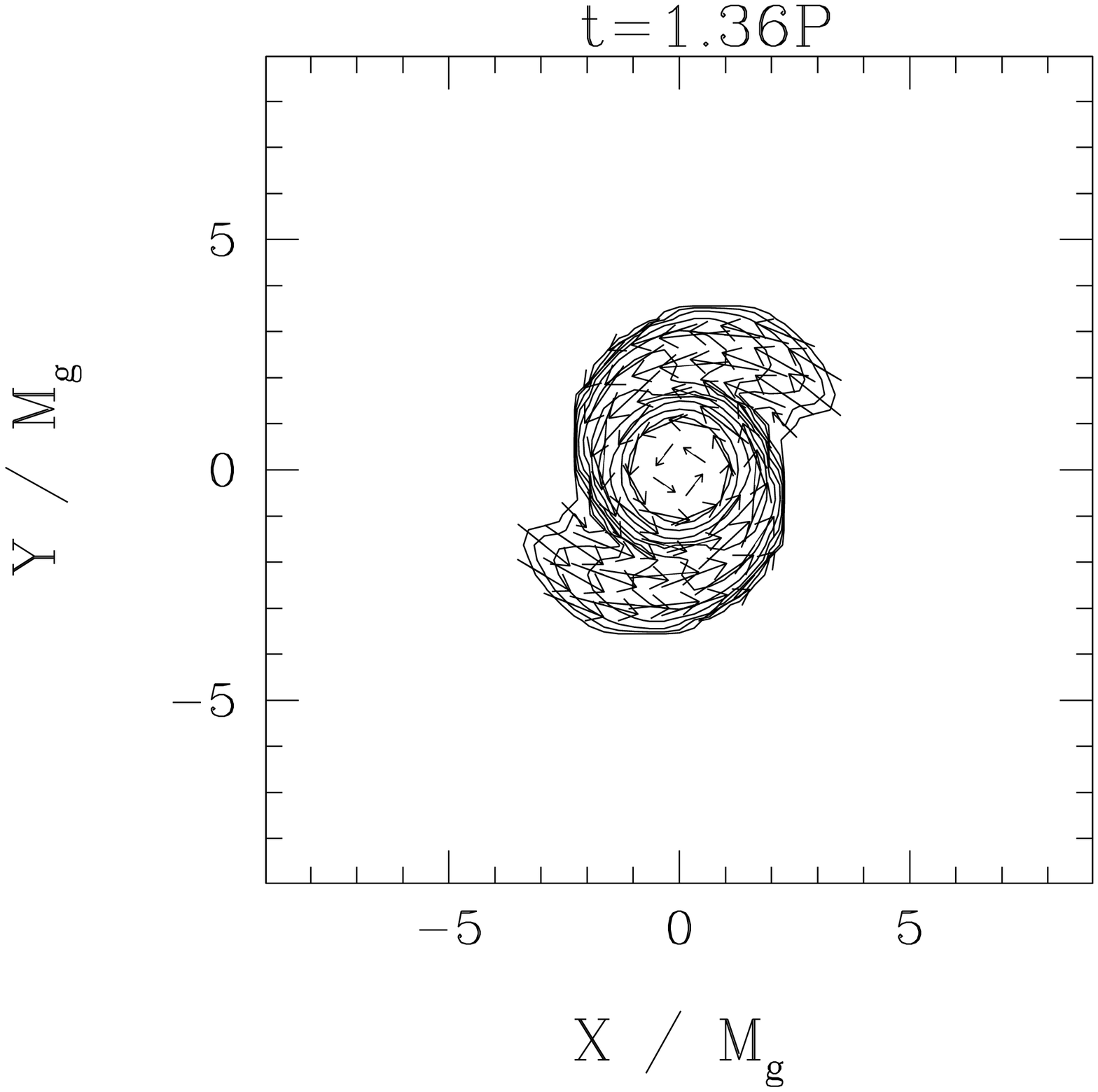}
\epsfxsize=2.2in
\leavevmode
\epsffile{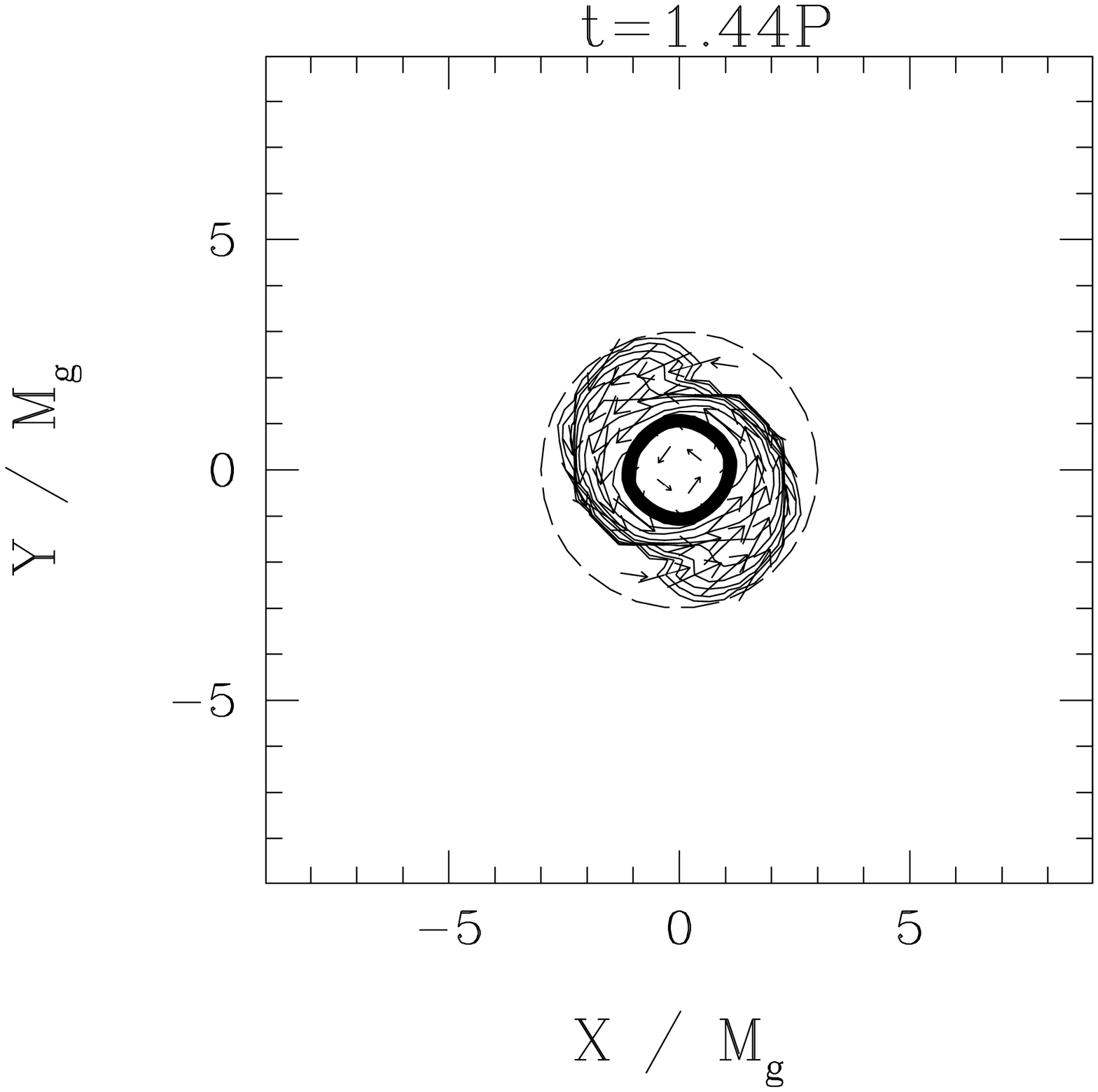} 
\caption{Same as Figure \ref{fig10.1}, but for a binary of total 
rest mass $\bar M_0 = 0.332$ (Model (I3) in \citet{su00a}).  Contour lines
denote densities $\rho_*/\rho_{*~{\rm max}}=10^{-0.3j}$ with
$\rho_{*~{\rm max}}=0.866$ and $j=0,1,2,\cdots,10$.  The dashed line
in the last snapshot is the circle $r = 3 M$ which encloses over 99\%
of the total rest mass.  The thick solid line at $r \approx M$ denotes
the location of the apparent horizon.  (Figure from
\citet{su00a}.)}
\label{fig10.3}
\end{center}
\end{figure}

In contrast to the coalescence of corotational binaries, no
significant spiral arms form during merger, and hardly any matter is
ejected.  It is quite surprising then that the remnant settles down to
a near-equilibrium neutron star and does not collapse to a black
hole\footnote{At least not on a dynamical timescale -- see below.},
even though its rest mass exceeds the maximum allowed rest mass of a
spherical, non-rotating star by about 45\%.  \citet{su00a} find only
small amounts of shock heating, which rules out thermal pressure as
providing the extra support.  As we have pointed out, the angular
momentum $J/M^2$ is smaller than the Kerr limit and therefore cannot
prevent black hole formation.  Uniform rotation can increase the
maximum allowed rest mass for $\Gamma = 2$ polytropes by only about
20\% \citep{cst94}. However,
\citet{su00a} find that the core is differentially rotating as opposed
to uniformly rotating, as shown in Figure \ref{fig10.2}.

\begin{figure}
\begin{center}
\epsfxsize=2.5in
\epsffile{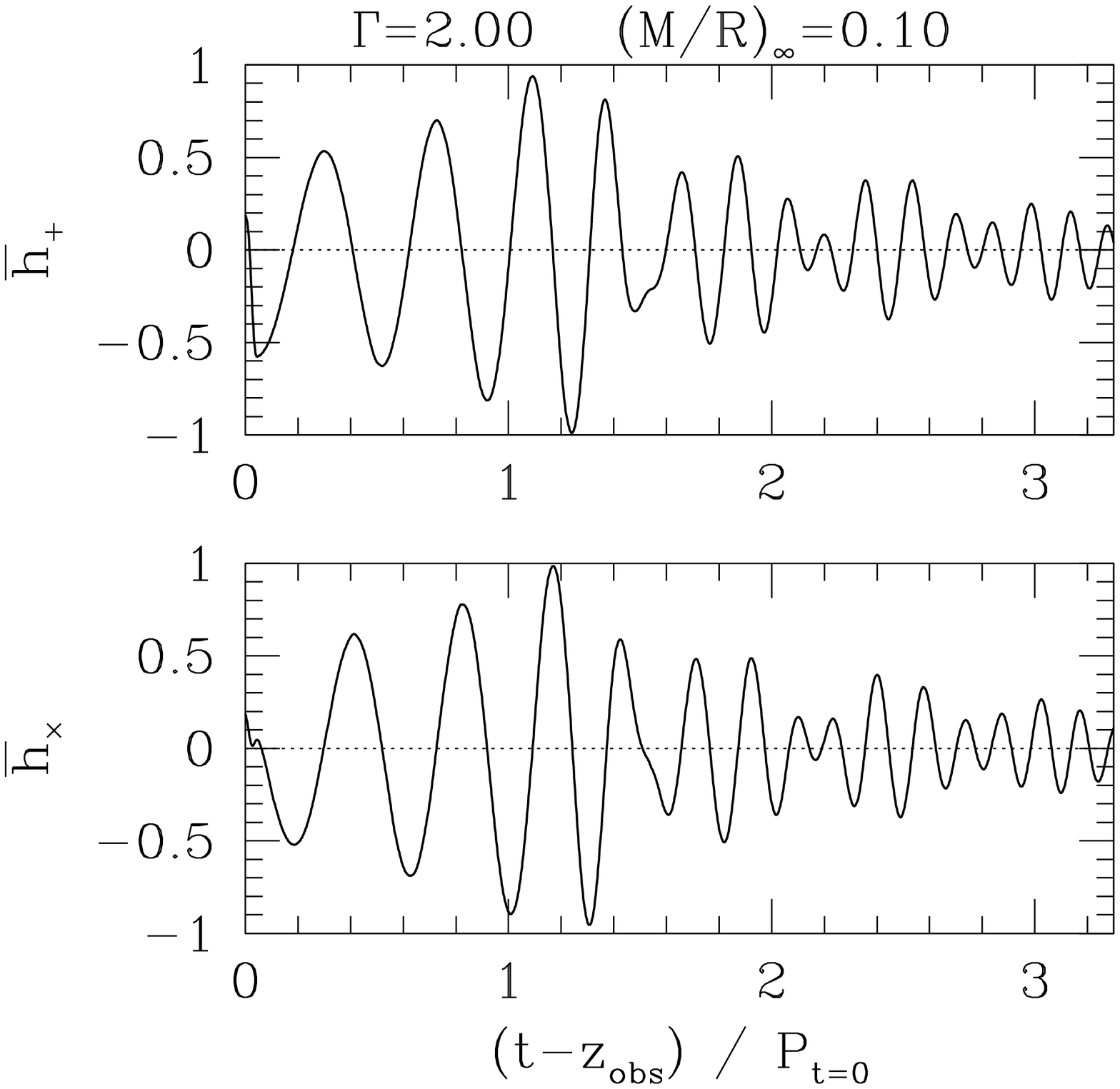}
\epsfxsize=2.5in
\epsffile{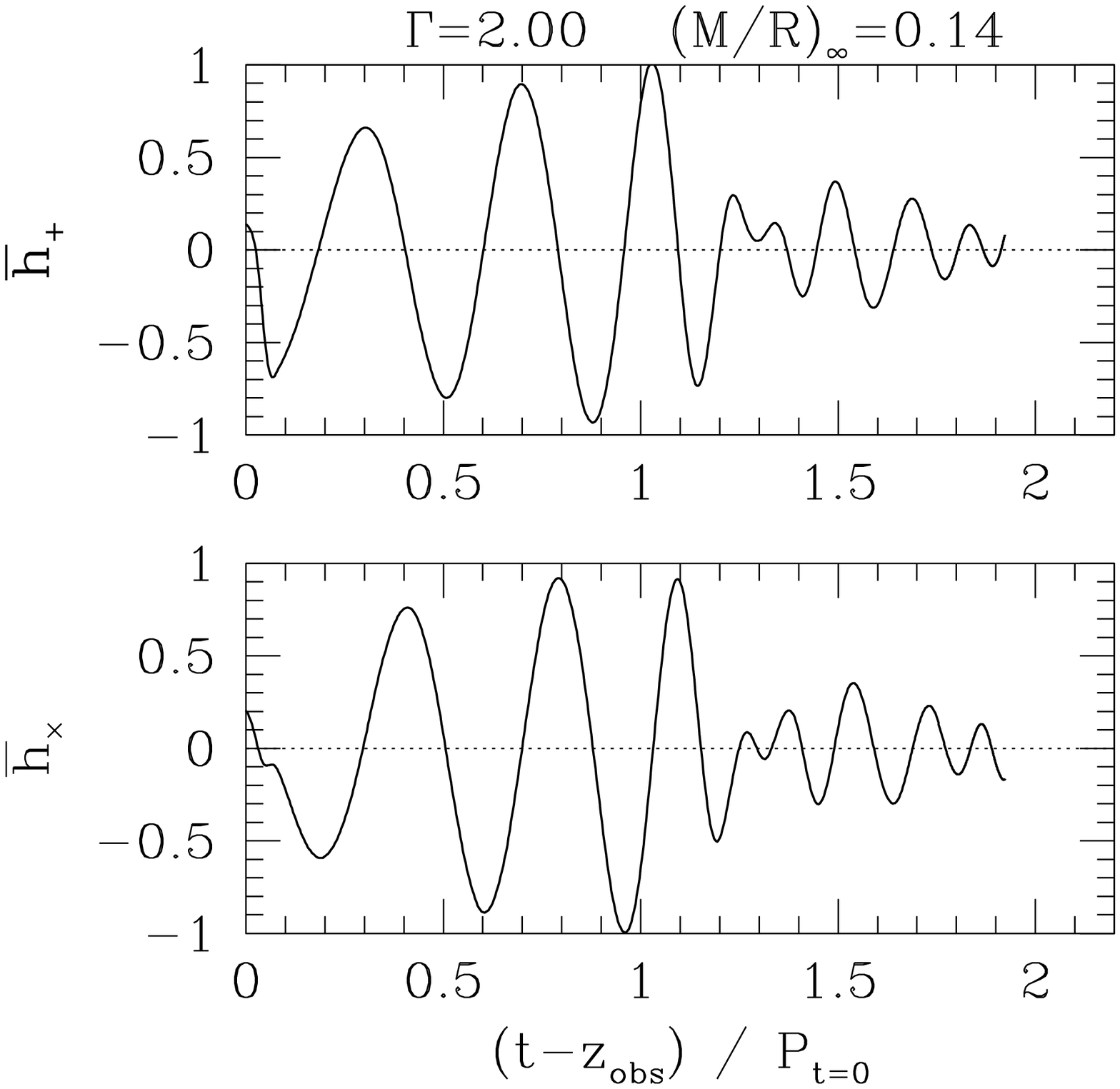}
\end{center}
\caption{Gravitational wave amplitudes $h_{+}$ and $h_{\times}$ as a 
function of retarded time for the irrotational binary neutron
star models (E) and (H) of \citet{su02}.  Model (E) corresponds to
a slightly smaller mass then (I1) and results in a differentially
rotating neutron star, while model (H) is similar to model (I2)
and results in a black hole.  (Figures from \citet{su02}.)}
\label{fig10.4}
\end{figure}

It is quite intuitive that differential rotation may significantly
increase the maximum mass \citep{bss00}.  For uniform rotation, the
angular velocity, and hence the centrifugal force which balances the
gravitational force to increase the maximum mass, is limited by the
Kepler limit at the equator, above which matter there would no longer
be gravitationally bound (the ``mass-shedding limit'').  For
differential rotation, the core may rotate faster than the equator,
and may further increase the maximum mass without violating the Kepler
limit at the equator.  \citet{bss00} have found that differential
rotation is very effective in increasing the maximum mass, and have
found mass increases of over 60\% even for very modest degrees of
differential rotation.

The remnant formed in this simulation is supported by differential
rotation, as the thermal pressure support is minimal while its rest
mass exceeds the maximum allowed rest mass by about 45\%.  Similar
results were found more recently by \citet{ort01}, who used a
completely independent numerical method (namely an SPH method to model
the hydrodynamics and the Wilson-Mathews approximation to model the
gravitational fields).  This result is quite surprising, since up to
then it was usually assumed that such massive neutron stars collapse
to black holes on a dynamical timescale.  (although earlier Newtonian
merger calculations already foreshadowed this result; see \citet{rs99}
and references therein).  Instead, it is possible that such
``hyper-massive'' remnants are dynamically stable and collapse to
black holes on a secular timescale, after some dissipative mechanism,
for example viscosity or, more likely, magnetic fields, has brought
the star into more uniform rotation \citep{bss00,s00}.

Both models (I2) and (I3), for which the total rest mass exceeds the
maximum allowed mass by 63\% and 85\% (see Table \ref{10_table1}),
form a black hole within a dynamical timescale upon merger.  In Figure
\ref{fig10.3} we show snapshots of the density contours and velocity
profiles for the more massive model (I3).  In the last frame of Figure
\ref{fig10.3} the thick solid line denotes the location of an apparent
horizon (see Section \ref{sec6.1}), indicating the formation of a
black hole.

\citet{su00a} find fairly similar results for corotational binary
models.  Probably the most significant difference is that corotational
binaries have more angular momentum in the outer parts of the binary,
which leads to the formation of spiral arms during the coalescence.
The spiral arms contain a few percent of the total mass, and may
ultimately form a disk around the central object.

In the simulations of \citet{su00a} one of the largest limitations on
the accuracy were the outer boundaries, which, because of limited
computational resources, had to be imposed well within a wavelength of
the gravitational radiation from the binary (at about $\lambda_{\rm
gw}/3$).  This means that the waves were extracted without being in
the radiation zone, which necessarily introduces error.  After having
gained access to a more powerful supercomputer, \citet{su02} therefore
repeated these calculations on computational grids that extend to
about a gravitational wavelength.  Qualitatively, these improved
results are very similar to their earlier ones, although the onset of
black hole formation shifts to slightly smaller masses.  Most strongly
affected by these improvements are the gravitational wave forms.  In
Fig \ref{fig10.4}, we therefore show examples from these improved
simulations for models that are similar to model (I1), leading to a
differentially rotating neutron star, and model (I2), leading to a
black hole.

While the simulations of \citet{su00a,su02} are pioneering, it would
be desirable to confirm their findings with independent simulations
with fully self-consistent codes.  Many aspects of the simulations
could also be improved in the future, including the setup of the
initial data (eliminating the need for an artificial reduction of the
angular momentum), the extraction of gravitational radiation, and the
handling of the hydrodynamics.  For example, coalescing irrotational
neutron stars form a vortex sheet at the contact surface that is
Kelvin-Helmholtz unstable on all wavelength (see, e.g.,
\citet{rs99} and references therein).  Reliably simulating such sheets
is quite challenging, and is likely to require more sophisticated
algorithms than artificial viscosity schemes (compare Section
\ref{sec10.1},  \citet{fmst00,fgimrssst01}).  Lastly it should be 
emphasized that these simulations assume polytropic stars governed by
a gamma-law equation of state, which is an idealization.  In reality,
a number of factors, including effects of a realistic nuclear equation
of state, magnetic fields and neutrino transport, may play an important
role in the coalescence of binary neutron stars.  It will probably be
a while until all these can be incorporated in fully dynamical and
self-consistent simulations.

\subsection{The Quasi-Adiabatic Inspiral of Binary Neutron Stars}
\label{sec10.4}

As we have discussed in Section \ref{sec1}, it is possible that other
means of modeling binary neutron stars, in particular PN point-mass
techniques, break down somewhat outside of the ISCO, when finite-size
and relativistic effects become important.  It is hard to imagine that
fully hydrodynamical numerical calculations would be able to follow
the inspiral reliably from such a point outside of the ISCO through
many orbital periods to the onset of instability at the ISCO, followed
by plunge and merger.  Such calculations would accumulate significant
amounts of numerical error and would be computational prohibitive.
This leaves a gap between the regimes that PN and fully numerical
calculations can model.  Filling this gap for the late inspiral,
immediately prior to plunge, therefore requires an alternative,
approximate approach (in the case of binary black holes, this problem
has been called the ``intermediate binary black hole'' problem
\citep{bct98}).

\begin{figure}
\begin{center}
\epsfxsize=4in
\epsffile{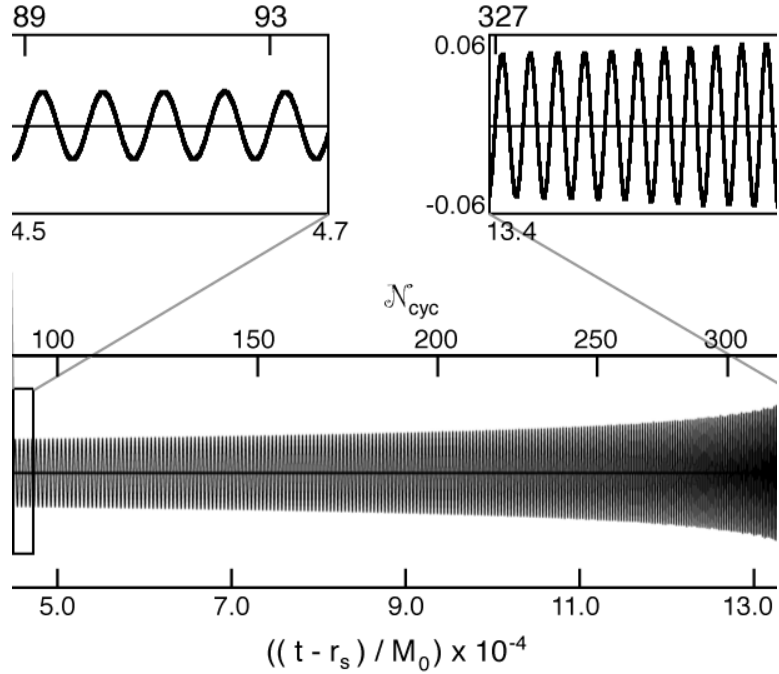}
\end{center}
\caption{The final hundreds of cycles of the inspiral waveform
$h_+$ or $h_{\times}$ on the axis of rotation as a function of
retarded time and cycle number for an irrotational binary neutron star
system.  Also indicated is the gravitational strain $h$ for a binary
of total rest mass $M_0 = 2 \times 1.5 M_{\odot}$ at a separation of
100 Mpc.  (Figure from
\citet{dbssu01}.)}
\label{fig10.5}
\end{figure}

Several different such approaches have been suggested 
\citep{bd92,bct98,l99,wr99,wkp00,dbs01,dbssu01,su01}.  Here we will
focus on the ``hydro-without-hydro'' approach (see also \citet{bhs99})
adopted by \citet{dbs01} and \citet{dbssu01} (see also
\citet{ybs01a}, who illustrated and calibrated this approach in
a scalar gravitation model problem).

This approach approximates the binary orbit outside of the ISCO as
circular, and treats the orbital decay as a small correction.  For
each binary separation, the matter distribution can then be determined
independently by the quasi-equilibrium methods of Section \ref{sec9}.
These matter profiles, which satisfy the equations of quasi-stationary
equilibrium, can then be inserted as matter sources into Einstein's
equations.  Evolving the gravitational fields yields the gravitational
wave signal and luminosity, and hence the rate at which the system
loses energy at each separation.  Combining this luminosity $dM/dt$
with the derivative of the binding energy $M$ with respect to
separation $r$ yields the inspiral rate
\begin{equation} \label{10_drdt}
\frac{dr}{dt} = \frac{dM/dt}{dM/dr}.
\end{equation}
Integrating this equation yields the separation as a function of time,
and accordingly the entire gravitational wave train.

\begin{figure}
\begin{center}
\epsfxsize=3in
\epsffile{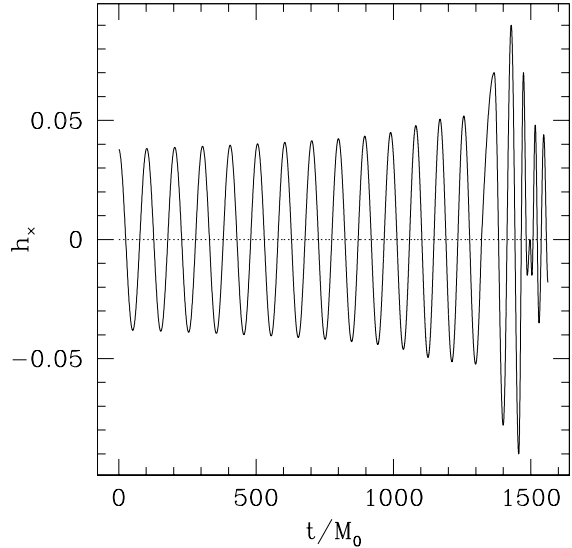}
\end{center}
\caption{A match of the late quasi-equilibrium inspiral wavetrain 
and the plunge and merger waveform for a binary of total rest mass
$M_0 = 1.62 M_0^{\rm max}$.  The merger of these binary results in
immediate black hole formation (compare Section \ref{sec10.3}).  Here
$h_{\times} = r_s/M_0 \bar \gamma_{xy},$ where $r_s$ is the distance
to the source.  For $M_0 = 2 \times 1.5 M_{\odot}$ and a distance to
the source of 100 Mpc, the maximum amplitude of the metric
perturbation is $\sim 5 \times 10^{-21}$. (Figure from
\citet{dbssu01}.)}
\label{fig10.6}
\end{figure}

\citet{dbs01} implemented such a scheme for corotational binaries,
based on the quasi-equilibrium models of \citet{bcsst98b} (see Section
\ref{sec9.2}).  In \citet{dbssu01}, these results were compared with
those for an irrotational sequence, based on the models of
\citet{use00} (see Section \ref{sec9.3}).  These simulations are 
illustrative only due to the small outer boundary radius which
necessitated gravitational read-off inside the wave zone (compare
Section \ref{sec10.3}).  However, the method is quite promising.  In
Figure \ref{fig10.5} we show such a gravitational wavetrain for an
irrotational binary.

For a given separation, the gravitational wave luminosity $dM/dt$ is
very similar for corotational and irrotational models\footnote{As one
might expect, since the gravitational wave emission is dominated by
the matter density, which is fairly similar for the corotational and
irrotational binaries, while matter current distributions play a less
important role.}, but since the binding energy of corotational models
includes the spin kinetic energy of the individual stars, the absolute
value of the corotational binding energy and hence $|dM/dr|$ is
smaller than that of irrotational binaries.  According to
(\ref{10_drdt}) this makes $|dr/dt|$ larger, meaning that the inspiral
of corotational binaries proceeds faster than that of irrotational
binaries.

\citet{dbssu01} also point out that the entire gravitational wavetrain,
from the slow inspiral to the ISCO and the subsequent plunge and
merger, can be constructed by matching results from an quasi-adiabatic
approximation of the inspiral and a dynamical simulation of the
coalescence.  An example, showing the last 13 orbits of the inspiral
together with the plunge and merger leading to black hole formation
(as obtained by \citet{su00a}), is shown in Figure \ref{fig10.6}.
\section{Summary and Outlook}
\label{sec11}

A search of the recent literature reveals the impressive progress that
numerical relativity and the modeling of compact binaries has made in
the past few years.  New formulations of the initial value problem
(Section \ref{sec3}) and the evolution equations (Section
\ref{sec4}), new coordinate conditions and their implementations
(Section \ref{sec5}) as well as new diagnostics (including the
horizon finders described in Section \ref{sec6}) have led to several
advances in the simulation of black holes and neutron stars.  Recent
breakthroughs include initial data for binary black holes (Section
\ref{sec7}) and neutron stars (Section \ref{sec9}), evolution
calculations for single and binary black holes (Section \ref{sec8})
and simulations of inspiraling and merging binary neutron stars
(Section \ref{sec10}).  Some results of astrophysical interest are
density profiles of relativistic binary polytropes, the location of
the ISCO in binary black holes and neutron stars, the stability
properties of neutron stars in close binaries, cusp formation in
irrotational binary neutron stars, the stabilizing effect of
differential rotation in remnants of binary neutron star coalescence,
and the first preliminary gravitational waveforms from coalescing
binary neutron stars and black holes.

One of the long-standing goals of numerical relativity is the
simulation of the coalescence of binary black holes.  While this goal
has not been achieved yet, some preliminary calculations have been
performed (Section \ref{sec8.2}).  Moreover, with recent progress in
the evolution of single black holes (Section \ref{sec8.1}), the
simulation of binary black holes seems much more feasible than it did
only a few years ago.

It is likely that most of the numerical results reported in this
article will be revisited and improved.  Some of these possible
improvements have already been discussed in the corresponding
Sections.  For binary black hole initial data (Section \ref{sec7}) it
would be desirable to gain a clearer understanding of why different
approaches lead to different values of the ISCO.  Specific
calculations, some of which have already been initiated, include
improvements of the thin-sandwich approach of \citet{ggb01b} and the
construction of binaries in circular orbit using Kerr-Schild or
post-Newtonian background data, which would avoid the assumption of
spatial conformal flatness.  Avoiding this assumption may also be
interesting for binary neutron star initial data (Section \ref{sec9}),
in addition to extending sequences of irrotational binaries beyond
cusp formation (see Section
\ref{sec9.3}).  The latter would provide better initial data for
dynamical simulations of binary neutron star coalescence (Section
\ref{sec10.3}), which could also be improved by using more
sophisticated hydrodynamics algorithms.  Other future improvements
will be the inclusion of more realistic nuclear equations of state,
magnetic fields, and possibly neutrino transport.  Most of the results
that we have discussed adopt the assumption of equal mass binaries,
which eventually will also have to be generalized\footnote{Note,
however, that the mass of most neutron stars in neutron star binaries
is close to 1.4 $M_{\odot}$ \citep{tc99}, so that the assumption of
equal mass seems quite adequate.}.

Almost all numerical results discussed in this article would benefit
from increased accuracy, and most will need to be significantly more
accurate before they can be useful in gravitational wave catalogs as
discussed in Section \ref{sec1}.  On uniform grids, two competing
sources of numerical error are finite difference errors (due to coarse
numerical grid resolution) and the location of outer boundaries (if
they are imposed at too small a separation).  Given the available
computational resources and hence the size of the computational grid
that can be afforded, a more or less suitable compromise has to be
chosen between grid resolution and distance to the outer boundaries.
In current simulations, this compromise still leads to significant
error, in particular for gravitational wave forms (see Sections
\ref{sec10.3} and \ref{sec10.4}).  Two improvements may be possible:
instead of uniform grids, nested grids or adaptive mesh refinement
(AMR) may be employed.  While such techniques have been used in many
other fields of computational physics as well as in lower-dimensional
numerical relativity (see, e.g.~\citet{c93}), they have been used
sparingly in $3+1$ numerical relativity so far
(e.g. \citet{b96,b99,nccmho00,jdkn02}).  In addition, the handling of
the outer boundaries and gravitational radiation extraction could be
refined.  Some recent efforts include
\citet{bghmpw96,aetal98,fn99,sgbw00,clt01,ssw02}.

Future effort will also be aimed in new directions.  For example, it
would be very desirable to model the late inspiral of binary black
holes with a quasi-adiabatic approximation, similar to the binary
neutron star calculations described in Section \ref{sec10.4}.  

Lastly, we have only discussed binaries containing two black holes or
two neutron stars, while black hole-neutron star binaries have so far
been neglected (but see\citet{m01}).  The inspiral, coalescence and
merger of black hole-neutron star binaries, including the possible
tidal disruption of the neutron star, seems like an extremely
interesting subject and a promising source of gravitational radiation,
and is likely to receive more attention in the future.

\begin{ack}
Over the years we have greatly benefitted from numerous discussions
with many colleagues, including Andrew Abrahams, Greg Cook, Matt Duez,
Mark Scheel, Masaru Shibata, Saul Teukolsky and Kip Thorne.  We would
like to thank the Visitors Program in the Numerical Simulation of
Gravitational Wave Sources at Caltech for extending their hospitality
while a portion of this review was being completed.  This work was
supported in part by NSF Grants PHY-0090310 and PHY-0205155 and NASA
Grant NAG5-10781 at the University of Illinois at Urbana-Champaign
and NSF Grant PHY-0139907 at Bowdoin College.
\end{ack}

\begin{appendix}

\section{Notation}
\label{appA}

We adopt the notation of \citet{w84}, which is based on that of
\citet{mtw73}.  By convention, we also adopt the ``fortran'' convention in
which the letters $a - h$ and $o - z$ are four-dimensional indices and
run from 0 to 3, while the letters $i - n$ are three-dimensional
indices that run from 1 to 3.  We use geometrized units in which $c =
G = 1$.

Unless stated differently, objects associated with the
four-dimensional metric $g_{ab}$ are denoted with a ${}^{(4)}$ in
front of the symbol, objects associated with the conformally
related (three-dimensional) metric $\bar
\gamma_{ij}$ carry a bar, and only objects related to the spatial
metric $\gamma_{ij}$ do not carry any decorations. For example,
$\Gamma^i_{jk}$ is associated with $\gamma_{ij}$, $\bar \Gamma^i_{jk}$
with $\bar \gamma_{ij}$, and ${}^{(4)}\Gamma^i_{jk}$ with $g_{ab}$.
The covariant derivative operator is denoted with $D_i$ and $\bar D_i$
when compatible with the spatial metric and the conformally related
metric, but with the nabla symbol $\nabla_a$ when compatible with the
four-dimensional metric $g_{ab}$.  We occasionally use $\Delta^{\rm
flat}$ for the flat scalar Laplace operator.

We denote the symmetric and antisymmetric parts of a tensor with
brackets $()$ and $[]$, for example
\begin{equation}
T_{(ab)} = \frac{1}{2} ( T_{ab} + T_{ba} )
\mbox{~~~and~~~}
T_{[ab]} = \frac{1}{2} ( T_{ab} - T_{ba} ).
\end{equation}
Finally, we write a flat four-dimensional metric as $\eta_{ab}$
(Minkowski spacetime) and a flat three-dimensional metric as
$\eta_{ij}$.

\section{Solving the Vector Laplacian}
\label{appB}

In this appendix we discuss two approaches to solving
the flat vectorial Poisson equation 
\begin{equation}
(\Delta_L^{\rm flat} W)^i = S^i,
\end{equation}
or, in Cartesian coordinates,
\begin{equation} \label{B_1}
\partial^j \partial_j W_i + \frac{1}{3} \partial_i \partial^j W_j 
= S_i.
\end{equation}
We have encountered this operator in the momentum constraint
(\ref{3_mom_1}) and (\ref{7_mom_1}), in the shift condition of the
thin sandwich approach (\ref{3_ts_md}), in the minimal distortion
condition (\ref{5_md_2}), and in the Gamma-freezing condition
(\ref{5_GF_2}).

\citet{by80} suggest writing the vector $W_i$ as a sum of a vector 
$V_i$ and a gradient of a scalar $U$,
\begin{equation}
W_i = V_i + \partial_i U.
\end{equation}
Inserting this into (\ref{B_1}) yields
\begin{equation} \label{B_2}
\partial^j \partial_j V_i + \frac{1}{3} \partial_i \partial^j V_j +
\partial^j \partial_j \partial_i U + 
\frac{1}{3} \partial_i \partial^j \partial_j U = S_i.
\end{equation} 
We can now choose $U$ such that the two $U$ terms in (\ref{B_2})
cancel the second $V_i$ term,
\begin{equation} \label{B_2a}
\partial^j \partial_j U = - \frac{1}{4} \partial_j V^j,
\end{equation}
so that (\ref{B_2}) reduces to
\begin{equation}
\partial^j \partial_j V_i = S_i.
\end{equation}
We have hence rewritten the vector Poisson equation (\ref{B_1}) as
a set of four scalar Poisson equation for $U$ and the components of 
$V_i$.

A second approach has been suggested by \citet[see also
\citet{ons97}, Appendix C]{s99c}, who used the ansatz
\begin{equation} \label{B_3}
W_i = \frac{7}{8} P_i - \frac{1}{8} 
\Big(\partial_i U + x^k \partial_i P_k \Big).
\end{equation}
Inserting this into (\ref{B_1}) yields
\begin{equation} \label{B_4}
\frac{5}{6} \partial^j \partial_j P_i
- \frac{1}{6} \partial_i \partial^j \partial_j U
- \frac{1}{6} x^k \partial_i \partial^j \partial_j P_k = S_i
\end{equation}
If we now choose $U$ so that it satisfies
\begin{equation} \label{B_5}
\partial^j \partial_j U = - S_j x^j,
\end{equation}
then (\ref{B_4}) reduces to
\begin{equation}
\frac{5}{6} \partial^j \partial_j P_i
+ \frac{1}{6} x^j \partial_i S_j 
- \frac{1}{6} x^j \partial_i \partial^k \partial_k P_j
= \frac{5}{6} S_i
\end{equation}
and is solved by
\begin{equation} \label{B_6}
\partial^j \partial_j P_i = S_i.
\end{equation}
The vector equation (\ref{B_1}) has again been reduced to a set of
four scalar Poisson equations, (\ref{B_5}) and (\ref{B_6}).  While the
approach of \citet{s99c} seems a little more complicated, it has the
advantage that the source terms in (\ref{B_5}) and (\ref{B_6}) are
non-zero only where $S_i$ is non-zero.  In some cases (for example for
the momentum constraint (\ref{3_mom_1})) this may lead to compact
sources, which can have advantages for numerical implementations
\citep{gbgm01}.  In the approach of \citet{by80}, on the other hand,
the source term of equation (\ref{B_2a}) is never compact.  A third
approach has been suggested by \citet{on99}.  Numerical
implementations (using spectral methods) of the three approaches have
been compared by \citet{gbgm01}.

\section{Conformally Flat or Not?}
\label{appC}

Many numerical calculations, especially for the construction of
initial data, assume the spatial metric to be conformally flat.
Simultaneously, many authors have pointed out the limitations of
conformal flatness, and have argued strongly against that simplification.
Not all arguments have been correct, and it may be useful to briefly
review those both in favor and against conformal flatness.

As we have seen in Section \ref{sec3}, conformal flatness greatly
simplifies the initial value equations.  Moreover, before the initial
value equations can be solved, some form of the conformally related
metric has to be chosen.  In some cases educated guesses can be made
(for example by choosing Kerr-Schild background data \citep{mm00} or
by adopting a post-Newtonian metric instead of a flat
conformally-related background metric as suggested in Section
\ref{sec7.3}).  However, in most cases it may not be clear what a
better choice than conformal flatness might be.

In Section \ref{sec3} we found that the dynamical degrees of freedom
of the gravitational fields can be identified with parts of the
conformally related spatial metric and the transverse-traceless part
of the extrinsic curvature.  This suggests that the assumption of
conformal flatness and vanishing of $\bar A_{ij}^{TT}$ may ``minimize
the gravitational radiation content'' of a spatial slice $\Sigma$.
This argument is not strictly true in general; for example it does not
even hold for single rotating black holes.  Rotating Kerr black holes,
for example, which are stationary and do not emit any gravitational
radiation, are not conformally flat\footnote{At least slices of
constant Boyer-Linquist time are not conformally flat, nor are
axisymmetric foliations that smoothly reduce to slices of constant
Schwarzschild time in the Schwarzschild limit \citep{gp00}}.
Similarly, conformally flat models of rotating black holes that are
constructed in the Bowen-York formalism do contain gravitational
radiation \citep[see also \citet{jdkn02}]{bs95a,bs95b,bs96,gnpp98}.

For rapidly rotating single neutron stars it has been shown that
conformal flatness introduces an error of at most a few percent
\citep{cst96}.  Similarly small discrepancies were found by
\citet{uue00,ue02}, who compared conformally flat binary neutron star
models with models constructed under different assumptions.  These
small deviations are not surprising, since differences between a
conformally flat metric and the ``correct'' metric appear at second
post-Newtonian order (e.g. \citet{rs96}), which are the order of a
few percent for neutron stars.  It is therefore quite likely that, at
least for neutron stars, the error arising from the conformal flatness
assumption may be less than other errors typically expected in current
simulations, including finite difference errors and effects from the
poor handling of the outer boundaries.

\citet[see also \citet{wmm96}]{wm95} adopted the conformal flatness 
approximation in their simulations of binary neutron stars, and found
that their neutron stars collapsed to black holes individually prior
to merging.  Since this result was very counter-intuitive, it was
suspected that this surprising result was erroneous and an artifact
caused by the assumption of conformal flatness.  It was later found
that these findings were indeed wrong and that they were caused by an
error in the derivation of the equations \citep{f99,mw00} and not by
conformal flatness.

It has been argued similarly that the disagreement between numerical
\citep{c94,b00} and post-Newtonian (e.g. \citet{djs00}) values for 
the ISCO of binary black holes could be caused by the assumption of
conformal flatness in the numerical calculations.  However, the more
recent numerical calculations of \citet{ggb01b} achieve much better
agreement with the post-Newtonian results, even though they also
assume conformal flatness (also \citet{b02,dgg02}).  As we discussed
in Section \ref{sec7.3}, it is more likely that the choice of initial
value decomposition, which affects the transverse parts of the
extrinsic curvature (see the discussion in \citet{pct02}), caused the
earlier discrepancies.  The good agreement between the numerical
results of \citet{ggb01b} and post-Newtonian results
\citep{b02,dgg02} may even suggest that the assumption of conformal
flatness is quite adequate for binary black hole models, at least as
long as the spin of the individual black holes is not too large.

\end{appendix}

\end{document}